# A Systematic Review on Affective Computing: Emotion Models, Databases, and Recent Advances

Yan Wang [a], Wei Song [c], Wei Tao [a], Antonio Liotta [d], Dawei Yang [a], Xinlei Li [a], Shuyong Gao [b], Yixuan Sun [a], Weifeng Ge [b], Wei Zhang [b], and Wenqiang Zhang [a,b,*]

[a] Academy for Engineering & Technology, Fudan University, Shanghai 200433; Shanghai Engineering Research Center of AI & Robotics, Shanghai 200433, China; and Engineering Research Center of AI & Robotics, Ministry of Education, Shanghai 200433, China;
yanwang19@fudan.edu.cn (Y.W.); 18110860008@fudan.edu.cn (W.T.); 18110860061@fudan.edu.cn (W.D.);
18110860019@fudan.edu.cn (X.L.); 1609271386@qq.com (Y.S.); wqzhang@fudan.edu.cn (W.Q.);

[b] Shanghai Key Laboratory of Intelligent Information Processing, Fudan University, Shanghai, China;
18110240022@fudan.edu.cn (S.G.); wfge@fudan.edu.cn (W.G.); weizh@fudan.edu.cn (W.Z.);
wqzhang@fudan.edu.cn (W.Q.)

[c] College of Information Technology, Shanghai Ocean University, Shanghai 201306, China;
wsong@shou.edu.cn (W.S.)

[d] Faculty of Computer Science, Free University of Bozen-Bolzano, Italy;
antonio.liotta@unibz.it (A.L.)

[*] Correspondence: wqzhang@fudan.edu.cn; Tel: +86-185-0213-9010

**Abstract:** Affective computing conjoins the research topics of emotion recognition and sentiment analysis, and can be realized with unimodal or multimodal data, consisting primarily of physical information (e.g., text, audio, and visual) and physiological signals (e.g., EEG and ECG). Physical-based affect recognition caters to more researchers due to the availability of multiple public databases, but it is challenging to reveal one's inner emotion hidden purposefully from facial expressions, audio tones, body gestures, etc. Physiological signals can generate more precise and reliable emotional results; yet, the difficulty in acquiring these signals hinders their practical application. Besides, by fusing physical information and physiological signals, useful features of emotional states can be obtained to enhance the performance of affective computing models. While existing reviews focus on one specific aspect of affective computing, we provide a systematical survey of important components: emotion models, databases, and recent advances. Firstly, we introduce two typical emotion models followed by five kinds of commonly used databases for affective computing. Next, we survey and taxonomize state-of-the-art unimodal affect recognition and multimodal affective analysis in terms of their detailed architectures and performances. Finally, we discuss some critical aspects of affective computing and its applications and conclude this review by pointing out some of the most promising future directions, such as the establishment of benchmark database and fusion strategies. The overarching goal of this systematic review is to help academic and industrial researchers understand the recent advances as well as new developments in this fast-paced, high-impact domain.

## 1. Introduction

Affective computing is an umbrella term for human emotion, sentiment, and feelings [1], emotion recognition, and sentiment analysis. Since the concept of affective computing [2] was proposed by Prof. Picard in 1997, it has been guiding computers to identify and express emotions and respond intelligently to human emotions [3]. In many practical applications, it is desired to build a cognitive, intelligent system [4] that can distinguish and understand people's affect, and meanwhile make sensitive and friendly responses promptly [5,6]. For example, in an intelligent vehicle system, the real-time monitoring of the driver's emotional state and the necessary response based on the monitored results can effectively reduce the possibility of accidents [7,8]. In social media, affective computing can avidly help to understand the opinions being expressed on different platforms [9]. Thus, many researchers [3,10] believe that affective computing is the key to promoting and advancing the development of human-centric AI and human intelligence.

Affective computing involves two distinct topics: emotion recognition and sentiment analysis [11–14]. To understand and compute the emotion or sentiment, psychologists proposed two typical theories to model human emotion: discrete emotion model (or categorical emotion model) [15] and dimensional emotion model [16]. The emotion recognition aims to detect the emotional state of human beings (i.e., discrete emotions or dimensional emotions) [17], and mostly focuses on visual emotion recognition (VER) [18], audio/speech emotion recognition (AER/SER) [19], and physiological emotion recognition (PER) [20]. In contrast, sentiment analysis mostly concentrates on textual evaluations and opinion mining [21] on social events, marketing campaigns, and product preferences. The result of sentiment analysis is typically positive, negative, or neutral [10,22]. Considering that one person in a happy mood typically has a positive attitude toward the surrounding environment, emotion recognition and sentiment analysis can be overlapped. For example, a framework based on the context-level inter-modal attention [23] was designed to predict the sentiment (positive or negative) and recognize expressed emotions (anger, disgust, fear, happiness, sadness, or surprise) of an utterance.

Recent advances in affective computing have facilitated the release of public benchmark databases, mainly consisting of unimodal databases (i.e., textual, audio, visual, and physiological databases) and multimodal databases. These commonly used databases have, in turn, motivated the development of machine learning (ML)-based and deep learning (DL)-based affective computing.

A study [24] revealed that human emotions are expressed mainly through facial expressions (55%), voice (38%), and language (7%) in daily human communication. Throughout this review, textual, audio, and visual signals are collectively referred to as physical data. Since people tend to express their ideas and thoughts freely on social media platforms and websites, a large amount of physical affect data can be easily collected. Based on these data, many researchers pay attention to identifying subtle emotions expressed either explicitly or implicitly [25–27]. However, physical-based affect recognition may be ineffective since humans may involuntarily or deliberately conceal their real emotions (so-called social masking) [20]. In contrast, physiological signals (e.g., EEG and ECG) are not subject to these constraints as spontaneous physiological activities associated with emotions are hardly changed by oneself. Thus, EEG-based or ECG-based emotion recognition can generate more objective predictions in real time and provide reliable features of emotional states [29, 30].

Human affect is a complex psychological and physiological phenomenon [30]. As human beings naturally communicate and express emotion or sentiment through multimodal information, more researches focus on multi-physical modality fusion for affective analysis [31]. For example, in a conversation scenario, a person's emotional state can be demonstrated by the words of speech, the tones of voice, facial expressions and emotion-related body gestures [32]. Textual, auditory and visual information together provide more information than they do individually [33], just like the brain validating events relies on several sensory input sources. With the rapid progress of physical-touching mechanisms or intrusive techniques such as low-cost wearable sensors, some emotion recognition methods are based on multimodal physiological signals (e.g., EEG, ECG, EMG and EDA). By integrating physiological modalities with physical modalities, physical-physiological affective computing can detect and recognize subtle sentiments and complex emotions [34,35]. It is worth mentioning that the suitable selection of the unimodal emotion data and multimodal fusion strategies [36] are the two key components of multimodal affective analysis systems, which often outperform the unimodal emotion recognition systems [37]. Therefore, this review not only provides a summary of the multimodal affective analysis, but also introduces an overview of unimodal affect recognition.

Although there exist many survey papers about affective computing, most of them focus on physical-based affect recognition. For facial expression recognition (FER), existing studies have provided a brief review of FER [38], DL-based FER systems [39], facial micro-expressions analysis (FMEA) [40], and 3D FER [41]. Besides, the work [42] discussed the research results of the sentiment analysis using transfer learning algorithms, whereas authors [43] surveyed the DL-based methods for SER. However, there are just a few review papers related to physiological-based emotion recognition and physical-physiological fusion for affective analysis in recent years. For example, Jiang et al. [17] discussed multimodal databases, feature extraction based on physical signals or EEG, and multimodal fusion strategies and recognition methods.

Several issues have not been thoroughly addressed in previous reviews: 1) Existing reviews take a specialist view and lack a broader perspective, for instance when classifying the various methods and advances, some reviews do not consider DL-based affect recognition or multimodal affective analysis; 2) Existing reviews do not provide a clear picture about the performance of state-of-the-art methods and the implications of their recognition ability. As the most important contribution of our review, we aim to cover different aspects of affective computing by introducing a series of research methods and results as well as discussions and future works.

To sum up, the major contributions of this paper are multi-fold:

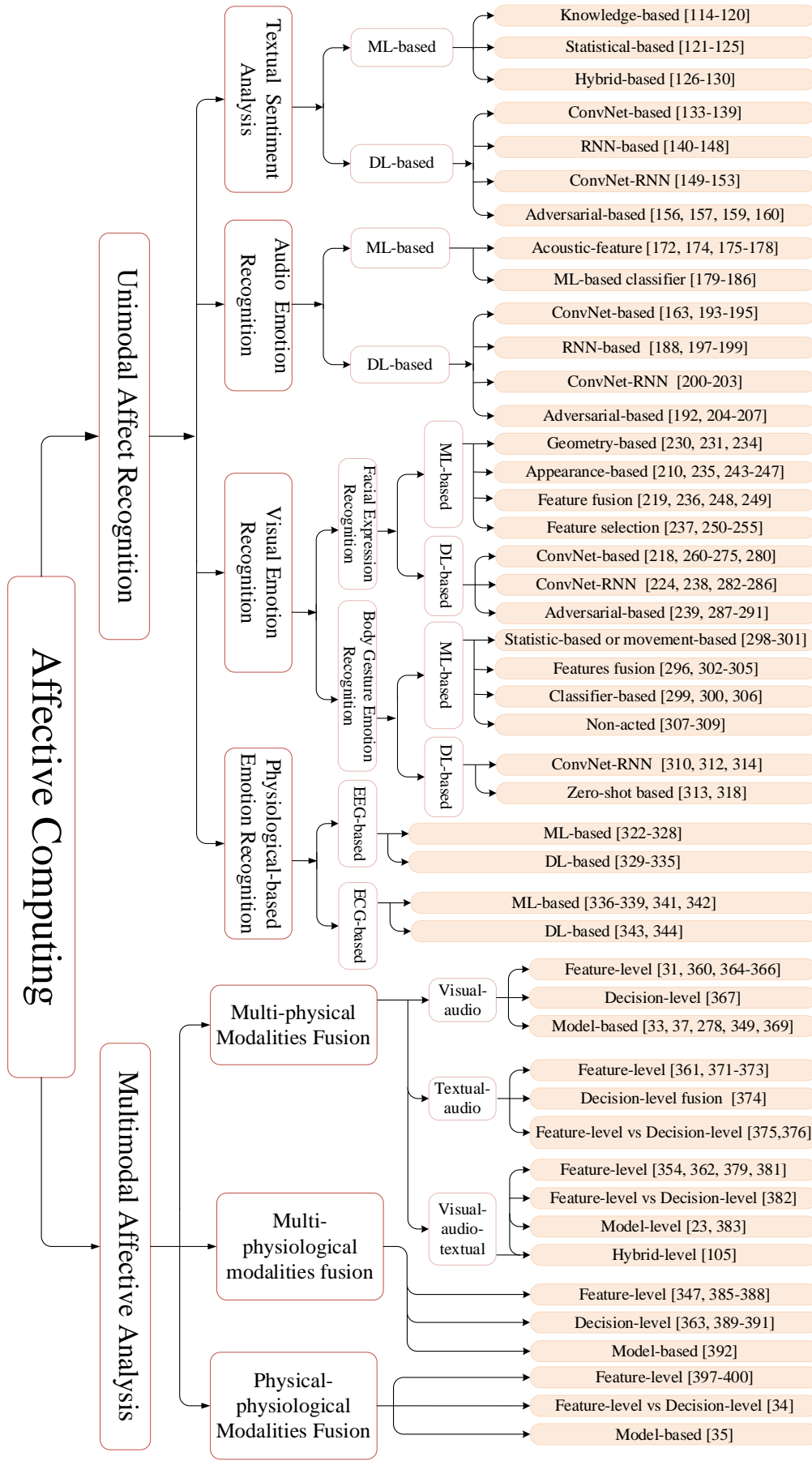

**Fig. 1.** Taxonomy of affective computing with representative examples.

1) To the best of our knowledge, this is the first review that categorizes affective computing into two broad classes, i.e., unimodal affect recognition and multimodal affective analysis, and further taxonomizes them based on the data modalities.
2) In the retrospect of 20 review papers released between 2017 and 2020, we present a systematic review of more than 380 research papers published in the past 20 years in leading conferences and journals. This leads to a vast body of works, as taxonomized in Fig. 1, which will help the reader navigate through this complex area.
3) We provide a comprehensive taxonomy of state-of-the-art (SOTA) affective computing methods from the perspective of either ML-based methods or DL-based techniques and consider how the different affective modalities are used to analyze and recognize affect.
4) Benchmark databases for affective computing are categorized by four modalities and video-physiological modalities. The key characteristics and availability of these databases are summarized. Based on publicly used databases, we provide a comparative summary of the properties and quantitative performance of some representative methods.
5) Finally, we discuss the effects of unimodal, multimodal, models as well as some potential factors on affective computing and some real-life applications of that, and further indicate future researches on emotion recognition and sentiment analysis.

The paper is organized as followed. Section 2 introduce the existing review works, pinpointing useful works and better identifying the contributions of this paper. Section 3 surveys two kinds of emotion models, the discrete and the dimensional one. Then Section 4 discusses in detail four kinds of databases, which are commonly used to train and test affective computing algorithms. We then provide an extensive review of the most recent advances in affective computing, including unimodal affect recognition in Section 5 and multimodal affective analysis in Section 6. Finally, discussions are presented in Section 7, while conclusions and new developments are stated in Section 8. Table 1 lists the main acronyms used herein for the reader's reference.

**Table 1**. Main acronyms.

| Acronym | Full Form | Acronym | Full Form |
|---|---|---|---|
| TSA | Textual Sentiment Analysis | SER | Speech Emotion Recognition |
| FER | Facial Expression Recognition | FMER | Facial Micro-Expression Recognition |
| 4D/3D FER | 4D/3D Facial Expression Recognition | EBGR | Emotional Body Gesture Recognition |
| EEG | Electroencephalogram | ECG | Electrocardiography |
| EMG | Electromyography | EDA | Electro-Dermal Activity |
| ML | Machine Learning | DL | Deep Learning |
| GMM | Gaussian Mixture Model | MLP | Multi-Layer Perceptron |
| NB | Naive Bayesian | LSTM | Long- Short-Term Memory |
| LDA | Linear Discriminant Analysis | DCNN | Deep Convolutional Neural Network |
| DT | Decision Tree | CNN | Convolutional Neural Network |
| KNN | K-Nearest Neighbors | RNN | Recurrent Neural Network |
| HMM | Hidden Markov Model | GRU | Gated Recurrent Unit |
| ANN | Artificial Neural Network | AE | Auto-encoder |
| PCA | Principal Component Analysis | GAN | Generative Adversarial Network |
| MLP | Multi-layer Perceptron | VGG | Visual Geometry Group |
| SVM | Support Vector Machine | DBN | Deep Belief Network |
| RBM | Restricted Boltzmann Machine | HAN | Hierarchical Attention Network |
| RBF | Radial Basis Function | ResNet | Residual Networks |
| FC | Full-connected | GAP | Global Average Pooling |
| MKL | Multiple Kernel Learning | AUs | Action Units |
| RF | Random Forest | AAM | Active Appearance Model |
| ICA | Independent Component Analysis | LFPC | Logarithmic Frequency Power Coefficient |
| BoW | Bag-of-Words | ROIs | Regions of Interest |
| LBP-TOP | Local Binary Pattern from Three Orthogonal Planes | MFCC | MEL Frequency Cepstrum Coefficient |

## 2. Related works

In this section, we introduce the most relevant reviews related to affective computing published between 2017 and 2020 in the perspective of affect modalities (i.e., physical modality, physiological modality, and physical-physiological modality). Table 2 shows an overview of these latest reviews, compared with our survey. The comparison includes publication year (Year), emotion model, database (DB), modality, multimodal fusion, method, and quantitative evaluation (QE). According to Table 2, our review systematically covers all aspects of affective computing.

5**Table 2.** Overview of the most relevant reviews related to affective computing published between 2017 and 2020 and our proposed review.

| Author | Year | [a] Emotion Model | | DB | [b] Modality | | | | [c] Multimodal Fusion | | | | | Method | | QE |
|---|---|---|---|---|---|---|---|---|---|---|---|---|---|---|---|---|
| | | Dis | Dim | | T | A | V | P | VA | TA | VAT | M-P | P-P | ML | DL | |
| *Reviews on physical-based Affect Recognition* | | | | | | | | | | | | | | | | |
| Ko [38] | 2018 | ✗ | ✗ | ✓ | ✗ | ✗ | ✓ | ✗ | ✗ | ✗ | ✗ | ✗ | ✗ | ✓ | ✓ | ✓ |
| Patel et al. [44] | 2020 | ✗ | ✗ | ✓ | ✗ | ✗ | ✓ | ✗ | ✗ | ✗ | ✗ | ✗ | ✗ | ✓ | ✓ | ✓ |
| Li et al. [39] | 2020 | ✗ | ✗ | ✓ | ✗ | ✗ | ✓ | ✗ | ✗ | ✗ | ✗ | ✗ | ✗ | ✗ | ✓ | ✓ |
| Alexandre et al. [41] | 2020 | ✗ | ✗ | ✓ | ✗ | ✗ | ✓ | ✗ | ✗ | ✗ | ✗ | ✗ | ✗ | ✓ | ✓ | ✓ |
| Merghani et al. [40] | 2018 | ✗ | ✗ | ✓ | ✗ | ✗ | ✓ | ✗ | ✗ | ✗ | ✗ | ✗ | ✗ | ✓ | ✓ | ✓ |
| Noroozi et al. [45] | 2018 | ✓ | ✓ | ✓ | ✗ | ✓ | ✓ | ✗ | ✓ | ✗ | ✗ | ✗ | ✗ | ✓ | ✓ | ✗ |
| Liu et al. [42] | 2019 | ✗ | ✗ | ✓ | ✓ | ✗ | ✗ | ✗ | ✗ | ✗ | ✗ | ✗ | ✗ | ✓ | ✓ | ✓ |
| Poria et al. [46] | 2019 | ✓ | ✓ | ✓ | ✓ | ✗ | ✗ | ✗ | ✗ | ✗ | ✗ | ✗ | ✗ | ✗ | ✓ | ✓ |
| Yue et al. [47] | 2019 | ✓ | ✓ | ✓ | ✓ | ✗ | ✗ | ✗ | ✗ | ✗ | ✗ | ✗ | ✗ | ✓ | ✓ | ✗ |
| Khalil et al. [43] | 2019 | ✗ | ✗ | ✓ | ✓ | ✗ | ✗ | ✗ | ✗ | ✗ | ✗ | ✗ | ✗ | ✗ | ✓ | ✓ |
| Wang et al. [48] | 2020 | ✓ | ✓ | ✓ | ✓ | ✗ | ✗ | ✗ | ✗ | ✗ | ✗ | ✗ | ✗ | ✓ | ✗ | ✗ |
| Han et al. [49] | 2019 | ✗ | ✗ | ✗ | ✓ | ✓ | ✓ | ✗ | ✗ | ✗ | ✗ | ✗ | ✗ | ✗ | ✓ | ✗ |
| Erik et al. [12] | 2017 | ✓ | ✓ | ✓ | ✓ | ✓ | ✓ | ✗ | ✓ | ✓ | ✓ | ✗ | ✗ | ✓ | ✓ | ✓ |
| *Reviews on physiological-based Emotion Recognition* | | | | | | | | | | | | | | | | |
| Bota et al. [50] | 2019 | ✓ | ✓ | ✓ | ✗ | ✗ | ✗ | ✓ | ✗ | ✗ | ✗ | ✗ | ✗ | ✓ | ✓ | ✓ |
| [1] G-M et al. [51] | 2019 | ✓ | ✓ | ✗ | ✗ | ✗ | ✗ | ✓ | ✗ | ✗ | ✗ | ✗ | ✗ | ✓ | ✗ | ✓ |
| [2] Ala. and Fon. [29] | 2019 | ✓ | ✓ | ✗ | ✗ | ✗ | ✗ | ✓ | ✗ | ✗ | ✗ | ✗ | ✗ | ✓ | ✗ | ✓ |
| *Reviews on physical-physiological fusion for affective analysis* | | | | | | | | | | | | | | | | |
| Rouast et al. [13] | 2019 | ✓ | ✓ | ✓ | ✗ | ✓ | ✓ | ✓ | ✓ | ✗ | ✗ | ✓ | ✗ | ✓ | ✓ | ✓ |
| Zhang et al. [20] | 2020 | ✓ | ✓ | ✓ | ✓ | ✓ | ✓ | ✓ | ✓ | ✗ | ✗ | ✓ | ✗ | ✓ | ✓ | ✓ |
| Jiang et al. [17] | 2020 | ✗ | ✗ | ✓ | ✓ | ✓ | ✓ | ✓ | ✓ | ✗ | ✓ | ✗ | ✓ | ✓ | ✓ | ✓ |
| Shoumy et al. [14] | 2020 | ✓ | ✓ | ✓ | ✓ | ✓ | ✓ | ✓ | ✓ | ✗ | ✓ | ✓ | ✓ | ✓ | ✓ | ✓ |
| *Proposed review* | | | | | | | | | | | | | | | | |
| Our proposed | 2021 | ✓ | ✓ | ✓ | ✓ | ✓ | ✓ | ✓ | ✓ | ✓ | ✓ | ✓ | ✓ | ✓ | ✓ | ✓ |

[a] Emotion Model: Dis=Discrete, Dim=Dimensional;

[1] G-M=Garcia-Martinez. [2] Ala. and Fon.=Alarcão and Fonseca;

[b] Modality: V=Visual (Facial expression, Body gesture), A=Audio (Speech), T=Textual, and P= Physiological (e.g., EEG, ECG, and EMG);

[c] Multimodal Fusion: VA=Visual-Audio, TA=Textual-Audio, VAT=Visual-Audio-Textual, M-P=Multi-Physiological, P-P=Physical-Physiological.

*2.1 Reviews on physical-based affect recognition*

The existing researches on physical-based affect recognition mainly used visual, textual, and audio modalities [12]. For visual modality, most studies surveyed FER [38,39,41,44] and FMEA [40], and 3D FER [41]. Besides, Noroozi et al. [45] reviewed representation learning and emotion recognition from the body gestures followed by multimodal emotion recognition on the basis of speech or face and body gestures. For textual modality, the works on sentiment analysis and emotion recognition [42,43,46–48] can be completed according to the implicit emotions in the conversation. Han et al. [49] presented DL-based adversarial training using three kinds of physical signals, aiming to address various challenges associated with emotional AI systems. In contrast, Erik et al. [12] considered multimodal fusion and provided a critical analysis of potential performance improvements with multimodal affect recognition under different fusion categories, compared to unimodal analysis.

All reviews on physical-based affect recognition listed in Table 2 have reviewed DL-based methods, which indicates a clear development trend in the application of deep learning for this domain. However, the existing works have not fully involved the latest research advances and achievements in DL-based affective computing, which is one of the objectives of our review.

*2.2 Reviews on physiological-based emotion recognition*

The emerging development of physiological emotion recognition has been made possible by the utilization of embedded devices for the acquisition of physiological signals. In 2019, Bota et al. [50] reviewed ML-based emotion recognition using different physiological signals, along with the key theoretical concepts and backgrounds, methods, and future developments. Garcia-Martinez et al. [51]

reviewed nonlinear EEG-based emotion recognition and identified some nonlinear indexes in future research. Alarcão and Fonseca [29] also reviewed works about EEG-based emotion recognition from 2009 to 2016, but focused on subjects, feature representation, classification, and their performances.

All reviews on physiological-based emotion recognition listed in Table 2 have reviewed the ML-based methods in discrete and dimensional emotion space, but only one work involved a few DL-based methods. In this review, we surveyed ML-based and DL-based works which have contributed to the advance of physiological-based emotion recognition.

*2.3 Reviews on physical-physiological fusion for affective analysis*

There is a clear trend of using both physical and physiological signals for affective analysis. In 2019, Rouast et al. [13] reviewed 233 DL-based human affective recognition methods that use audio-visual and physiological signals to learn spatial, temporal, and joint feature representations. In 2020, Zhang et al. [20] introduced different feature extraction, feature reduction, and ML-based classifiers in terms of the standard pipeline for multi-channel EEG emotion recognition, and discussed the recent advances of multimodal emotion recognition based on ML or DL techniques. Jiang et al. [17] summarized the current development of multimodal databases, the feature extraction based on EEG, visual, audio and text information, multimodal fusion strategies, and recognition methods according to the pipeline of the real-time emotion health surveillance system. Shoumy et al. [14] reviewed different frameworks and lasted techniques using textual, audio, visual and physiological signals, and extensive analysis of their performances. In the end, various applications of affective analysis were discussed followed by their trends and future works.

All these recent works listed in Table 2 have reviewed general aspects related to affective computing including emotion models, unimodal affect recognition and multimodal fusion for affective analysis as well as ML-based and DL-based models. However, the existing works have not fully elaborated their comparisons in unimodal and multimodal affective analysis, which is one of the objectives of our review.

**3. Emotion models**

The definition of the emotion or affect is essential to establish a criterion for affective computing. The basic concept of emotions was first introduced by Ekman [52] in the 1970s. Although psychologists attempt to classify emotions in different ways in the multidisciplinary fields of neuroscience, philosophy, and computer science [53], there are no unanimously accepted emotion models. However, there are two types of generic emotion models in affective computing, namely discrete emotion model [52] and dimensional emotion model (or continuous emotion model) [16,54].

*3.1 Discrete emotion model*

The discrete emotion model, also called as categorical emotion model, defines emotions into limited categories. Two widely used discrete emotion models are Ekman's six basic emotions [52] and Plutchik's emotional wheel model [55], as shown in Fig.2 (a) and Fig. 2 (b), respectively.

Ekman's basic emotion model and its variants [56,57] are widely accepted by the emotion recognition community [58,59]. Six basic emotions typically include anger, disgust, fear, happy, sad, and surprise. They were derived with the following criteria [52]: 1) Basic emotions must come from human instinct; 2) People can produce the same basic emotions when facing the same situation; 3) People express the same basic emotions under the same semantics; 4) These basic emotions must have the same pattern of expression for all people. The development of Ekman's basic emotion model is based on the hypothesis that human emotions are shared across races and cultures. However, different cultural backgrounds may have different interpretations of basic emotions, and different basic emotions can be mixed to produce complex or compound emotions [15].

In contrast, Plutchik's wheel model [55] involves eight basic emotions (i.e., joy, trust, fear, surprise, sadness, anticipation, anger, and disgust) and the way of how these are related to one another (Fig. 2 (b)). For example, joy and sadness are opposites, and anticipation can easily develop into vigilance. This wheel model is also referred to as the componential model, where the stronger emotions occupy the centre, while the weaker emotions occupy the extremes, depending on their relative intensity levels. These discrete emotions can be generally categorized into three kinds of polarity (positive, negative, and neutral), which are often used for sentiment analysis. To describe fine-grained sentiments, ambivalent sentiment handling [60] is proposed to analyze the multi-level sentiment and improve the performance of binary classification.

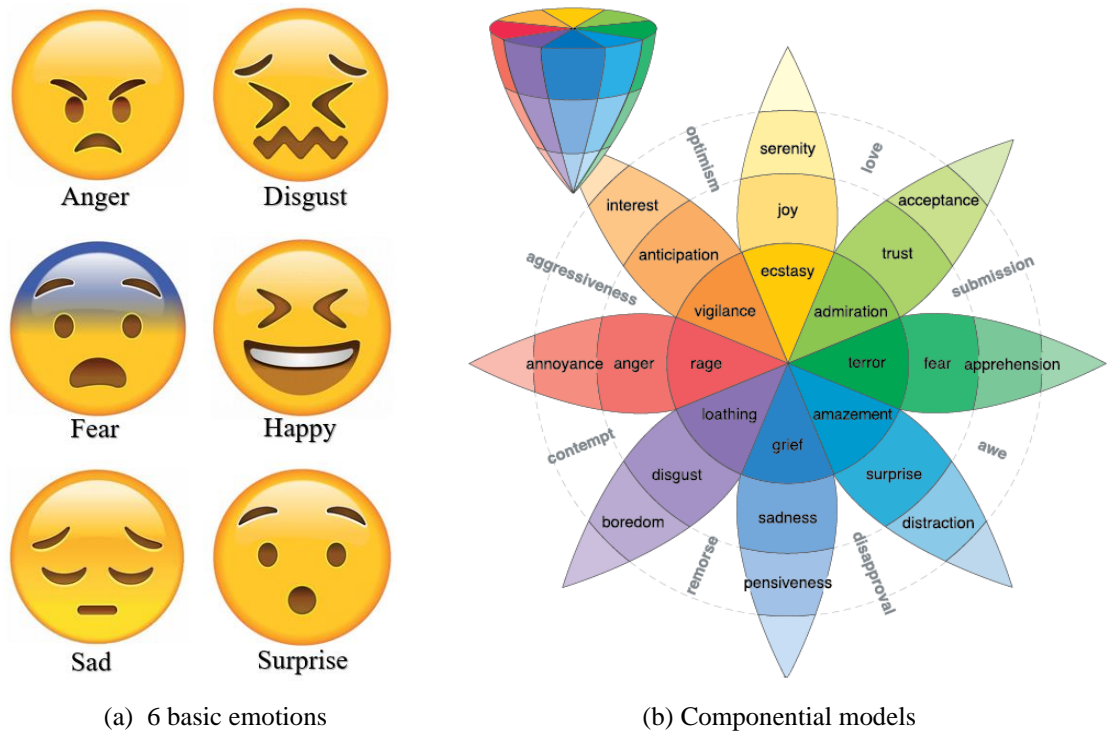

(a) 6 basic emotions  (b) Componential models

**Fig. 2.** Two discrete emotion models for affective computing. (a) 6 basic emotion models [15] shown in emoji types and (b) componential models (Plutchik's emotional wheel model) [55].

*3.2 Dimensional emotion model*

To overcome the challenges confronted by the discrete emotion models, many researchers have adopted the concept of a continuous multi-dimensional model. One of the most recognized models is the Pleasure-Arousal-Dominance (PAD) [16], as shown in Fig. 3 (a).

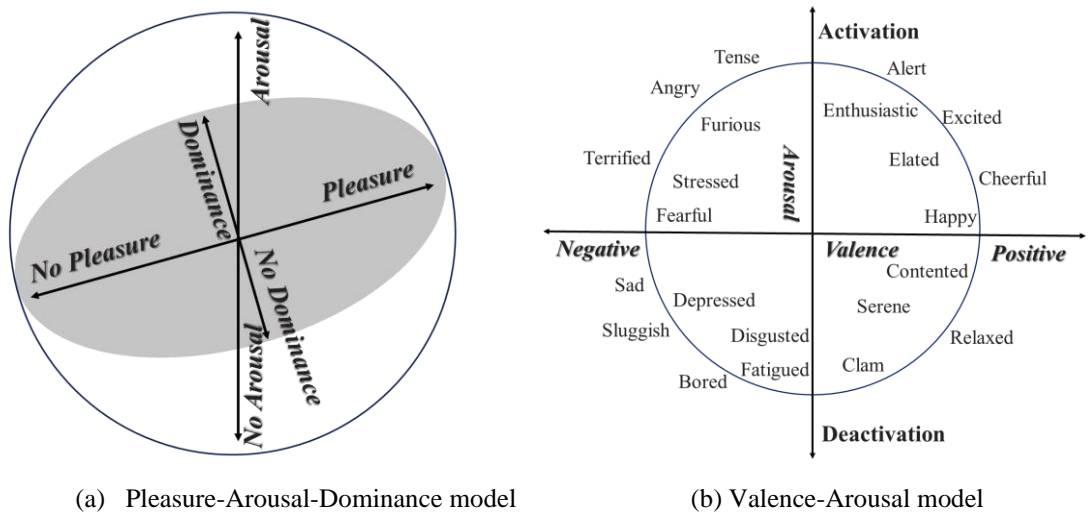

(a) Pleasure-Arousal-Dominance model  (b) Valence-Arousal model

**Fig. 3.** Dimensional emotion models. (a) Pleasure-Arousal-Dominance (PAD) model reproduced based on [61] and (b) Valence-Arousal (V-A) model reproduced based on [54].

Similar to Mehrabian´s three-dimensional space theory of emotion [61], the PAD model has three-dimensional spaces: 1) Pleasure (Valence) dimension, representing the magnitude of human joy from distress extreme to ecstasies; 2) Arousal (Activation) dimension, measuring physiological activity and psychological alertness level; 3) Dominance (Attention) dimension, expressing the feeling of influencing the surrounding environment and other people, or of being influenced by the surrounding environment and others.

Since two dimensions of Pleasure and Arousal in the PAD model could represent the vast majority of different emotions [62], Russell [54] proposed a Valence-Arousal based circumplex model to represent complex emotions. This model defines a continuous, bi-dimensional emotion space model with the axes of Valence (the degree of pleasantness or unpleasantness) and Arousal (the degree of activation or deactivation), as shown in Fig. 3 (b). The circumplex model consists of four quadrants. The first quadrant, activation arousal with positive valence, shows the feelings associated with happy emotions; And the third quadrant, with low arousal and negative valence, is associated with sad emotions. The second quadrant shows angry emotions within high arousal and negative valence; And the fourth quadrant shows calm emotion within low arousal and positive valence [63].

## 4. Databases for affective computing

Databases for affective computing can be classified into textual, speech/audio, visual, physiological, and multimodal databases according to the data modalities. The properties of these databases have a far-reaching influence on the model design and network architecture for affective computing.

*4.1 Textual databases*

Databases for TSA consist of the text data in different granularities (e.g., word, sentence, and document), labelled with emotion or sentiment tags (positive, negative, neutral, emphatic, general, sad, happy, etc). The earliest textual sentiment database is **Multi-domain sentiment (MDS)** [64,65]**,** which contains more than 100,000 sentences of product reviews acquired from Amazon.com. These sentences are labelled with both two sentiment categories (positive and negative) and five sentiment categories (strong positive, weak positive, neutral, weak negative, strong negative).

Another widely-used large database for binary sentiment classification is **IMDB** [66]. It provides 25,000 highly polar movie reviews for training and 25,000 for testing. **Stanford sentiment treebank (SST)** [67] is the semantic lexical database annotated by Stanford University. It includes fine-grained emotional labels of 215,154 phrases in a parse tree of 11,855 sentences, and it is the first corpus with fully labelled parse trees.

*4.2 Speech/Audio databases*

Speech databases can be divided into two types: non-spontaneous (simulated and induced) and spontaneous. In the early stage, non-spontaneous speech databases were mainly generated from professional actors' performances. Such performance-based databases are regarded as reliable ones because they can perform well-known emotional characteristics in professional ways. **Berlin Database of Emotional Speech (Emo-DB)** [68] contains about 500 utterances spoken by 10 actors (5 men and 5 women) in a happy, angry, anxious, fearful, bored and disgusting way. However, these non-spontaneous emotions can be exaggerated a little more than the real emotions. To narrow this gap, spontaneous speech databases have been developed recently. The **Belfast Induced Natural Emotion (Belfast)** [69] was recorded from 40 subjects (aged between 18 and 69, 20 men and 20 women) at Queen University in Northern Ireland, UK. Each subject took part in five tests, each of which contains short video recordings (5 to 60 seconds in length) with stereo sound, and related to one of the five emotional tendencies: anger, sadness, happiness, fear, and neutrality.

*4.3 Visual databases*

Visual databases can also be divided into two categories: facial expression databases and body gesture emotion databases.

4.3.1   *Facial expression databases*

Table 3 provides an overview of facial expression databases, including the main reference (access), year, samples, subject, and expression category. The early FER databases are derived from the emotions purposely performed by subjects in the laboratory (In-the-Lab). For example, **JAFFE** [70] released in 1998, includes 213 images of 7 facial expressions, posed by 10 Japanese female models. To construct the **extended Cohn-Kanade (CK+)** [71], an extension of **CK** [72], subjects were instructed to perform 7 facial expressions. The facial expression images were recorded and analyzed to provide protocols and baseline results for facial feature tracking, action units (AUs), and emotion recognition. Different from **CK+, MMI** [73] consists of onset-apex-offset sequences. **Oulu-CASIA NIR-VIS** (**Oulu-CASIA**) [74] released in 2011, includes 2,880 image sequences captured with one of two kinds of imaging systems under three kinds of illumination conditions.

**TABLE 3.** Overview of facial expression databases
SBE = Seven Basic Emotions (anger, disgust, fear, happy, sad, surprise, and neutral).

| Database | Year | Samples | Subject | Expression Category |
|---|---|---|---|---|
| **In-the-Lab** | | | | |
| [1] JAFFE [70] | 1998 | 219 images | 10 | SBE |
| [2] CK [72] | 2000 | 1,917 images | 210 | SBE plus Contempt |
| [2] CK+ [71] | 2010 | 593 sequences | 123 | SBE plus Contempt |
| [3] MMI [73] | 2010 | 740 images, 2,900 videos | 25 | SBE |
| [5] Oulu-CASIA [74] | 2011 | 6 kinds of 480 sequences | 80 | SBE |
| [6] BU-3DFE [75] | 2006 | 2,500 3D images | 100 | SBE |
| [6] BU-4DFE [76] | 2008 | 606 3D sequences | 101 | SBE |
| [6] BP4D [77] | 2014 | 328 3D + 2D sequences | 41 | Six basic expressions plus Embarrassed |
| [7] 4DFAB [78] | 2018 | 1.8 million+ 3D images | 180 | Six basic expressions |
| [8] SMIC [79] | 2013 | 164 sequences | 16 | Positive, Negative, Surprise |
| [9] CASME II [80] | 2014 | 255 sequences | 35 | Six basic expressions plus Others |
| SAMM [81] | 2018 | 159 sequences | 32 | Six basic expressions plus Others |
| **In-the-Wild** | | | | |
| [10] FER2013 [82] | 2013 | 35,887 gray images | / | SBE |
| [11] SFEW 2.0 [83] | 2015 | 1694 images | | Six basic expressions |
| [12] EmotioNet [84] | 2016 | 1,000,000 images | / | Compound expressions |
| [13] ExpW [85] | 2016 | 91,793 images | / | SBE |
| [14] AffectNet [86] | 2017 | 450,000 images | / | SBE plus Contempt Valence and Arousal |
| [15] RAF-DB [87] | 2017 | 29,672 images | / | SBE Compound expressions |
| [16] DFEW [88] | 2020 | 12059 clips | / | SBE |

[1] kasrl.org/jaffe; [2] jeffcohn.net/Resources/; [3] mmifacedb.eu/; [4] socsci.ru.nl:8180/RaFD2/RaFD; [5] oulu.fi/cmvs/node/41316;

[6] cs.binghamton.edu/~lijun/Research/3DFE/3DFE_Analysis; [7] eprints.mdx.ac.uk/24259/; [8] oulu.fi/cmvs/node/41319;

[9] fu.psych.ac.cn/CASME/casme2-en.php; [10] kaggle.com/c/challenges-in-representation-learning-facial-expression-recognition-challenge;

[11] cs.anu.edu.au/few/emotiw2015; [12] cbcsl.ece.ohio-state.edu/dbform_emotionet; [13] mmlab.ie.cuhk.edu.hk/projects/socialrelation/;

[14] mohammadmahoor.com/affectnet/; [15] whdeng.cn/raf/model; [16] dfew-database.github.io.

There are various 3D/4D databases designed for multi-view and multi-pose FER. **Binghamton University 3D Facial Expression (BU-3DFE)** [75] contains 606 facial expression sequences captured from 100 people with one of six facial expressions. **BU-4DFE** [76] is developed based on the dynamic 3D space, which includes 606 high-resolution 3D facial expression sequences. **BP4D** [77] released in 2014, is a well-annotated 3D video database consisting of spontaneous facial expressions, elicited from 41 participants (23 women, 18 men), by well-validated emotion inductions. **4DFAB** [78] released in 2018, contains at least 1,800,000 dynamic high-resolution 3D faces captured from 180 subjects in four different sessions spanning.

Similar to **MMI** [73], all micro-expression databases consist of onset-apex-offset sequences. **Spontaneous Micro-expression (SMIC)** [79] contains 164 micro-expression video clips elicited from 16 participants. **CASME II** [80] has videos with relatively high temporal and spatial resolution. The participants' facial expressions have been elicited in a well-controlled laboratory environment and proper illumination. **Spontaneous Micro-Facial Movement (SAMM)** [81] is a currently-public database that contains sequences with the highest resolution, and its participants are from diverse ethnicities and the widest range of ages.

Acted facial expression databases are often constructed in a specific environment. Another way to build the databases is collecting facial expression images/videos from the Internet, which we refer to as In-the-Wild. **FER2013** [82] is a firstly public large-scale and unconstrained database that contains 35,887 grey images with $48 \times 48$ pixels, collected automatically through the Google image search API. **Static Facial Expressions In-the-Wild (SFEW 2.0)** [83] is divided into three sets, including Train (891 images) and Val (431 images), labelled as one of six basic expressions (anger, disgust, fear, happiness, sadness and surprise), as well as the neutral and Test (372 images) without expression labels. **EmotioNet** [84] consists of one million images with 950,000 automatically annotated AUs and 25,000 manually annotated AUs. **Expression in-the-Wild (ExpW)** [85] contains 91,793 facial images, manually annotated as one of seven basic facial expressions. Non-face images were removed in the annotation process. **AffectNet** [86] contains over 1,000,000 facial images, of which 450,000 images are manually annotated as one of eight discrete expressions (six basic expressions plus neutral and contempt), and the dimensional intensity of valence and arousal. **Real-world Affective Face Database (RAF-DB)** [87] contains 29,672 highly diverse facial images downloaded from the Internet, with manually crowd-sourced annotations (seven basic and eleven compound emotion labels). **Dynamic Facial Expression in the Wild (DFEW)** [88] consists of over 16,000 video clips segmented from thousands of movies with various themes. Professional crowdsourcing is applied to these clips, and 12,059 clips have been selected and labelled with one of 7 expressions (six basic expressions plus neutral).

4.3.2 *Body gesture emotion databases*

Although studies in emotion recognition focused mainly on facial expressions, a growing number of researchers in affective neuroscience demonstrates the importance of the full body for unconscious emotion recognition. In general, bodily expressive cues are easier to be perceived than subtle changes in the face. To capture natural body movements, body gesture emotion databases contain a corpus of video sequences collected from either real life or movies. Table 4 provides an overview of body gesture emotion databases, as described next.

**EmoTV** [89] contains the interview video sequences from French TV channels. It has multiple types of annotations but is not publicly available. To our best knowledge, **FAce and BOdy database (FABO)** [90] is the first publicly available bimodal database containing both face and body gesture. The recordings for each subject take more than one-hour store rich information of various affect statements. Moreover, **THEATER Corpus** [91] consists of sections from two movie versions which are coded with eight affective states corresponding to the eight corners of PAD space [92]. **GEneva Multimodal Emotion Portrayals (GEMEP)** [93] is a database of body postures and gestures collected from the perspectives of both an interlocutor and an observer. The GEMEP database is one of the few databases that have frame-by-frame AU labels [94]. **Emotional body expression in daily actions database (EMILYA)** [95] collected body gestures in daily motions. Participants were trained to be aware of using their bodies to express emotions through actions. **EMILYA** includes not only videos of facial and bodily emotional expressions, but also 3D data of the whole-body movement.

**TABLE 4.** Overview of body gesture emotion databases.

| Database | Year | Modality | Sources | Sub. | Category |
|---|---|---|---|---|---|
| EmoTV [89] | 2005 | Face, Body Gaze | 51 images | / | 14 emotions |
| [1] FABO [90] | 2006 | Face, Body | 1900 videos | 23 | 10 emotions |
| THEATRE | 2009 | Body | / | / | 8 emotions |
| [2] GEMEP [93] | 2010 | Face, Body | 7000+ | 10 | 18 emotions |
| EMILYA [95] | 2014 | Face, Body | / | 11 | 8 emotions |

[1] cl.cam.ac.uk/~hg410/fabo/; [2] affective-sciences.org/gemep.

4.4 *Physiological databases*

Physiological signals ((e.g., EEG, RESP, and ECG) are not affected by social masking compared to textual, audio, and visual emotion signals, and thus are more objective and reliable for emotion recognition. Table 5 provides an overview of physiological-based databases, as described next.

**TABLE 5.** Overview of physiological-based databases.

| Database | Year | Components | | | | | | | Subjects | Emotion Categories |
|---|---|---|---|---|---|---|---|---|---|---|
| | | EEG | EOG | EMG | GSR | RESP | ECG | Others | | |
| [1] DEAP [96] | 2012 | ■ | ■ | ■ | ■ | ■ | / | Plethysmograph | 32 | Valence and Arousal |
| [2] SEED [97,98] | 2015 | ■ | / | / | / | / | / | / | 15 | Positive, Neutral, and Negative |
| DSdRD [7] | 2005 | / | / | ■ | ■ | / | ■ | / | 24 | Low, medium, and high stress |
| [3] AMIGOS [99] | 2017 | ■ | / | / | ■ | / | ■ | Frontal HD video, both RGB & depth full body videos | 40 | Valence and Arousal |
| WESAD [100] | 2018 | / | / | ■ | / | ■ | ■ | EDA, blood volume pulse & temperature, AAC | 15 | Neutral, Stress, Amusement |

[1] eecs.qmul.ac.uk/mmv/databases/deap/; [2] bcmi.sjtu.edu.cn/~seed/; [3] eecs.qmul.ac.uk/mmv/databases/amigos/.

**DEAP** [96] comprises a 32-channel EEG, a 4-channel EOG, a 4-channel EMG, RESP, plethysmograph, Galvanic Skin Response (GSR) and body temperature, collected from 32 subjects. Immediately after watching each video, subjects were required to rate their truly-felt emotion from five dimensions: valence, arousal, dominance, liking and familiarity. **SEED** [97,98] contains EGG recordings from 15 subjects. In their study, participants were asked to experience three EEG recording sessions, with an interval of two weeks between two successive recording sessions. Within each session, each subject was exposed to the same sequence of fifteen movie excerpts, each one approximately four-minute-long, to induce three kinds of emotions: positive, neutral, and negative.

Some databases are task-driven. For example, **Detecting Stress during Real-World Driving Tasks (DSdRD)** [7] is used to determine the relative stress levels of drivers. It contains various signals from 24 volunteers while having a rest for at least 50 minutes after their driving tasks. These volunteers are asked to fill out questionnaires which are used to map their state into low, medium, and high-stress levels. **AMIGOS** [99] was designed to collect participants' emotions in two social contexts: individual and group. AMIGOS was constructed in 2 experimental settings: 1) 40 participants watch 16 short emotional videos; 2) they watch 4 long videos, including a mix of lone and group sessions. These emotions were annotated with self-assessment of affective levels and external assessment of valence and arousal. Wearable devices help to bridge the gap between lab studies and real-life emotions. **Wearable Stress and Affect Detection (WESAD)** [100] is built for stress detection, providing multimodal, high-quality data, including three different affective states (neutral, stress, amusement).

*4.5 Multimodal databases*

In our daily life, people express and/or understand emotions through multimodal signals. Multimodal databases can be mainly divided into two types: multi-physical and physical-physiological databases. Table 6 provides an overview of multimodal databases, as described next.

**Table 6.** Overview of multimodal databases. Five basic sentiments = Strongly Positive, Weakly Positive, Neutral, Strongly Negative and Weakly Negative.

| Name | Year | Components | | | | Subjects | Type | Emotion Categories |
| --- | --- | --- | --- | --- | --- | --- | --- | --- |
| | | Text | Speech | Visual | Psych. | | | |
| [1] IEMOCAP [101] | 2008 | ■ | ■ | ■ | / | 10 | Acted | Happiness, Anger, Sad, Frustration and Neutral Activation-Valence-Dominance |
| [2] CreativeIT [102,103] | 2010 | ■ | ■ | ■ | / | 16 | Induced | Activation-Valence-Dominance |
| [3] HOW [104] | 2011 | ■ | ■ | ■ | / | / | Natural | Positive, Negative and Neutral |
| ICT-MMMO [105] | 2013 | ■ | ■ | ■ | / | / | Natural | Five basic sentiments |
| [4] CMU-MOSEI [106] | 2018 | ■ | ■ | ■ | / | / | Natural | Six basic emotions, Five basic sentiments |
| [5] MAHNOB-HCI [107] | 2012 | / | ■ | ■ | ■ | 27 | Induced | Arousal-Valence-Dominance-Predictability Disgust, Amusement, Joy, Fear, Sadness, Neutral |
| [6] RECOLA [108] | 2013 | / | ■ | ■ | ■ | 46 | Natural | Arousal-Valence, Agreement, Dominance, Engagement, Performance and Rapport |
| [7] DECAF [109] | 2015 | / | ■ | ■ | ■ | 30 | Induced | Arousal-Valence-Dominance, Six basic expressions Amusing, Funny and Exciting |

[1] sail.usc.edu/iemocap/iemocap_release.htm; [2] sail.usc.edu/CreativeIT/ImprovRelease.htm;
[3] ict.usc.edu/research/; [4] github.com/A2Zadeh/CMU-MultimodalSDK;
[5] mahnob-db.eu/hci-tagging/; [6] diuf.unifr.ch/diva/recola; [7] mhug.disi.unitn.it/wp-content/DECAF/DECAF.

**Interactive Emotional Dyadic Motion Capture (IEMOCAP)** [101] is constructed by the Speech Analysis and Interpretation Laboratory. During recording, 10 actors are asked to perform selected emotional scripts and improvised hypothetical scenarios designed to elicit 5 specific types of emotions. The face, head, and hands of actors are marked to provide detailed information about their facial expressions and hand movements while performing. Two famous emotion taxonomies are employed to label the utterance level: discrete categorical-based annotations and continuous attribute-based annotations. Afterwards, **CreativeIT** [102,103] contains detailed full-body motion visual-audio and text description data collected from 16 actors, during their affective dyadic interactions ranging from 2-10 minutes each. Two kinds of interactions (two-sentence and paraphrases exercises) are set as improvised. According to the video frame rate, the annotator gave the values of each actor's emotional state in the three dimensions. **Harvesting Opinions from the Web database (HOW)** [104] contains 13 positive, 12 negative and 22 neutral videos captured from YouTube. **Institute for Creative Technologies Multimodal Movie Opinion database (ICT-MMMO)** [105] contains 308 YouTube videos and 78 movie review videos from ExpoTV. It has five sentiment labels: strongly positive, weakly positive, neutral, strongly negative, and weakly negative. As far as we know, **Multimodal Opinion Sentiment and Emotion Intensity (CMU-MOSEI)** [106] is the largest database for sentiment analysis and emotion recognition, consisting of 23,453 sentences and 3,228 videos collected from more than 1,000 online YouTube speakers. Each video contains a manual transcription that aligns audio and phoneme grades.

**MAHNOB-HCI** [107] is a video-physiological database. Using 6 video cameras, a head-worn microphone, an eye gaze tracker, and physiological sensors, it is constructed by monitoring and recording the emotions of 27 participants while watching 20 films. **Remote Collaborative and Affective Interactions (RECOLA)** [108] consists of a multimodal corpus of spontaneous interactions from 46 participants (in French). These participants work in pairs to discuss a disaster scenario escape plan and reach an agreement via remote video conferencing. The recordings of the participants' activities are annotated by 6 annotators with two continuous emotional dimensions: arousal and valence, as well as social behaviour labels on five dimensions. **DECAF** [109] is a Magnetoencephalogram-based database for decoding affective responses of 30 subjects while watching 36 movie clips and 40 one-minute music video clips. DECAF contains a detailed analysis of the correlations between participants' self-assessments and their physiological responses, single-trial classification results for valence, arousal and dominance dimensions, performance evaluation against existing data sets, and time-continuous emotion annotations for movie clips.

## 5. Unimodal affect recognition

In this section, we systematically summarize the unimodal affect recognition methods from the perspective of affect modalities: physical modalities (e.g., textual, audio, and visual) [12] and physiological modalities (e.g., EEG and ECG) [29].

*5.1 Textual sentiment analysis*

With the rapid increase of online social media and e-commerce platforms, where users freely express their ideas, a huge amount of textual data are generated and collected. To identify subtle sentiment or emotions expressed explicitly or implicitly from the user-generated data, textual sentiment analysis (TSA) was introduced [110]. Traditional approaches of TSA [111,112] often rely on the process known as "feature engineering" to find useful features that are related to sentiment. This is a tedious task. DL-based models can realize an end-to-end sentiment analysis from textual data.

Table 7 shows an overview of some representative methods for TSA, where the publication (or preprint) year, analysis granularity, feature representation, classifier, database, and performance are presented. Next, the works of TSA are described in detail.

*5.1.1 ML-based TSA*

TSA based on the traditional ML methods mainly relies on knowledge-based techniques or statistical methods [113]. The former requires thesaurus modelling of large emotional vocabularies, and the latter assumes the availability of large databases, labelled with polarity or emotional labels.

**Knowledge-based TSA.** Knowledge-based TSA is often based on lexicons and linguistic rules. Different lexicons, such as WordNet, WordNet-Affect, SenticNet, MPQA, and SentiWordNet [114], often contain a bag of words and their semantic polarities. The lexicon-based approaches can classify a given word into positive or negative, but perform poorly without linguistic rules. Hence, Ding et al. [115] adopted a lexicon-based holistic approach that combines external evidence with linguistic conventions in natural language to evaluate the semantic orientation in reviews. Melville et al. [116] developed a framework for domain-dependent sentiment analysis using lexical association information.

To better understand the orientation and flow of sentiment in natural language, Poria et al. [117] proposed a new framework combining computational intelligence, linguistics and common-sense computing [118]. They found that the negation played an important role in judging the polarity of the overall sentence. Jia et al. [119] also verified the importance of negation on emotional recognition through extensive experiments. Blekanov et al. [120] utilized specifics of the Twitter platform in their multi-lingual knowledge-based approach of sentiment analysis. Due to the limitations of knowledge itself, knowledge-based models are limited to understanding only those concepts that are typical and strictly defined.

**Statistical-based TSA.** Statistical-based TSA relies more on an annotated dataset to train an ML-based classifier by using prior statistics or posterior probability. Compared with lexicon-based approaches [121,122], statistical-based TSA approaches are more suitable for sentiment analysis due to their ability to deal with large amounts of data. Mullen and Collier [123] used the semantic orientation of words to create a feature space that is classified by a designed SVM. Pak et al. [124] proposed a new sub-graph-based model. It represents a document as a collection of sub-graphs and inputs the features from these sub-graphs into an SVM classifier. Naïve Bayes (NB) is another powerful and widely-used classifier, which assumes that dataset features are independent. It can be used to filter out the sentences that do not support comparative opinions [125].

**Hybrid-based TSA**. By integrating knowledge with statistic models, hybrid-based TSA can take full advantage of both [126]. For example, Xia et al. [127] utilized SenticNet and a Bayesian model for contextual concept polarity disambiguation. Due to the ambiguity and little information of neutral between positive and negative, Valdivia et al. [128] proposed consensus vote models and weighted aggregation to detect and filter neutrality by representing the vague boundary between two sentiment polarities. According to experiments, they concluded that detecting neutral can help improve the performance of sentiment analysis. Le et al. [129] proposed a novel hybrid method in the combination of word sentiment score calculation, text pre-processing, sentiment feature generation and an ML-based classifier for sentiment analysis. The hybrid method [129] performed much better than popular lexicon-based methods in Amazon, IMDb, and Yelp, and achieved an average accuracy of 87.13%. Li et al. [130] utilized lexicons with one of the ML-based methods including NB and SVM. This hybrid method is more effective in detecting expressions that are difficult to be polarized into positive-negative categories.

*5.1.2 DL-based TSA*

DL-based techniques have a strong ability to automatically learn and discover discriminative feature representations from data themselves. DL-based TSA has been proved to be successful with the success of word embeddings [131] and the increase of the training data with multi-class classification [132]. Various DL-based approaches for TSA include deep convolutional neural network (ConvNet) learning, deep RNN learning, deep ConvNet-RNN learning and deep adversarial learning, as detailed next.

**Deep ConvNet learning for TSA.** CNN-based methods have been applied to different levels of TSA including document-level [133], sentence-level [134], and aspect-level (or word-level) [135] by using different filters to learn local features from the input data. Yin et al. [136] proposed a framework of sentence-level sentiment classification based on the semantic lexical-augmented CNN (SCNN) model, which makes full use of word information. Conneau et al. [137] applied a very deep CNN (VDCNN), which learns the hierarchical representations of the document and long-range dependencies to text processing. To establish long-range dependencies in documents, Johnson and Zhang [138] proposed a word-level deep pyramid CNN (DPCNN) model, which stacked alternately the convolutional layer and the max-pooling downsampling layer to form a pyramid to reduce computing complexity. The DPCNN with 15 weighted layers outperformed the previous best models on six benchmark databases for sentiment classification and topic categorization. For aspect-level sentiment analysis, Huang and Carley also [139] proposed a novel aspect-specific CNN by combining parameterized filters and parametrized gates.

**Deep RNN learning for TSA.** RNN-based TSA is capable of processing long sequence data. For example, Mousa and Schuller [140] designed a novel generative approach, contextual Bi-LSTM with a language model (cBi-LSTM LM), which changes the structure of Bi-LSTM to learn the word's contextual information based on its right and left contexts. Moreover, Wang et al. [141] proposed a model of recursive neural conditional random field (RNCRF) by integrating the RNN-based dependency tree of the sentence and conditional random fields.

The attention mechanism contributes to prioritizing relevant parts of the given input sequence according to a weighted representation at a low computational cost. In the paradigm of attention-based LSTM for TSA, LSTM helps to construct the document representation, and then attention-based deep memory layers compute the ratings of each document. Chen et al. [142] designed the recurrent attention memory (RAM) for aspect-level sentiment analysis. Specifically, a multiple-attention mechanism was employed to capture sentiment features separated by a long distance, the results of which were non-linearly combined with RNN. Inspired by the capability of human eye-movement behavior, Mishra et al. [143] introduced a hierarchical LSTM-based model trained by cognition grounded eye-tracking data, and they used the model to predict the sentiment of the overall review text. More recently, some researchers [144,145] suggested that user preferences and product characteristics should be taken into account.

For document-level sentiment classification, Dou [145] proposed a deep memory network combining LSTM on account of the influence of users who express the sentiment and the products that are evaluated. Chen et al. [144] designed a hierarchical LSTM with an attention mechanism to generate sentence and document representations, which incorporates global user and product information to prioritize the most contributing items. Considering the irrationality of encoding user information and product information as one representation, Wu et al. [146] designed an attention LSTM-based model, which executed hierarchical user attention and product attention (HUAPA) to realize sentiment classification.

For multi-task classification (e.g., aspect category and sentiment polarity detection), J et al. [147] proposed convolutional stacked Bi-LSTM with a multiplicative attention network concerning global-local information. In contrast, to fully exploit contextual affective knowledge in aspect-level TSA, Liang et al. [148] proposed GCN-based SenticNet to enhance graph-based dependencies of sentences. Specifically, LSTM layers were employed to learn contextual representations, and GCN layers were built to capture the relationships between contextual words in specific aspects.

**Deep ConvNet-RNN learning for TSA.** Although ConvNet-based or RNN-based models have been extensively employed to generate impressive results in TSA, more researchers [149,150] have tried to combine both ConvNets and RNNs to improve the performance of TSA by getting the benefits offered by each model.

As the CNN is proficient in extracting local features and the BiLSTM is skilled in a long sequence, Li et al. [150] combined CNN and BiLSTM in a parallel manner to extract both types of features, improving the performance of sentiment analysis. To decrease the training time and complexity of LSTM and attention mechanism for predicting the sentiment polarity, Xue et al. [151] designed a gated convolutional network with aspect embedding (GCAE) for aspect-category sentiment analysis (ACSA) and aspect-term sentiment analysis (ATSA). The GCAE uses two parallel CNNs, which output results combined with the gated unit and extended with the third CNN, extracting contextual information of aspect terms.

To distinguish the importance of different features, Basiri et al. [152] proposed an attention-based CNN-RNN deep model (ABCDM), which utilized bidirectional LSTM and GRU layers to capture temporal contexts and apply the attention operations on the discriminative embeddings of outputs generated by two RNN-based networks. In addition, CNNs were employed for feature enhancement (e.g., feature dimensionality reduction and position-invariant feature extraction). All the above works focused on detecting sentiment or emotion, but it is important to predict the intensity or degree of one sentiment in the description of human intimate emotion. To address the problem, Akhtar et al. [153] proposed a stacked ensemble method by using an MLP to ensemble the outputs of the CNN, LSTM, GUR, and SVR.

**Deep adversarial learning for TSA.** Deep adversarial learning with the ability to regularize supervised learning algorithms was introduced to text classification [154]. Inspired by the domain-adversarial neural network [155], Li et al. [156] constructed an adversarial memory network model that contains sentiment and domain classifier modules. Both modules were trained together to reduce the sentiment classification error and allowed the domain classifier not to separate both domain samples. The attention mechanism is incorporated into the deep adversarial based model to help the selection of the pivoted words, which are useful for sentiment classification and shared between the source and target domains. Li et al. [157] initiated the use of GANs [158] in sentiment analysis, reinforcement learning and recurrent neural networks to build a novel model, termed category sentence generative adversarial network (CS-GAN). By combining GANs with RL, the CS-GAN can generate more category sentences to improve the capability of generalization during supervised training. Similarly, to tackle the sentiment classification in low-resource languages without adequate annotated data, Chen et al. [159] proposed an adversarial deep averaging network (ADAN) to realize cross-lingual sentiment analysis by transferring the knowledge learned from labelled data on a resource-rich source language to low-resource languages where only unlabeled data existed. Especially, the ADAN was trained with labelled source text data from English and unlabeled target text data from Arabic and Chinese. More recently, Karimi et al. [160] fine-tuned the general-purpose BERT and domain-specific post-trained BERT using adversarial training, which showed promising results in TSA.

**Table 7.** Overview of the representative methods for TSA.

| Pub. | Year | Granularity | Feature Representation | Classifier | Database | Performance |
|---|---|---|---|---|---|---|
| *ML-based SER* | | | | | | |
| [117] | 2014 | Concept-level | Multi-feature fusion | SVM-ELM | Movie Review | 2 classes: 86.21 |
| [120] | 2018 | Aspect-level | Multilingual features | Metric-based | Twitter | 2 classes: 66.00, 60.00 |
| [125] | 2008 | Aspect-level | CDM | NB | Reuters-2157 | 10 classes: 85.62 |
| [130] | 2019 | Aspect-level | Features fusion | LML | Amazin; IMDB | 2 classes: 89.7; 10 classes: 85.2 |
| *DL-based SER* | | | | | | |
| [133] | 2020 | Document-level | CNN | HieNN-DWE | IMDB etc. | 10 classes: 47.8 |
| [137] | 2017 | Document-level | VDCNN | ReLU | Amazon etc. | 2 classes: 95.07 |
| [139] | 2018 | Aspect-level | PF-CNN | Softmax | Laptops; Restaurants | 2 classes: 86.35; 90.15 |
| [140] | 2017 | Document-level | cBLSTM | Decision Rule | IMDB | 2 classes: 92.83 |
| [141] | 2016 | Aspect-level | DT-RNN, CRF | RNCRF | Laptops; Restaurants | 2 classes: 79.44; 84.11 |
| [146] | 2018 | Document-level | HUAPA | Softmax | IMDB etc. | 10 classes: 55.0 |
| [150] | 2020 | Document-level | Word2Vec | CNN-LSTM | SST etc. | 5 classes: 50.6812 |
| [151] | 2018 | Aspect-level | Word Embeddings | GCAE | Laptops; Restaurants | 4 classes: 69.14; 85.92 |
| [157] | 2018 | Sentence-level | CS-GAN | Softmax | Amazon-5000 | 2 classes: 86.43 |
| [159] | 2018 | Aspect-level | ADAN | Softmax | TARGET | 5 classes: 42.49; 3 classes: 54.54 |

*5.2 Audio emotion recognition*

Audio emotion recognition (also called SER) detects the embedded emotions by processing and understanding speech signals [161]. Various ML-based and DL-based SER systems have been carried out on the basis of these extracted features for better analysis [162,163]. Traditional ML-based SER concentrates on the extraction of the acoustic features and the selection of the classifiers. However DL-based SER constructs an end-to-end CNN architecture to predict the final emotion without considering feature engineering and selection [164].

Table 8 shows an overview of representative methods for SER, including the most relevant papers, their publication (or preprint) year, feature representation, classifier, database, and performance (from best or average reported results). These works are next described in detail.

5.2.1  *ML-based SER*

The ML-based SER systems include two key steps: the strong features representation learning for emotional speech and an appropriate classification for final emotion prediction [19]. Different kinds of acoustic features can be fused to get the mixed features for a robust SER. Although prosodic features and spectral features are more frequently used in SER systems [165], in some cases, voice-quality features and other features are sometimes more important [166]. OpenSMILE [167] is a popular audio feature extraction toolkit that extracts all key features of the speech. The commonly used classifiers for SER systems encompass HMM, GMM, SVM, RF and ANN. In addition to these classifiers, the improved conventional interpretable classifiers and ensemble classifiers are also adopted for SER. In this subsection, we divided the ML-based SER systems into acoustic-feature based SER and interpretable-classifier based SER [168], [169]. Note that different combinations of models and features result in obvious differences in the performance of SER [170].

**Acoustic-feature based SER.** Prosodic features (e.g. intonation and rhythm) have been discovered to convey the most distinctive properties of emotional content for SER. The prosodic features consist of fundamental frequency (rhythmical and tonal characteristics) [171], energy (volume or the intensity), and duration (the total of time to build vowels, words and similar constructs). Voice quality is determined by the physical properties of the vocal tract such as jitter, shimmer, and harmonics to noise ratio. Lugger and Yang [172] investigated the effect of prosodic features, voice quality parameters, and different combinations of both types on emotion classification. Spectral features are often obtained by transforming the time-domain speech signal into the frequency-domain speech signal using the Fourier transform [173]. Bitouk et al. [174] introduced a new set of fine-grained spectral features which are statistics of Mel Frequency cepstrum coefficients (MFCC) over three phoneme type classes of interest in the utterance. Compared to prosodic features or utterance level spectral features, the fine-grained spectral features can yield results with higher accuracy. In addition, the combination of these features and prosodic features also improves accuracy. Shen et al. [175] utilized SVM to evaluate the performance of using energy and pitch, linear prediction cepstrum coefficients (LPCC), MFCC, and their combination. The experiment demonstrated the superior performance based on the combination of various acoustic features.

In order to improve the recognition performance of speaker-independent SER, Jin et al. [176] designed feature selection with L1-normalization constraint of Multiple kernel learning (MKL) and feature fusion with Gaussian kernels. This work [176] achieved an accuracy of 83.10% on the Berlin Database, which is 2.10% higher than the model that used the sequential floating forward selection algorithm, and a GMM [177]. Wang et al. [178] proposed a new Fourier parameter-based model using the perceptual content of voice quality, and the first-order and second-order differences for speaker-independent SER. Experimental results revealed that the combination of Fourier parameters and MFCC could significantly increase the recognition rate of SER.

**ML-based classifier for SER.** The HMM classifier is extensively adopted for SER due to the production mechanism of the speech signals. In 2003, Nwe et al. [179] designed an HMM with log frequency power coefficients (LFPC) to detect human stress and emotion. It achieved the best recognition accuracy of 89%, which was higher than 65.8% of human recognition. The GMM and its variants can be considered as a special continuous HMM and are appropriate for global-feature based SER. For example, Navas et al. [180] proposed a GMM-based baseline method by using prosodic, voice quality, and MFCC features.

Different from HMM and GMM, SVM maps the emotion vector to a higher dimensional space by using a kernel function and establishes the maximum interval hyperplane in the high-dimensional space for optimal classification. Milton et al. [181] designed a three-stage hierarchical SVM with linear and RBF kernels to classify seven emotions using MFCC features on the Berlin EmoDB. There are some works [182,183] using different SVM classifiers with various acoustic features and their combinations. Ensemble learning has been proven to give superior performance compared to a single classifier. Yüncü et al. [184] designed the SVM with Binary DT by integrating the tree architecture with SVM. Bhavan et al. [185] proposed a bagged ensemble comprising of SVM with different Gaussian kernels as a viable algorithm. Considering differences among various categories of human beings, Chen et al. [186] proposed the two-layer fuzzy multiple RF by integrating the decision trees with Bootstraps to recognize six emotional states.

### 5.2.2 *DL-based SER*

DL-based SER systems can understand and detect contexts and features of emotional speech without designing a tailored feature extractor. The CNNs with auto-encoder [187] are regarded as commonly used techniques for DL-based SER [187]. The RNNs [188] and their variants (e.g., Bi-LSTM) [189] are widely introduced to capture temporal information. The hybrid deep learning for SER includes ConvNets and RNNs, as well as attention mechanisms [190,191]. For the issues of limited data amount and low quality of the databases, adversarial learning can be used for DL-based SER [192] by augmenting trained data and eliminating perturbations.

**ConvNet learning for SER**. Huang et al. [163] utilized the semi-CNN to extract features of spectrogram images, which are fed into SVM with different parameters for SER. Badshah et al. [193] proposed an end-to-end deep CNN with three convolutional layers and three fully connected layers to extract discriminative features from spectrogram images and predict seven emotions. Zhang et al. [194] designed AlexNet DCNN pre-trained on ImageNet with discriminant temporal pyramid matching (DTPM) strategy to form a global utterance-level feature representation. The deep ConvNet-based models are also used to extract emotion features from raw speech input for multi-task SER [195].

**RNN learning for SER.** RNNs can process a sequence of speech inputs and retain its state while processing the next sequence of inputs. They learn the short-time frame-level acoustic features and aggregate appropriately these features over time into an utterance-level representation. Considering the different emotion states that may exist in a long utterance, RNN and its variants (e.g. LSTM and Bi-LSTM) can tackle the uncertainty of emotional labels. Extreme learning machine (ELM), a single-hidden layer neural network, was regarded as the utterance-level classifier [196]. Lee and Tashev [188] proposed the Bi-LSTM with ELM to capture a high-level representation of temporal dynamic characteristics. Ghosh et al. [197] pre-trained a stacked denoising autoencoder to extract low-dimensional distributed feature representation and trained Bi-LSTM-RNN for final emotion classification of speech sequence.

The attention model is designed to ignore irrelevantly emotional frames and other parts of the utterance. Mirsamadi et al. [198] introduced an attention model with a weighted time-pooling strategy into the RNN to more emotionally salient regions. Because there exists some irrelevantly emotional silence in the speech, the silence removal needs to be employed before BLSTMs, incorporated with the attention model used for feature extraction [190]. Chen et al. [199] proposed the attention-based 3D-RNNs to learn discriminative features for SER. Experimental results show that the Attention-3D-RNNs can exceed the performance of SOTA SER on IEMOCAP and Emo-DB in terms of unweighted average recall.

The attention model is designed to ignore irrelevant emotional frames in the utterance. Mirsamadi et al. [198] introduced an attention model with a weighted time-pooling strategy into the RNN to highlight more emotionally salient regions. Since there exists some irrelevant emotional silence in the speech, it is necessary to combine the attention model and silence removal for feature extraction [190]. Chen et al. [199] proposed the attention-based 3D-RNNs to learn discriminative features for SER, which achieved outstanding performances on IEMOCAP and Emo-DB in terms of unweighted average recall.

**ConvNet-RNN learning for SER.** As both ConvNets and RNNs have their advantages and limitations, the combination of CNNs and RNNs enables the SER system to obtain both frequency and temporal dependency [191]. For example, Trigeorgis et al. [200] proposed an end-to-end SER system consisting of CNNs and LSTMs to automatically learn the best representation of the speech signal directly from the raw time representation. In contrast, Tzirakis et al. [201] proposed an end-to-end model comprising of CNNs and LSTM networks, which show that a deeper network is consistently more accurate than one shallow structure, with an average improvement of 10.1% and 17.9% of Arousal and Valence (A/V) on RECOLA. Wu et al. [202] designed the capsule networks (CapsNets) based SER system by integrating with the recurrent connection. The experiments employed on IEMOCAP demonstrated that CapsNets achieved 72.73% and 59.71% of weighted accuracy (WA) and unweighted accuracy (UA), respectively. Zhao et al. [203] designed a spatial CNN and an attention-based BLSTM for deep spectrum feature extraction. These features were concatenated and fed into a DNN to predict the final emotion.

**Adversarial learning for SER.** In SER systems, the classifiers are often exposed to training data that have a different distribution from the test data. The difference in data distributions between the training and testing data results in severe misclassification. Abdelwahab and Busso [192] proposed the domain adversarial neural network to train gradients coming from the domain classifier, which makes the source and target domain representations closer. Sahu et al. [204] proposed an adversarial AE for the domain of emotion recognition while maintaining the discriminability between emotion classes. Similarly, Han et al. [205] proposed a conditional adversarial training framework to predict dimensional representations of emotion while distinguishing the difference between generated predictions and the ground-truth labels.

GAN and its variants have been demonstrated that the synthetic data generated by generative models can enhance the classification performance of SER [206]. To augment the emotional speech information, Bao et al. [207] utilized Cycle consistent adversarial networks (CycleGAN) [208] to generate synthetic features representing the target emotions, by learning feature vectors extracted from unlabeled source data.

**Table 8.** Overview of the representative methods for SER.

| Pub. | Year | Feature Representation | Classifier | Database | Performance |
|---|---|---|---|---|---|
| *ML-based SER* | | | | | |
| [174] | 2010 | Acoustic-feature | SVM | LDC/ Emo-DB | 6 classes: 43.8/79.1 |
| [176] | 2014 | Feature selection and fusion | MKL | Emo-DB | 7 classes: 83.1 |
| [181] | 2013 | MFCC | SVM | Emo-DB | 7 classes: 68 |
| [186] | 2020 | Acoustic features | TL-FMRF | CASIA/Emo-DB | 6 classes: 81.75/77.94 |
| *DL-based SER* | | | | | |
| [163] | 2013 | Semi-CNN | SVM | SAVEE/Emo-DB DES/ MES | 7 classes: 89.7/93.7 7 classes: 90.8/90.2 |
| [193] | 2017 | CNN | FC | Emo-DB | 7 classes: 84.3 |
| [197] | 2014 | Bi-LSTM-RNN | RNN | USC-IEMOCAP | 4 classes: 51.86 |
| [198] | 2017 | DNN/RNN | Pooling | IEMOCAP | 7 classes: 58.8 |
| [199] | 2018 | Attention-3D-RNN | FC | IEMOCAP Emo-DB | 4 classes: 64.74 7 classes: 82.82 |
| [200] | 2016 | CNNs-LSTM | LSTM | RECOLA | A/V: 74.1/ 32.5 |
| [203] | 2018 | CNN/Attention-BLSTM | DNN | IEMOCAP | 7 classes: 68.0 |
| [204] | 2018 | Adversarial AE | SVM | IEMOCAP | 7 classes: 58.38 |
| [205] | 2018 | Conditional adversarial | CNN | RECOLA | A/V: 73.7/44.4 |

*5.3 Visual emotion recognition*

Visual emotion recognition [209,12] can be primarily categorized into facial expression recognition (FER) and body gesture emotion recognition (also known as emotional body gesture recognition, or EBGR). The following section looks at FER and EBGR in a great deal of detail, capturing a vast body of research (well-over 100 references).

5.3.1 *Facial expression recognition*

FER is implemented using images or videos containing facial emotional cues [210, 211]. According to whether static images or dynamic videos are to be used for facial expression representation, FER systems can be divided into static-based FER [212] and dynamic-based FER [213]. When it comes to the duration and intensity of facial expression [214,215], FER can be further divided into macro-FER and micro-FER (or FMER) [216–218]. Based on the dimensions of the facial images, macro-FER can be further grouped into 2D FER [219–221], and 3D/4D FER [222–224]. Note that since facial images or videos suffer from a varied range of backgrounds, illuminations, and head poses, it is essential to employ pre-processing techniques (e.g., face alignment [225], face normalization [226], and pose normalization [227]) to align and normalize semantic information of face region.

In this sub-section, we distinguish FER methods (shown in Fig. 4) via the point of whether the features are hand-crafted features based ML models [228] or high-level features based on DL-based models [229]. Table 9 provides an overview of representative FER methods. The first column gives the publication reference (abbreviated to Pub.), followed by the publication (or preprint) year in the second column. The feature representations and classifiers of the referenced method are given in the third and fourth columns, respectively. The last six columns show the best or average results (%) with the given databases. These works are next described in detail.

*5.3.1.1 ML-based FER*

The ML-based FER methods mainly rely on hand-crafted feature extraction and feature post-processing. Generally, hand-crafted facial features can be categorized into geometry-based features explaining the shape of the face and its components, and appearance-based features defining facial texture. The feature post-processing can be divided into two types: feature fusion and feature selection [18].

**Geometry-based FER.** Ghimire and Lee [230] (Fig. 4 (a)) utilized 52 facial landmark points (FLP), which represent features of geometric positions and angles, for automatic FER in facial sequences. Sujono and Gunawan [231] used the Kinect motion sensor to detect the face region based on depth information and active shape model (AAM) [232]. The change of key features in AAM and a fuzzy logic model is utilized to recognize facial expression based on prior knowledge derived from the facial action coding system (FACS) [233]. As the geometry of different local structures with distortions has unstable shape representations, the local prominent directional pattern descriptor (LPDP) [234] was proposed by using statistical information of a pixel neighbourhood to encode more meaningful and reliable information.

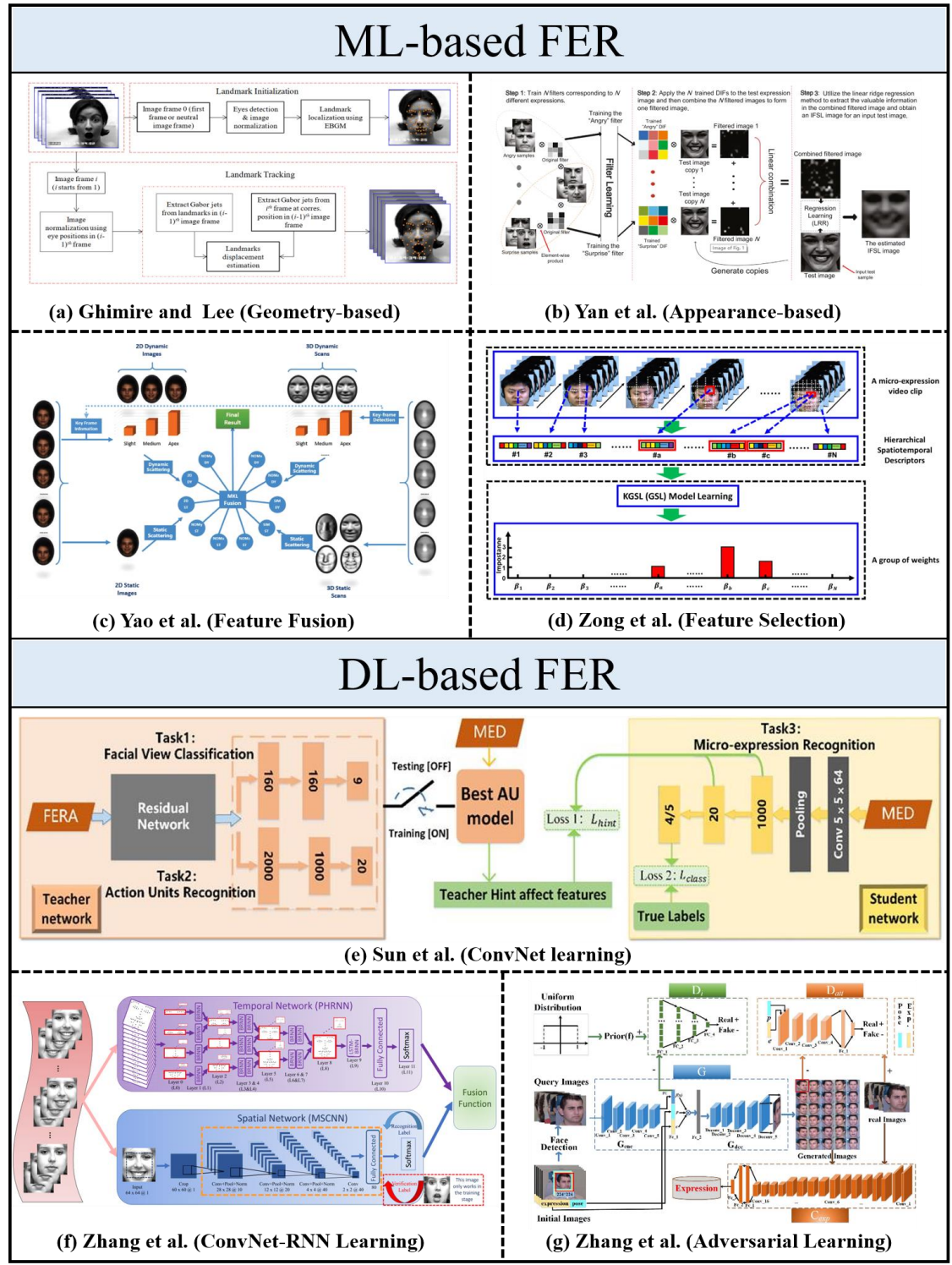

**Fig. 4.** Taxonomy of representative FER methods based on ML techniques or DL models. (a) Geometry-based FER adopted from [230]; (b) Appearance-based FER adopted from [235]; (c) Feature fusion for 3D FER adopted from [236]; (d) Feature selection for FMER adopted from [237]; (e) ConvNet learning for FMER adopted from [218]; (f) ConvNet-RNN learning for FER adopted from [238]; (g) Adversarial learning for 3D FER adopted from [239].

**Appearance-based FER.** Appearance-based approaches often extract and analyze spatial information or spatial-temporal information from the whole or specific facial regions [240]. Tian et al. [210] designed an automatic face analysis system that captured fine-grained changes of facial expression into AUs [241] of FACS by using the permanent and transient facial features extracted by the Gabor wavelet, SIFT and local binary pattern (LBP) [242]. Gu et al. [243] divided one facial image into several local regions by grids, then applied multi-scale Gabor-filter operations on local blocks, and finally encoded the mean intensity of each feature map. Yan et al. [235] (Fig. 4 (b)) proposed a framework of low-resolution FER based on image filter-based subspace learning (IFSL), including deriving discriminative image filters (DIFs), their combination, and an expression-aware transformation matrix.

The local binary pattern from three orthogonal planes (LBP-TOP) [244] is an extension of LBP [242] computes over three orthogonal planes at each bin of a 3D volume formed by stacking the frames. The LBP-TOP and its variants have shown a promising performance on dynamic FER. For example, Wang et al. [245] designed two kinds of feature extractors (LBP-six intersection points (LBP-SIP) and super-compact LBP-three mean orthogonal planes (LBP-MOP)) based on the improved LBP-TOP to preserve the essential patterns while reducing the redundancy. Davison et al. [246] employed LBP-TOP features and Gaussian derivatives features for FEMR. Similarly, Liong et al. [247] designed a framework of FMER based on two kinds of feature extractors, consisting of optical strain flow (OSF) and block-based LBP-TOP.

**Feature fusion for FER.** It has been proved to fuse different types of geometry-based features and appearance-based features to enhance the robustness of FER [236]. For example, Majumder et al. [219] designed an automatic FER system based on the deep fusion of geometric features and LBP features using autoencoders. For the FMER task, Zhang et al. [248] proposed the aggregating local spatiotemporal patterns (ALSTP), which adopts cascaded fusion of local LBP-TOP and LOF extracted from 9 representative local regions of the face. For 3D/4D facial expression images, Zhen et al. [249] computed spatial facial deformations using a Riemannian based on dense scalar fields (DSF) and magnified them by a temporal filtering technique. For 2D/3D facial expression images, Yao et al. [236] (Fig. 4 (c)) used the MKL fusion strategy to combine 2D texture features, 3D shape features, and their corresponding Fourier transform maps of different face regions.

**Feature selection for FER.** Although more 3D/4D features [250] or dynamic features [251] have different effects on FER [252], excessive features may break the predictive model. The feature selection is to choose a relevant and useful subset of the given set of features while identifying and removing redundant attributes. To overcome the high-dimensionality problem of 3D facial features, Azazi et al. [253] firstly transformed the 3D faces into 2D planes using conformal mapping, and then proposed the differential evolution (DE) to select the optimal facial feature set and SVM classifier parameters, simultaneously. Savran and Sankur [254] investigated the model-free 3D FER based on the non-rigid registration by selecting the most discriminative feature points from 3D facial images. As different facial regions contributed different to micro-expressions, Chen et al. [255] utilized weighted 3DHOG features and weighted fuzzy classification for FMER. Different from the spatial division with fixed grid, a hierarchical spatial division scheme (HSDS) [237] (Fig. 4 (d)) was proposed to generate multiple types of gradually denser grids and designed kernelized group sparse learning (KGSL) to learn a set of importance weights.

*5.3.1.2 DL-based FER*

The backbone networks of DL-based FER are mostly derived from well-known pre-trained ConvNets such as VGG [256], VGG-face [257], ResNet [258], and GoogLeNet [259]. Thus, we divide DL-based FER into ConvNet learning for FER, ConvNet-RNN learning for FER, and adversarial learning for FER considering the difference of network architectures.

**ConvNet learning for FER.** ConvNet-based FER often design transform learning or loss function [221] to overcome overfitting when using relatively small facial expression databases. For example, Yang et al. [260] proposed a de-expression residue learning (DeRL) to recognize facial expressions by extracting expressive information from one facial expression image. For FMER, the transform-learning based model is pre-trained on the ImageNet and several popular macro-expression databases with the original residual network [261]. Su et al. [218] (Fig. 4 (e)) proposed a novel knowledge transfer technique, which comprised a pre-trained deep teacher neural network and a shallow student neural network. Specifically, the AU-based model is trained on the residual network, which is then distilled and transferred for FMER. Liu et al. [262] developed an identity-disentangled FER by integrating the hard negative generation with the radial metric learning (RML). Specifically, the RML module combined inception-structured convolutional groups with an expression classification branch for the final emotion recognition by minimizing both the cross-entropy loss and RML loss. To alleviate variations introduced by personal attributes, Meng et al. [263] proposed an identity-aware convolutional neural network (IACNN) consisting of two identical sub-CNNs with shared weights. Besides, they designed two losses for identity-invariant FER: expression-sensitive loss and identity-sensitive loss.

To highlight the most helpful information of facial images, various attention mechanisms [264,265] are proposed to discriminate distinctive features. For example, Fernandez et al. [266] proposed an attention network and embedded it into an encoder-decoder architecture for the facial expression representation. Similarly, Xie et al [267] proposed an attention-based salient expressional region descriptor (SERD) to locate the most expression-related regions that are beneficial to FER. To reduce the uncertainties of FER, Wang et al. [268] proposed the self-cure network consisting of the self-attention importance weighting module, the ranking regularization module, and the relabeling module. Zhu et al. [269] proposed a novel discriminative attention-based CNN, where the attention module was used to emphasize the unequal contributions of features for different expressions, and a dimensional distribution loss was designed to model the inter-expression relationship. To provide local-global attention across the channel and spatial location for feature maps, Gera and Balasubramanian [270] proposed the spatial-channel attention net (SCAN). Besides, the complementary context information (CCI) branch is proposed to enhance the discriminating ability by integrating with another channel-wise attention.

Except for the efficient network architectures with attention networks or loss functions [271], there is a fast and light manifold convolutional neural network (FLM-CNN) based on the multi-scale encoding strategy [272] or the deep fusion convolutional neural network [273]. Due to the facial expression database biases, conditional probability distributions between source and target databases are often different. To implement a cross-database FER, Li and Deng [274] proposed a novel deep emotion-conditional adaption network to learn domain-invariant and discriminative feature representations. Another challenge for FER is the class imbalance of in-the-wild databases. Li et al. [275] proposed an adaptive regular loss function named AdaReg loss, which can re-weight category importance coefficients, to learn class-imbalanced expression representations.

The 3D ConvNet (or C3D) [276] has been universally used for dynamic-based FER by learning and representing the spatiotemporal features of videos or sequences. For example, the C3D was first used to learn local spatiotemporal features [277] and then these features were cascaded with the multimodal deep-belief networks (DBNs) [278]. Besides, the C3D can be combined with the global attention module to represent Eulerian motion feature maps generated based on the Eulerian video magnification (EVM) [279]. Lo et al. [280] utilized a 3D CNN to extract AU features and applied a graph convolutional network (GCN) to discover the dependency of AU nodes.

**ConvNet-RNN learning for FER.** For dynamic facial sequences or videos, the temporal correlations of consecutive frames should be considered as important cues. RNNs and their variants (LSTMs) can robustly derive temporal characteristics of the spatial feature representation. In contrast, spatial characteristics of the representative expression-state frames can be learned with CNNs [281]. Based on the architecture of ConvNet-RNN networks, many studies have proposed cascaded fusion [224,282] or the ensemble strategy [238] to capture both spatial and temporal information for FER.

The standard pipeline of ConvNet-RNN based FER using the cascaded fusion is to cascade outputs of ConvNets into RNNs to extract temporal dynamics. For example, Kim et al. [283] proposed an end-to-end FMER framework by cascading the spatial features extracted by the CNN into the LSTM to encode the temporal characteristics. Xia et al. [284] proposed a deep spatiotemporal recurrent convolutional network (STRCN) with a balanced loss that can capture the spatiotemporal deformations of the micro-expression sequence. For dimensional emotion recognition, Kollias and Zafeiriou [285] proposed RNN subnets to explore the temporal dynamics of low-, mid- and high-level features extracted from the trained CNNs. Liu et al. [286] proposed a framework of dynamic FER based on the siamese action-units attention network (SAANet). Specifically, the SAANet is a pairwise sampling strategy, consisting of CNNs with global-AU attention modules, a BiLSTM module and an attentive pooling module. For 4D FER, Behzad et al. [224] utilized CNNs to extract deep features of multi-view and augmented images, and then fed these features into the Bi-LSTM to predict 4D facial expression. On the basis of the work [224], sparsity-aware deep learning [282] was further proposed to compute the sparse representations of multi-view CNN features.

The standard pipeline of ConvNet-RNN based FER using the ensemble strategy is to fuse outputs of two streams. For example, Zhang et al. [238] (Fig. 4 (f)) proposed a deep evolutional spatial-temporal network, which consists of a part-based hierarchical bidirectional recurrent neural network (PHRNN) and a multi-signal convolutional neural network (MSCNN), for analyzing temporal facial expression information and still appearance information, respectively.

**Adversarial learning for FER.** As GANs can generate synthetic facial expression images under different poses and views, GAN-based models are used for pose/view-invariant FER [287] or identity-invariant FER [288]. For pose/view-invariant FER, Zhang et al. [239] proposed the GAN using AE structure to generate more facial images with different expressions under arbitrary poses (Fig. 4 (g)) and [287] further took the shape geometry into consideration. Since the slight semantic perturbations of the inputs often affected prediction accuracy, Fu et al. [289] proposed a semantic neighbourhood aware (SNA) network, which formulated the semantic perturbation based on the asymmetric AE with additive noise. For identity-invariant FER, Ali and Hughes [290] proposed a novel disentangled expression learning GAN (DE-GAN) by untangling the facial expression representation from identity information. Yu et al. [291] investigated a framework of facial micro-expression recognition and synthesis based on the identity-aware and capsule-enhanced GAN (ICE-GAN), which consisted of an AE-based generator for identity-aware expression synthesis, and a capsule-enhanced discriminator (CED) for discriminating the real/fake images and recognizing micro-expressions.

**Table 9.** Overview of the representative FER methods (publicly used database).

| Pub. | Year | Feature Representation | Classifier | CK+ | Oulu-CASIA | SFEW | BU-4DFE | CASME II | SMIC |
|---|---|---|---|---|---|---|---|---|---|
| *ML-based FER* | | | | | | | | | |
| [230] | 2013 | FLP | SVM | 97.35 | / | / | / | / | / |
| [234] | 2019 | LPDP | SVM | 94.50 | / | / | 73.40 | / | / |
| [243] | 2012 | MSGF | KNN | 91.51 | / | / | / | / | / |
| [235] | 2020 | IFSL | SVM | 98.70 | / | 46.50 | / | / | / |
| [247] | 2018 | OSF+LBP-TOP | SVM | / | / | / | / | [2]41.74/[1]61.82 | [2]50.79/[1]62.74 |
| [245] | 2015 | LBP-MOP | SVM | / | / | / | / | [2]45.75/[1]66.80 | [2]50.61/[1]60.98 |
| [245] | 2015 | LBP-SIP | SVM | / | / | / | / | [2]44.53/[1]66.40 | [2]50.00/[1]64.02 |
| [236] | 2018 | 2D+3D maps | MKL | / | / | / | 90.12 | / | / |
| [249] | 2019 | DSF | HMM | / | / | / | 95.13 | / | / |
| [253] | 2015 | DE | SVM | / | / | / | 85.81/[3]84 | / | / |
| [216] | 2016 | MDMO | SVM | / | / | / | / | [2]67.37 | [2]80.00 |
| [237] | 2018 | HSDS | KGSL | / | / | / | / | [2]65.18 | [2]66.46 |
| *DL-based FER* | | | | | | | | | |
| [272] | 2018 | FLM-CNN | FC | / | / | / | [4]83.32 | / | / |
| [280] | 2020 | GCN-CNN | FC | / | / | / | / | [2]42.7 | / |
| [263] | 2017 | IACNN | FC | 95.37 | / | 54.30 | / | / | / |
| [270] | 2021 | SCAN | FC | 97.31 | 89.11 | 58.93 | / | / | / |
| [238] | 2017 | CNN-RNN | Fusion | 98.50 | 86.25 | / | / | / | / |
| [286] | 2020 | CNN-LSTM | Pooling | 99.54 | 88.33 | 54.56 | / | / | / |
| [282] | 2021 | CNN-LSTM | [6]Col-Cla | / | / | / | 99.69 | / | / |
| [284] | 2020 | STRCN | Pooling | / | / | / | / | [1]84.10 | [1]75.80 |
| [290] | 2019 | DE-GAN | SVM/MLP | 97.28 | 89.17 | / | / | / | / |
| [289] | 2020 | SNA | FC | 98.58 | 87.60 | / | / | / | / |
| [291] | 2020 | ICE-GAN | CED | / | / | / | / | [5]86.8 | [5]79.10 |

[1] Leave-one-subject-out (LOSO); [2] Leave-one-video-out (LOVO); [3] BU-3DFE/ Bosphorus; [4] The average accuracy of BU-3DFE under both the protocols, P1 and P2; [5] Unweighted Average Recall (UAR); [6] Col-Cla=Collaborative Classification.

### 5.3.2 Body gesture emotion recognition

Most studies of visual emotion recognition focus on FER due to the prominent advantages of distinguishing human emotions. However, FER will be unsuitable when the dedicated sensors fail to capture facial images or just capture low-resolution facial images in some environments. EBGR [45] aims to expose one's hidden emotional state from full-body visual information (e.g., body postures) and body skeleton movements or upper-body visual information (e.g., hand gestures, head positioning and eye movements) [292,293]. The general pipeline of EBGR includes human detection [294,258,295] (regarded as pre-processing), feature representation, and emotion recognition. In the view of whether the process of feature extraction and emotion recognition is performed in an end-to-end manner, EBGR systems are categorized into ML-based EBGR and DL-based EBGR.

#### 5.3.2.1 ML-based EBGR

In existing ML-based EBGR systems, the input is an abstraction of the human body gestures (or their dynamics) through an ensemble body parts or a kinematic model [296]; and the output emotion is distinguished through ML-based methods or statistical measures to map the input into the emotional feature space, where the emotional state can be recognized by an ML-based classifier [297].

**Statistic-based or movement-based EBGR**. The ways of feature extraction can be grouped into statistic-based analysis [298–300] and movement-based analysis [301]. For example, Castellano et al. [298] proposed an emotional behavior recognition method based on the analysis of body movement and gesture expressivity. Different statistic qualities of dynamic body gestures are used to infer emotions with designed indicators describing the dynamics of expressive motion cues. Similarly, Saha et al. [299] and Maret et al. [300] utilized low-level features for EBGR, which were calculated based on the statistical analysis of the 3-D human skeleton generated by a Kinect sensor. Besides, Senecal et al. [301] proposed a framework of continuous emotional recognition based on the gestures and full-body dynamical motions using the laban movement analysis-based feature descriptors.

**Feature fusion for EBGR**. Fusing multiple body posture features can enhance the generalization capability and the robustness of EBGR [296]. By analyzing postural and dynamic expressive gesture features, Glowinski et al. [302] proposed a framework of upper-body based EBGR, using the minimal representation of emotional displays with a reduced amount of visual information related to human upper-body movements. Razzaq et al. [303] utilized skeletal joint features from a Kinect v2 sensor, to build mesh distance features and mesh angular features for upper-body emotion representation. Santhoshkumar and Geetha successively proposed two EBGR methods, by using histogram of orientation gradient (HOG) and HOG-KLT features from the sequences [304], or by calculating and extracting four kinds of geometric body expressive features [305]. They achieved the best recognition accuracies of 95.9% and 93.1% on GEMEP, respectively.

**Classifier-based EBGR**. The common classifiers for ML-based EBGR include decision tree, ensemble decision tree, KNN, SVM, etc. For example, Kapur et al. [306] proposed gesture-based affective computing by using five different classifiers to analyze full-body skeletal movements captured by the Vicon system. Different from full-body based EBGR, Saha et al. [299] focused on the upper-body based EBGR by employing some comparisons, using five classic ML-based classifiers in terms of the average classification accuracy and computation time. The experimental results showed that the ensemble decision tree achieved the highest recognition rate of 90.83%, under an acceptable execution efficiency. Maret et al. [300] also utilized five commonly used classifiers, whereby the genetic algorithm was invoked to search the optimal parameters of the recognition process.

**Non-acted EBGR**. While the above works show good results on acted data, they fail to address the more difficult non-acted scenario due to the exaggerated displays of emotional body motions. Kleinsmith et al. [307] used low-level posture descriptions and feature analysis using non-acted body gestures to implement MLP-based emotion and dimension recognition. Volkova et al. [308] further investigated whether emotional body expressions could be recognized when they were recorded during natural scenarios. To explore the emotion recognition from daily actions, Fourati et al. [309] recorded varieties of emotional body expressions in daily actions and constructed a new database.

*5.3.2.2 DL-based EBGR*

Although DL-based EBGR systems do not require to design of a tailored-feature extractor, they often pre-processed the input data based on the commonly-used pose estimation models or low-level feature extractors [310]. High-level features can be learned in spatial, temporal or spatial-temporal dimensions through the CNN-based network [311], the LSTM-based network [310] or the CNN-LSTM based network [312]. Recently, many studies have demonstrated the advantages of effectively combining different DL-based models and the attention mechanism [313] to improve the performance of EBGR.

**ConvNet-RNN learning for EBGR.** Ly et al. [312] first utilized the hashing model to detect keyframes of upper-body videos, and then applied a CNN-LSTM network to extract sequence information. The model employed on FABO achieved recognition accuracy of 72.5%. Avola et al. [314] investigated a framework of non-acted EBGR based on 3D skeleton and DNNs, which consist of MLP and N-stacked LSTMs. Shen et al. [310] proposed full-body based EBGR by fusing RGB features of optical flow extracted by the temporal segment network [315], and skeleton features extracted by spatial-temporal graph convolutional networks [316].

**Zero-shot based EBGR.** Due to the complexity and diversity of human emotion through body gestures, it is difficult to enumerate all emotional body gestures and collect enough samples for each category [317]. Therefore, the existing methods fail to determine which emotional state a new body gesture belongs to. In order to recognize unknown emotions from seen body gestures or to know emotions from unseen body gestures, Banerjee et al. [318] and Wu et al. [313] introduced the generalized zero-shot learning framework, including CNN-based feature extraction, autoencoder-based representation learning, and emotion classifier.

*5.4 Physiological-based emotion recognition*

Facial expressions, text, voice, and body gesture from a human being can be easily collected. As the reliability of physical information largely depends on the social environment and cultural background, and personalities of testers, their emotions are easy to be forged [20]. However, the changes in physiological signals directly reflect the changes in human emotions, which can help humans recognize, interpret and simulate emotional states [30,319]. Therefore, it is highly objective to learn human emotions through physiological signals [320].

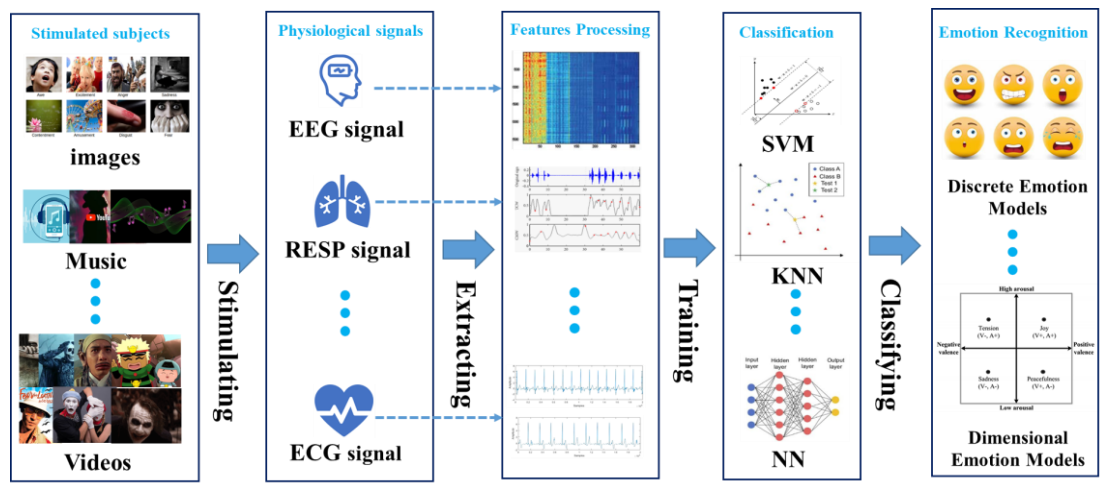

**Fig. 5.** The diagram of emotion recognition via physiological signals.

The diagram of physiological-based emotion recognition, as shown in Fig. 5, typically includes the following five aspects: 1) Stimulating subjects' emotions with images, music and videos; 2) Recording physiological signals that mainly include EEG, skin conductance, RESP, heart rate, EMG, and ECG; 3) Extracting features through physiological signals pre-processing, feature analysis, feature selection and reduction; 4) Training the classification model such as SVM, KNN, LDA, RF, NB and NN, etc.; and 5) Emotion recognition based on a discrete emotion model or a dimensional emotion model.

Among the above physiological emotion signals, EEG or ECG can provide simple, objective, and reliable data for identifying emotions [321], and is most frequently used for sentiment analysis and emotion recognition. Afterwards, we review EEG-based and ECG-based emotion recognition in this subsection. Table 10 shows an overview of representative methods for physiological-based emotion recognition, as detailed next.

5.4.1   *EEG-based emotion recognition*

Compared with other peripheral neuro-physiological signals, EEG can directly measure the changes of brain activities, which provides internal features of emotional states [20]. Besides, EEG with a high temporal resolution makes it possible to monitor a real-time emotional state. Therefore, various EEG-based emotion recognition techniques [29,322] have been developed recently.

5.4.1.1   *ML-based EEG emotion recognition*

The performance of ML-based EEG-based emotion recognition [20] depends on how to properly design feature extraction, feature dimensionality reduction (or feature selection), and classification methods.

The core objective of feature extraction is to extract important EEG features, which contain time-domain, frequency-domain, and time–frequency-domain features. The fast Fourier transform (FFT) analysis is often used to transform the EEG into the power spectrum [323]. Feature dimensionality reduction is an important step in EEG-based emotion recognition due to the redundancy of EEG. Yoon and Chung [323] used the FFT analysis for feature extraction and the Pearson correlation coefficient (PCC) for feature selection. Yin et al. successively designed a novel transfer recursive feature elimination [324] and the dynamical recursive feature elimination [325] for EEG feature selection, which determined sets of robust EEG features. Puk et al. [326] investigated an alternating direction method of multipliers ADMM-based sparse group lasso (SGL), with hierarchical splitting for recognizing the discrete states of three emotions. He et al. [327] designed a firefly integrated optimization algorithm (FIOA) to realize the optimal selection of the features subset and the classifier, without stagnating in the local optimum for the automatic emotion recognition. The FIOA evaluated on DEAP can gradually regulate the balance between ACC and feature number in the whole optimization process, which achieves an accuracy of 95.00%.

The commonly used ML-based classifier for EEG-based emotion recognition is SVM or its variations. For example, Atkinson and Campos [328] combined the mutual information-based EEG feature selection approach and SVM to improve the recognition accuracy. Yin et al. successively designed the linear least square SVM [324] and the selected least square SVM [325] for EEG-based emotion recognition.

*5.4.1.2 DL-based EEG emotion recognition*

Different from the pipeline of ML-based EEG-based emotion recognition, Gao et al. [329] utilized CNNs and restricted Boltzmann machine (RBM) with three layers, to simultaneously learn the features and classify EEG-based emotions. The CNN-based emotion recognition with subject-tied protocol achieves an accuracy of 68.4%. The affective computing team directed by Prof. Zheng [330–332] from Southeast University, China, has proposed various EEG-based emotion recognition networks such as bi-hemispheres domain adversarial neural network (BiDANN) [330], instance-adaptive graph network [331] and variational pathway reasoning [332].

As the biological topology among different brain regions can capture both local and global relations among different EEG channels, Zhong et al. [333] designed a regularized graph neural network (RGNN), which consists of both regularization operators of node-wise domain adversarial training and emotion-aware distribution learning. Gao et al. [334] proposed a channel-fused dense convolutional network for EEG-based emotion recognition. Considering the spatial information from adjacent channels and symmetric channels, Cui et al. [335] proposed the regional-asymmetric CNN (RACNN), including temporal, regional and asymmetric feature extractors. An asymmetric differential layer is introduced into three feature extractors to capture the discriminative information, by considering the asymmetry property of emotion responses. The RACNN achieves prominent results with average accuracies of 96.88% and 96.28% on DEAP [96] and DREAMER [321], respectively.

*5.4.2 ECG-based emotion recognition*

ECG records the physiological changes of the human heart in different situations through the autonomous nervous system activity. With the change of human emotion or sentiment state, ECG will detect the corresponding waveform transformation [336], which can provide enough information in emotion recognition. Next, we introduce ECG-based emotion recognition using ML models or DL models.

*5.4.2.1 ML-based ECG emotion recognition*

Following the diagram of physiological-based emotion recognition, Hsu et al. [336] first constructed a music-induced ECG emotion database, and then developed a nine-stage framework for automatic ECG-based emotion recognition, including 1) Signal preprocessing; 2) R-wave detection; 3) Windowing ECG recording; 4) Noisy epoch rejection; 5) Feature extraction based on the time-, and frequency-domain and nonlinear analysis; 6) Feature normalization; 7) Feature selection using sequential forward floating selection-kernel-based class separability; 8) Feature reduction based on the generalized discriminant analysis, and 9) Classifier construction with LS-SVM. Note that not all ML-based ECG emotion recognition methods follow the abovementioned steps, but the steps of feature extraction, feature selection and classifier are indispensable.

The ECG features can be directly extracted in the time domain. Bong et al. [337] extracted three time-domain features: heart rate, mean R peak amplitude, and mean R-R intervals to detect human emotional stress detection. Another common way is to transform time-domain ECG features into those in other domains. For example, Jerritta et al. [338] applied FFT, Discrete Wavelet Transform (DWT), and Hilbert Huang Transform (HHT) to transform ECG signals into frequency-domain features, and then utilized PCA and Tabu search to select key features in low-, high- and total (low and high together) frequency range. Note that both [338] and [337] used the SVM to implement the final emotion classification.

It may be beneficial to combine different types of ECG features. Cheng et al. [339] computed linear-derived features, nonlinear-derived features, time-domain features, and time-frequency domain features from ECG and its derived heart rate variability, and then fused them for SVM-based negative emotion detection. The experiments implemented on BioVid Emo DB [340] show that [339] achieves an accuracy of 79.51%, with a minor time cost of 0.13ms in the classification of positive and negative emotion states.

Statistic ECG features are also useful. Selvaraj et al. [341] proposed a non-linear Hurst feature extraction method by combining the rescaled range statistics and the finite variance scaling with higher-order statistics. They further investigated the performances of four conventional classifiers of NB, RT, KNN and fuzzy KNN for emotion classification. A novel Hurst feature and fuzzy KNN achieved recognition accuracy of 92.87%. Ferdinando et al. [342] investigated the effect of feature dimensionality reduction in ECG-based emotion recognition. A bivariate empirical mode decomposition was employed to compute features for KNN based on the statistical distribution of dominant frequencies.

5.4.2.2 *DL-based FER ECG emotion recognition*

Chen et al. [343] proposed a novel EmotionalGAN-based framework to enhance the generalization ability of emotion recognition by incorporating the augmented ECG samples generated by EmotionalGAN. Compared with that using only original data, around 5% improvement of average accuracy shows the significance of GAN-based models on the emotion recognition task. Sarkar and Etemad [344] introduced one self-supervised approach to training the signal transformation recognition network (STRN) to learn spatiotemporal features and abstract representations of the ECG. The weights of convolutional layers in the STRN are frozen and then train two dense layers to classify arousal and valence.

**Table 10.** Overview of the representative methods for physiological-based emotion recognition.

| Publication | Year | Feature Representation | Classifier | Database | Performance |
|---|---|---|---|---|---|
| *EEG-based Emotion Recognition* | | | | | |
| [328] | 2016 | mRMR-based | SVM | DEAP | A/V: 73.06/73.14 |
| [326] | 2019 | ADMM-based | Multi-class SGL | MAHNOB | 3 classes: 74.0 |
| [330] | 2018 | BiDANN | LSTM+Softmax | SEED | 3 classes: 92.38 |
| [334] | 2020 | Spatial-temporal | Pooling+Softmax | SEED | 3 classes: 90.63 |
| | | | | DEAP | A/V: 92.92/92.24 |
| *ECG-based Emotion Recognition* | | | | | |
| [342] | 2017 | Feature Fusion+LLN | KNN | Mahnob-HCI | A/V: 66.1/64.1 |
| [339] | 2017 | Feature Fusion | SVM | BioVid EmoDB | 2 classes: 79.15 |
| [343] | 2019 | EmotionalGAN | SVM | DECAF | A/V: 58.6/59.4 |
| [344] | 2020 | Self-supervised CNN | FC+Sigmoid | SWELL | A/V: 96.0/96.3 |
| | | | | AMIGOS | A/V: 85.8/83.7 |

## 6. Multimodal affective analysis

We have reviewed the relevant studies of unimodal feature extraction and emotion classification, in this section we then describe how to integrate multiple unimodal signals to develop a framework of multimodal affective analysis [345], which can be regarded as the fusion of different modalities [346], aiming to achieve a more accurate result and more comprehensive understanding than unimodal affect recognition [347,348].

Nowadays, most reviews of multimodal affective analysis [12,14,17] focus on multimodal fusion strategies and classify them into feature-level fusion (or early fusion), decision-level fusion (or late fusion), model-level fusion, and hybrid-level fusion. However, the multimodal affective analysis can be also varied with combinations of different modalities. Therefore, we categorize multimodal affective analysis into multi-physical modality fusion for affective analysis, multi-physiological modality fusion for affective analysis, and physical-physiological modality fusion for affective analysis, and further classify them based on four kinds of fusion strategies. Fig. 6 illustrates prominent examples of using different fusion strategies:

1. Feature-level fusion combines features extracted from the multimodal inputs to form one general feature vector, which is then sent into a classifier. Fig. 6 (a), (b) and (c) show examples based on feature-level fusion for visual-audio modalities, text-audio modalities, and visual-audio-text modalities, respectively.
2. Decision-level fusion connects all decision vectors independently generated from each modality into one feature vector. Fig. 6 (d) shows one example based on decision-level fusion for multi-physiological modalities of EGG, ECG and EDA.
3. Model-level fusion discovers the correlation properties between features extracted from different modalities and uses or designs a fusion model with relaxed and smooth types such as HMM and two-stage ELM [349]. Fig. 6 (e) and (f) are two examples based on model-level fusion for physical-physiological modalities and visual-audio-text modalities, respectively.
4. Hybrid fusion combines feature-level fusion and decision-level fusion. Fig. 6 (g) shows one example based on hybrid fusion for visual-audio-text modalities.

*6.1 Multi-physical modality fusion for affective analysis*

In light of common manners of modality combinations, we categorize multi-physical modalities fusion for affective analysis into visual-audio emotion recognition [350,31], text-audio emotion recognition [351,352], and visual-audio-text emotion recognition [353,354]. Table 11 shows an overview of representative methods for multi-physical affective analysis, as detailed next.

6.1.1 *Visual-audio emotion recognition*

Visual and audio signals are the most natural and affective cues to express emotions when people communicate in daily life [355]. Many research works [356–359,350] show that visual-audio emotion recognition outperforms visual or audio emotion recognition.

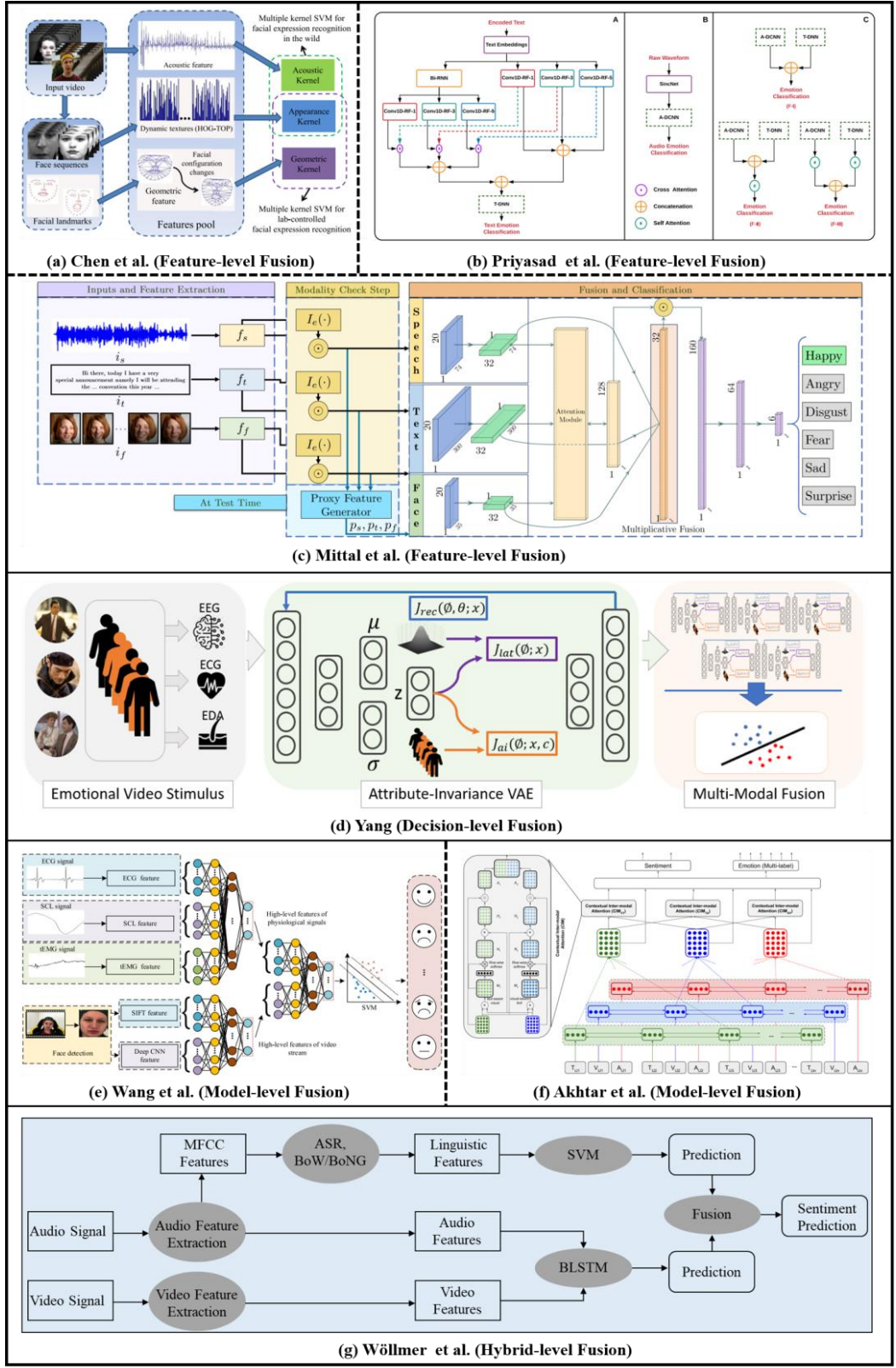

**Fig. 6.** Taxonomy of multimodal affective analysis. (a) Feature-level fusion for visual-audio emotion recognition adopted from [360]; (b) Feature-level fusion for text-audio emotion recognition adopted from [361]; (c) Feature-level fusion for visual-audio-text emotion recognition adopted from [362]; (d) Decision-level fusion for multi-physiological affective analysis adopted from [363]; (e) Model-level fusion for physical-physiological affective analysis adopted from [35]; (f) Model-level fusion for visual-audio-text emotion recognition adopted from [23]; (g) Hybrid-level fusion for visual-audio-text emotion recognition adopted from [105].

**Feature-level fusion**. Chen et al. [360] (Fig. 6 (a)) proposed ML-based visual-audio emotion recognition by fusing dynamic HOG-TOP texture features and acoustic/ geometric features. These two kinds of features are then sent into a multiple kernel SVM for FER both under the wild and lab-controlled environments. Tzirakis et al. [31] adopted a CNN and a deep residual network to extract audio and visual features, respectively; and then concatenated these visual-audio features to feed into a 2-layer LSTM to predict Arousal-Valence values.

Various attention mechanisms have been successfully applied for visual-audio emotion recognition [364,365]. For example, Zhang et al. [365] introduced an embedded attention mechanism to obtain the emotion-related regions from their respective modalities. To deeply fuse video-audio features, the factorized bilinear pooling (FBP) fusion strategy was proposed in consideration of feature differences in expressions of video frames. The FBP achieves a recognition accuracy of 62.48% on AFEW of the audio-video sub-challenge in EmotiW2018. Zhao et al [364] proposed a novel deep visual-audio attention network (VAANet) with specific attention modules and polarity-consistent cross-entropy loss. Specifically, spatial, channel-wise, and temporal attentions are integrated with a 3D CNN [366] for video frame segments, and spatial attention is integrated with a 2D CNN (ResNet-18) for audio MFCC segments.

**Decision-level fusion**. Hao et al. [367] proposed an ensemble visual-audio emotion recognition framework based on multi-task and blending learning with multiple features. Specifically, SVM classifiers and CNNs for handcraft-based and DL-based visual-audio features generated four sub-models, which are then fused to predict the final emotion based on the blending ensemble algorithm. The work [367] achieved the average accuracies of 81.36% (speaker-independent) and 78.42% (speaker-dependent) on eNTERFACE [368].

**Model-level fusion.** It requires an ML-based model (e.g., HMM, Kalman filters and DBN) to construct the relationships of different modalities to make decisions. Lin et al. [33] proposed a semi-coupled HMM (SC-HMM) to align the temporal relations of audio-visual signals, followed by a Bayesian classifier with an error weighted scheme. The SC-HMM with Bayesian achieves prominent average accuracies of 90.59% (four-class emotions) and 78.13% (four quadrants) on the multimedia human-machine communication posed database and public SEMAINE database, respectively. Glodek et al. [369] designed Kalman filters based on a Markov model to combine temporally ordered classifier decisions with the reject option to recognize the affective states. For audio-visual data, the unimodal feature extractors and base classifiers are fused based on a Kalman filter and confidence measures.

Zhang et al. [37] utilized CNNs to extract audio-visual features and developed a deep fusion method (DBNs) for feature fusion. A linear SVM classifier achieved average accuracies of 80.36%, 54.57% and 85.97% on RML, eNTERFACE05 and BAUM-1s, respectively. Similarly, Nguyen et al. [278] proposed a score-level fusion approach to compute all likelihoods of DBNs trained on spatiotemporal information of the audio-video streams. Hossain and Muhammad [349] used a 2D CNN and a 3D CNN to extract high-level representations of the pre-processed audio-video signals. A two-stage ELM based fusion model with an SVM classifier was designed to estimate different emotional states.

6.1.2 *Text-audio emotion recognition*

Although SER [26,163,178] and TSA [156,370] have achieved significant progress, performing the two tasks separately makes it hard to achieve compelling results [361]. Text-audio emotion recognition approaches use linguistic content and speech clues to enhance the performance of the unimodal emotion recognition system [352,371].

**Feature-level fusion**. Yoon et al. [372] proposed a deep dual recurrent neural network for encoding audio-text sequences and then concatenated their outputs to predict the final emotion. It achieves an accuracy of 71.8% (four classes) on IEMOCAP. Afterwards, Cai et al. [373] designed an improved CNN and Bi-LSTM to extract spatial features and capture their temporal dynamics. Considering the phonemes effect on emotion recognition, Zhang et al. [371] utilized a temporal CNN to investigate the acoustic and lexical properties of phonetic information based on unimodal, multimodal single-stage fusion, and multimodal multi-stage fusion systems. To deeply exploit and fuse text-acoustic features for emotion classification, Priyasad et al. [361] (Fig. 6 (b)) designed T-DNN (DCNNs and Bi-RNN followed by other DCNNs) and A-DCNN (SincNet with band-pass filters followed by a DCNN) to extract textual features and learning acoustic features, respectively. Textual-acoustic features are fused with different attention strategies (self-attention or no attention) to predict four emotions on IEMOCAP.

**Decision-level fusion.** Wu et al. [374] fused acoustic-prosodic information based on a meta decision tree (MDT) with multiple base classifiers (e.g. GMM, SVM, and MLP), and employed a maximum entropy model (MaxEnt) to establish the relationship between emotional states and emotion association rules in semantic labels for speech emotion recognition and text emotion recognition, respectively. Using the weighted product fusion strategy, AP-based and SL-based emotion confidences are fused to predict final emotions.

**Feature-level fusion versus decision-level fusion.** To verify which feature-level fusion or decision-level fusion is more effective in text-audio emotion recognition, Jin et al. [375] firstly generated new lexical features and different acoustic features and then utilized two fusion strategies to achieve four-class recognition accuracies of 55.4% and 69.2% on IEMOCAP, respectively. Considering the emotional dialogue composed of sound and spoken content, Pepino et al. [376] exploited multiple dual RNNs to encode audio-text sequences. The feature-level fusion and decision-level fusion approaches in 3 different ways are explored to compare their performances.

6.1.3 *Visual-audio-text emotion recognition*

The vocal modulation, facial expression, and context-based text provide important cues to better identify the true affective states of the opinion holder [377,378]. For example, when a lady is proposed and then she says "I do" in tears, none of textual-based, audio-based, or visual-based emotion recognition models can predict confident results. While visual-audio-text emotion recognition leads to a better solution.

**Feature-level fusion.** Veronica et al. [379] used BoW, OpenEAR [380], and vision software to extract linguistic, audio and visual features, respectively. Compared with three unimodal and three bimodal models, the multimodal (text-audio-visual) model can significantly improve the model performance over the individual use of one modality or fusion of any two modalities. In addition, Poria et al. [354] utilized a standard RNN, a deep CNN, and openSMILE to capture temporal dependence of visual data, spatial information of textual data and low-level descriptors of audio data, respectively. The MKL is further designed for feature selection of different modalities to improve the recognition results. Mittal et al. [362] (Fig. 6 (c)) proposed the multiplicative multimodal emotion recognition (M3ER): firstly, feature vectors are extracted from the raw three modalities; then, these features are transferred into modality check step to retain the effective features and discard the ineffectual ones which are used to regenerate proxy feature vectors; finally, selected features are fused to predict six emotions based on the multiplicative feature-level fusion combined with attention module. The M3ER achieved better recognition accuracies of 82.7% and 89.0% on IEMOCAP and CMU-MOSEI, respectively. Different from language-independent approaches for English or German sentiment analysis, Chinese sentiment analysis not only understands symbols with explicit meaning, but also captures phonemic orthography (tonal language) with implicit meaning. Based on this assumption, Peng et al. [381] proposed reinforcement learning based disambiguate intonation for sentiment analysis (DISA) which consists of policy network, embedding lookup, loss computation, and feature-level fusion of three modalities. By integrating phonetic features with textual and visual representations, multimodal Chinese sentiment analysis significantly outperforms than unimodal models.

**Feature-level fusion versus decision-level fusion.** Poria et al. [382] utilized a CNN to extract textual features, calculated handcrafted features from visual data, and generated audio features by the openSMILE. These three kinds of feature vectors were fused and then used to train a classifier based on the MKL. Feature-level fusion and decision-level fusion with feature selection are conducted to implement unimodal, bimodal and multimodal emotion recognition employed on HOW [104]. Experimental results show that textual emotion recognition achieves the best performance among three unimodal emotion recognition methods, and in feature-level fusion, visual-audio-text emotion recognition outperforms other unimodal and bimodal emotion recognition. Besides, the accuracy of feature-level fusion is significantly higher than that of decision-level fusion at the expense of computational speed.

**Model-level fusion.** Considering the interdependencies and relations among the utterances of a video [378], Poria et al. [383] introduced some variants of the contextual LSTM into the hierarchical architecture to extract context-dependent multimodal utterance features. The visual-audio-text features are concatenated and then fed into a contextual LSTM to predict emotions, reaching 80.3%, 68.1% and 76.1% on MOSI, MOUD and IEMOCAP, respectively. Akhtar et al. [23] (Fig. 6 (f)) proposed an end-to-end DL-based multi-task learning for multimodal emotion recognition and sentiment analysis. Due to the unequal importance of three modalities, a context-level inter-modal (CIM) attention module is designed to learn the joint association between textual-audio-visual features of utterances captured by three bi-directional gated recurrent unit (biGRU) networks. These features of three CIMs and three individual modalities are concatenated to generate high-level feature representation to predict sentiments and multi-label emotions.

**Hybrid-level fusion.** To take full advantage of feature-based fusion and decision-based fusion, and overcome the disadvantages of both, Wöllmer et al. [105] (Fig. 6 (g)) used BoW or Bag-of-N-Gram (BoNG) features with SVM for linguistic sentiment classification, and fused audio-video features with Bi-LSTM for audio-visual emotion recognition. And then these two prediction results were fused to obtain the final emotion under a strategy of weighted fusion.

**Table 11.** Overview of representative methods for multi-physical affective analysis.

| Publication | Year | Feature Representation | Classifier | Fusion Strategy | Database | Performance |
|---|---|---|---|---|---|---|
| *Visual-audio Emotion Recognition* | | | | | | |
| [360] | 2018 | Acoustic, Geometric and HOG-TOP | Multiple kernel SVM | Feature-level | CK+ <br> AFEW | [1] 7 classes: 95.7 <br> [1] 7 classes: 45.2 |
| [31] | 2017 | CNN, Resnet | LSTM | Feature-level | RECOLA | A/V: 78.8/73.2 |
| [364] | 2020 | 2D ResNet+Attention <br> 3D ResNet+Attention | FC | Feature-level | VideoEmotion-8 <br> Ekman-6 | 8 classes: 54.50 <br> 6 classes: 55.30 |
| [367] | 2020 | Multitask CNN | Meta-Classifier | Decision-level | eNTERFACCE | 6 classes: 81.36 |
| [278] | 2017 | C3D + DBN | Score-level Fusion | Model-based | eNTERFACE | 6 classes: 89.39 |
| [349] | 2019 | 2D CNN, 3D CNN | ELM-based fusion, SVM | Model-based | Big Data <br> eNTERFACE | 3 classes: 91.3 <br> 6 classes: 78.42 |
| *Text-audio Emotion Recognition* | | | | | | |
| [361] | 2020 | A-DCNN, T-DNN <br> Self-attention | FC | Feature-level | IEMOCAP | [2] 4 classes: 80.51 <br> [3] 4 classes: 79.22 |
| [374] | 2011 | Acoustic-prosodic <br> Semantic labels | Base classifiers, MDT, MaxEnt | Decision-level | 2033 utterances | 4 classes: 83.55 <br> [4] 4 classes: 85.79 |
| [376] | 2020 | Acoustic features <br> Word embeddings | Pooling <br> Scalar weight fusion | Feature-level <br> Decision-level | IEMOCAP/MSP-PODCAST | [5] 65.1/[5] 58.2 <br> [5] 63.9/[5] 58.0 |
| *Visual-audio-text Emotion Recognition Emotion Recognition* | | | | | | |
| [362] | 2020 | Proxy and Attention <br> Multiplicative fusion | FC | Feature-level | IEMOCAP <br> CMU-MOSEI | 4 classes[1]: 82.7 <br> 6 classes[1]: 89.0 |
| [382] | 2015 | CNN, handcrafted, CFS, PCA | MKL | Feature-level <br> Decision-level | HOW | 3 classes: 88.60 <br> 3 classes: 86.27 |
| [23] | 2019 | Three Bi-GRU | CIM-attention | Model-based | CMU-MOSEI | [2] Multi-label: 62.8 <br> [2] 2-class: 80.5 |
| [105] | 2013 | Facial movement, MFCC | SVM, BiLSMT | Hybrid-level | SEMAINE | [5] 65.2 |

[1] LOSO; [2] WA; [3] UA; [4] The recognition accuracy considering the individual personality trait for personalized application; [5] The median values evaluated on IEMOCAP/MSP-PODCAST databases; [6] Mean WA of Arousal, Expectation, Power and Valence.

*6.2 Multi-physiological modality fusion for affective analysis*

With the enhancement and refinement of wearable technologies, automatic affective analysis based on multi-physiological modalities has attracted more attention [384]. However, due to the complexity of emotion and significant individual differences in physiological responses [348], it is difficult to achieve satisfactory prediction performance with EEG-based or ECG-based emotion recognition. In this sub-section, we review multi-physiological modality fusion for affective analysis. Table 12 provides an overview of representative methods for multi-physiological affective analysis, as detailed next.

**Feature-level fusion.** Li et al. [385] collected multi-physiological emotion signals induced via music, picture or video, and extracted low-level descriptors and statistical features from the signal waveforms. Then, they fused these features for emotion prediction using ML-based classifiers with a Group-based IRS (Individual Response Specificity) model. The combination of RF with Group-based IRS achieved a higher accuracy of about 90% in the imbalance of the emotion database. Similarly, Nakisa et al. [386] pre-processed, extracted and fused time-frequency features of EEG and BVP, and then fed them into an LSTM network, whose hyperparameters were optimized by differential evolution (DE). On their self-collected dataset, its overall accuracy was 77.68% for the four-quadrant dimensional emotions.

Using ECG, GSR, ST, BVP, RESP, EMG, and EOG selected from DEAP [96], Verma and Tiwary [347] extracted multi-resolution features based on DWT and used them to estimate the valence-arousal-dominance emotions. Hossain et al. [387] extracted 9 statistical features, 9 power spectral density features, and 46 DBN features of EDA, Photo plethysmogram, and zEMG. These features were fused to train a fine gaussian SVM to recognize 5 basic emotions. Ma et al. [388] designed a multimodal residual LSTM network for learning the dependency of high-level temporal-feature EEG, EOG, and EMG to predict emotions. It achieves the classification accuracies of 92.87% and 92.30% for arousal and valence, respectively.

**Decision-level fusion.** Wei et al. [389] selected EEG, ECG, RESP and GSR from MAHNOB-HCI [107], and designed a linear fusing weight matrix to fuse the outputs from multiple SVM classifiers to predict 5 basic emotions. Its highest average accuracy of recognition was 84.6%. To explore the emotional physiological-based temporal features, Li et al. [390] transformed EEG, ECG and GSR selected from AMIGOS [99] into spectrogram images to represent their time-frequency information. The attention-based Bi-LSTM-RNNs was designed to automatically learn the best temporal features, which were further fed into a DNN to predict the probability of unimodal emotion. The final emotion state was computed based on either an equal weights scheme or a variable weights scheme. As personality-specific human beings often show different physiological reactions after being stimulated by emotional elements, Yang and Lee [363] (Fig. 6 (d)) proposed an attribute-invariance loss embedded variational autoencoder to learn personality-invariant representations. With EEG, ECG and EDA selected from AMIGOS, different features were extracted and then fed into different SVM classifiers to predict unimodal classification results. Dar et al. [391] designed a 2D-CNN for EEG and combined LSTM and 1D-CNN for ECG and GSR. By using majority voting based on decisions made by multiple classifiers, the framework achieved the overall highest accuracy of 99.0% and 90.8% for AMIGOS [99] and DREAMER [321], respectively.

**Model-level fusion.** Yin et al. [392] first pre-processed and extracted 425 salient physiological features of 7 signals selected from DEAP. Separate deep hidden neurons in stacked-AEs were then investigated to extract higher-level abstractions of these salient features. An ensemble of deep classifiers with an adjacent graph-based hierarchical feature fusion network was designed for recognizing emotions based on a Bayesian model.

**Table 12.** Overview of some representative methods for multi-physiological affective analysis.

| Publication | Year | Signal | Architecture | Fusion Strategy | Database | Performance |
| --- | --- | --- | --- | --- | --- | --- |
| [347] | 2014 | EEG, GSR, ST, BVP, RESP, EMG, EOG | DWT, Multiple kernel SVM | Feature-level | DEAP | 13 classes: 85.00 |
| [387] | 2019 | EDA, BVP, zEMG | DBN, FGSVM | Feature-level | DEAP | 5 classes: 89.53 A/V: 65.1/61.8 |
| [363] | 2019 | EEG, ECG, EDA | AILE-VAE, SVM | Decision-level | AMIGOS | A/V: 68.8/ 67.0 |
| [390] | 2020 | EEG, ECG, GSR | Attention-based LSTM-RNNs, DNN | Decision-level | AMIGOS | A/V: 83.3/79.4 |
| [392] | 2017 | EEG, EOG, EMG, ST, GSR, BVP, RESP | Deep stacked AE, Bayesian model | Model-based | DEAP | A/V: 77.19/76.17 |

*6.3 Physical-physiological modality fusion for affective analysis*

Since the change of human emotions is a complicated psycho-physiological activity, the research on affective analysis is related to many cues (e.g., behavioral, physical, and psychological signals) [393,394]. Researchers have focused on the physical-physiological modality fusion for affective analysis via mining the strong external expression of physical data and undisguisable internal changes of physiological signals [395–397]. Table 13 provides an overview of representative methods for physical-physiological affective analysis, as detailed next.

**Feature-level fusion.** To fully leverage the advantages of the complementary property between EEG and eye movement, Liu et al. [398] proposed a bimodal deep AE based on an RBM to recognize emotions on SEED and DEAP. Soleymani et al. [399] designed a framework of video-EEG based emotion detection using an LSTM-RNN and continuous conditional random fields. Wu et al. [400] proposed a hierarchical LSTM with a self-attention mechanism to fuse the facial features and EEG features to calculate the final emotion. Yin et al. [397] proposed an efficient end-to-end framework of EDA-music fused emotion recognition, denominating it as a 1-D residual temporal and channel attention network (RTCAN-1D). Specially, the RTCAN consists of shallow feature extraction, residual feature extraction, the attention module stacked by a signal channel attention module, and a residual non-local temporal attention module. It achieved outstanding performances in AMIGOS, DEAP and PMEmo [401].

**Feature-level fusion versus decision-level fusion.** Huang et al. [34] proposed video-EEG based multimodal affective analysis by fusing external facial expression of spatiotemporal local monogenic binary pattern and discriminative spectral power of internal EEG on feature-level and decision-level aspects. According to their experiments, the multimodal affective analysis achieves a better performance than the unimodal emotion recognition; and when it comes to fusion strategies, decision-level fusion outperforms feature-level fusion in the recognition of valence and arousal on MAHNOB-HCI.

**Model-level fusion.** Wang et al. [35] (Fig. 6 (e)) designed a multimodal deep belief network for fusing and optimizing multiple psycho-physiological features, a bimodal DBN (BDBN) for representing discriminative video-based features, and another BDBN to extract high multimodal features of both video and psycho-physiological modalities. The SVM was employed for emotion recognition after the features of all modalities were integrated into a unified dimension.

**Table 13.** Overview of the representative methods for physical-physiological affective analysis.

| Publication | Year | Signal | Architecture | Fusion Strategy | Database | Performance |
| --- | --- | --- | --- | --- | --- | --- |
| [397] | 2020 | EDA, Music | RTCAN-1D | Feature-level | PMEmo | Arousal: 82.51/Valence: 77.3 |
| [400] | 2020 | EEG, Video | LSTM, Self-attention | Model-level | DEAP | [2]AAVA: 85.9 |
| [395] | 2015 | EEG, Eye Movements | Feature extraction Fuzzy integral fusion | Feature-level Decision-level | SEED | 3 classes: 83.70 [1]3 classes: 87.59 |
| [35] | 2020 | ECG, SCL, tEMG, Video | Inception-ResNet-v2, BDBN, SVM | Model-level | BioVid Emo DB | 5 classes: 80.89 [3]AAVAL: 84.29 |

[1]The fuzzy integral fusion strategy; [2]AAVA = Average accuracy of Valence-Arousal; [3]AAVAL = Average accuracy of Valence-Arousal-Liking.

## 7. Discussions

In this review, we have involved emotion models and databases commonly used for affective computing, as well as unimodal affect recognition and multimodal affective analysis. In this section, we mainly discuss the following aspects:

1) Effects of different signals (textual, audio, visual, or physiological) on unimodal affect recognition [152,194,244,289,327,331];
2) Effects of modality combinations and fusion strategies on multimodal affective analysis [37,371,375,383,388,400];
3) Effects of ML-based techniques [127,184,249,305,326] or DL-based methods [146,203,273,283,316,335] on affective computing;
4) Effects of some potential factors (e.g., released databases and performance metrics) on affective computing;
5) Applications of affective computing in real-life scenarios.

*7.1 Effects of different signals on unimodal affect recognition*

According to unimodal affect recognition based on the text [126,402], audio [178,194,199], visual [253,281,403], EEG [404], or ECG [336], we can find that the most widely used modality is the visual signal, mainly consisting of facial expressions and body gestures. The number of visual-based emotion recognition systems is comparable to the sum of that of systems based on other modalities since the visual signals are easier to capture than other signals and emotional information in visual signals is more helpful than other signals in recognizing the emotion state of human beings. Visual-based emotion recognition is more effective than audio-based emotion recognition because audio signals are susceptible to noise [405]. However, a study [23] reveals that textual-based affective analysis achieves the highest accuracy in emotion recognition and sentiment analysis. Although the physiological signals collected by wearable sensors are more difficult to obtain than physical signals, numerous EEG-based [334] or ECG-based [336] emotion recognition methods have been investigated and proposed due to their objective and reliable outcomes.

*7.2 Effects of modality combinations and fusion strategies on multimodal affective analysis*

The combination of different modalities and the fusion strategy are two key aspects of the multimodal affective analysis. Multimodal combinations are divided into multi-physical modalities, multi-physiological modalities, and physical-physiological modalities. The fusion strategies consist of feature-level fusion [361,387], decision-level fusion [375], hybrid-level fusion [105], and model-level fusion.

In the multi-physical modalities [406], there are three kinds of combinations of different modalities, consisting of visual-audio, text-audio and visual-audio-text. Integrating visual and audio information can enhance performance over unimodal affect recognition [407]. There are similar results in other combinations of multi-physical modalities [375], in which the text modality plays the most vital role in multimodal sentiment analysis [382,362]. In studies of multi-physiological modality fusion for affective analysis, in addition to EEG and ECG, other types of physiological signals (e.g., ECG, EOG, BVP, GSR, and EMG) are jointly combined to interpret emotional states [388,390]. The visual modality (facial expression, voice, gesture, posture, etc.) may also be integrated with multimodal physiological signals for visual-physiological affective analysis [397,400].

Two basic fusion strategies for multimodal affective analysis are feature-level fusion [361,387] and decision-level fusion [375]. The concatenation [362] or factorized bilinear pooling [365] of feature vectors is commonly used for feature-level fusion. The majority/average voting is often used for decision-level fusion. Linear weighted computing [374] can be utilized for both feature-level and decision-level fusion, by employing sum or product operators to fuse features or classification decisions of different modalities. According to the multimodal affective analysis, we find that feature-level fusion [388] is strikingly more common than decision-level fusion. The performance of an affect classifier based on feature-level fusion is significantly influenced by the time scales and metric levels of features coming from different modalities. On the other hand, in decision-level fusion, the input coming from each modality is modelled independently, and these results of unimodal affect recognition are combined in the end. Compared with feature-level fusion, decision-level fusion [362] is performed easier, but ignores the relevance among features of different modalities. Hybrid-level fusion [105] aims to make full use of the advantages of feature-based fusion and decision-based fusion strategies as well as overcome the disadvantages of either one. Unlike the above three fusion strategies, model-level fusion uses HMM [33] or Bayesian networks [392] to establish the correlation between features of different modalities and one relaxed fusion mode. The selection and establishment of HMM or Bayesian have a fatal effect on the results of the model-level fusion, which is often designed for one specific task.

*7.3 Effects of ML-based and DL-based models on affective computing*

The majority of the early works on affective computing have employed ML-based techniques [19,10,18]. The ML-based pipeline [124,181,249,309,326] consists of pre-processing of raw signals, hand-crafted feature extractor (feature selection if possible), and well-designed classifiers. Although various types of hand-crafted features have been designed for different modalities, ML-based techniques for affective analysis are hard to be reused across similar problems on account of their task-specific and domain-specific feature descriptors. The commonly used ML-based classifiers are SVM [182,253], HMM [179], GMM [180], RF [305], KNN [337] and ANN [374], of which the SVM classifier is the most effective one, and is indeed used in most tasks of ML-based affective computing. These ML-based classifiers are also used for final classification when the DL-based model is only designed for unimodal feature extraction [290] or multimodal feature analysis [35,387].

Nowadays, DL-based models have become hot spots and outperformed ML-based models in most areas of affective computing [13,135,43,17,39] due to their strong ability of feature representation learning. For static information (e.g., facial and spectrogram images), CNNs and their variants are designed to extract important and discriminative features [137,194,289]. For sequence information (e.g., physiological signals and videos), RNNs and their variants are designed for capturing temporal dynamics [143,201,310]. The CNN-LSTM models can perform the deep spatial-temporal feature extraction. Adversarial learning is widely used to improve the robustness of models by augmenting data [206,291] and cross-domain learning [156,192]. Besides, different attention mechanisms [142,199,203,279] and autoencoders [363,289] are integrated with DL-based techniques to improve the overall performance. It seems that DL-based methods have an advantage in automatically learning the most discriminative features. However, DL-based approaches have not yet had a huge impact on physiological emotion recognition, if compared with ML-based models [20].

*7.4 Effects of some potential factors on affective computing*

Throughout this review, we have consistently found that advances in affective computing are driven by various database benchmarks. For example, there are few video-physiological emotion recognition methods due to the limitations of physical-physiological emotion databases. In contrast, the rapid development of FER is inseparable from various baseline databases, which are publicly available and can be freely downloaded. Besides, some large-scale visual databases such as BU-4DFE and BP4D can be employed to pre-train the target model to recognize facial expressions [260] or micro-expressions [408]. However, there are significant discrepancies in size, quality, and collection conditions [274] across different databases. For example, most body gesture emotion databases contain only several hundred samples with limited gesture categories. What is worse, samples are typically collected in a laboratory environment, which is often far away from real-world conditions. Furthermore, the size and quality of databases have a more obvious effect on DL-based emotion recognition than on ML-based emotion recognition. Many studies have concluded that the reduced size of the available databases is a key factor in the quest for high-performance affective analysis [409,207]. To tackle this problem, the pre-trained DL-based models [410,411] may be transferred into task-based models specialized for affective analysis.

Although the representation of the natural affective states has no consensus, most affective analyses are trained and evaluated based on two types of emotion models: discrete models and dimensional models. In building the affective databases, either discrete or dimensional labels are typically chosen alternatively to fit the raw signals. For example, emotional images or sequences are typically matched with a discrete affective state (basic emotions or polarity). Affective recognition can be divided into classification (emotions, dimensional or polarity) and continuous dimensional regression (Pleasure, Arousal, Dominance Expectation, or Intensity) [412,413]. The metrics of accuracy, precision and recall are generally adopted for categorical or componential emotion classification. When the databases are imbalanced, the F-Measure (or F1-Score) seems to be the best choice out of the existing evaluation metrics of the emotional classification, across 10-fold cross validation and LOSO. The weighted average recall/F1-score (WAR/WF1) and the unweighted average recall/F1-score (UAR/UF1) are best suited for the classification performance of visual, audio or multimodal affective analysis [291]. On the other hand, MSE and RMSE are commonly used for the evaluation of continuous dimensional emotion prediction [414]. In order to describe the degree of coincidence, by integrating PCC and MSR, the coefficient concordance correlation coefficient is advised to assess the baseline performance assessment [415,416].

*7.5 Applications of affective computing in real-life scenarios*

In recent years, more and more research teams have shifted their focus to applications of affective computing in real-life scenarios [417,418]. In order to detect emotions and sentiments from the textual information, the SenticNet directed by Erik Cambria of NTU applied the research outputs of affective computing [106,150,419] and sentiment analysis [22,420–424] into many aspects of daily life, including HCI [425], finance [426] and social media monitoring and forecasting [132,427]. TSA is often used for recommender systems, by integrating diverse feedback information [428] or microblog texts [429]. The applications of visual emotion recognition include course teaching [430], smarter decision aid [431], HCI [432], dynamic quality adaption to the players in games [433–435], depression recognition [436] and for helping medical rehabilitation children affected by the autism spectrum condition [437]. In particular, audio and physiological signals are often used for detecting clinical depression and stress [100,166,438] due to the reliability and stability of audio/speech emotion signals and the accessibility of physiological signals from wearable devices [439]. As multimodal affective analysis can enhance the robustness and performance of unimodal affect recognition, more researches have begun to transform them into various real-life applications [440,441], making it a promising research avenue.

## 8. Conclusion and new developments

This review has comprehensively surveyed more than 400 papers, including an overview of the recent reviews on affective computing in Section 2, and built the taxonomy of affective computing with representative examples in Section 1. In Section 3, we categorize current emotion models based on psychological theories into discrete models and dimensional models, which determine the category of the output of the affective analysis. These recognition results via either classification or regression are evaluated by a range of corresponding metrics. More importantly, the development of affective computing requires benchmark databases for training and computational models for either DL-based or ML-based affective understanding. In Section 4, we survey five kinds of the commonly adopted baseline databases for affective computing, which are classified into textual, audio, visual, physiological, and multimodal databases. Most methods for affective analysis benefit from these released databases.

In Section 5 and Section 6, we introduced recent advances of affective computing, which are mainly grouped into unimodal affect recognition and multimodal affective analysis, and further divide them into ML-based techniques and DL-based models. The unimodal affect recognition systems are further divided into textual sentiment analysis, speech emotion recognition, visual emotion recognition (FER and EBGR) and physiological emotion recognition (EEG-based and ECG-based). Traditional ML-based unimodal affect recognition mostly investigates hand-crafted feature extractors or pre-defined rules and interpretable classifiers. In contrast, DL-based unimodal affect recognition further improves the ability of the feature representation and classification by designing deeper network architectures or task-specific network modules and learning objectives. Generally, in addition to employing different strategies (feature-level, decision-level, model-level, or hybrid fusion strategies), the multimodal affective analysis is divided into multi-physical approaches (visual-audio, text-audio and visual-audio-text modalities), multi-physiological approaches, and physical-physiological approaches. The performance of multimodal affective analysis is mainly affected by both modality combination and fusion strategy.

In Section 7, we discuss some important issues related to affective computing including effects of textual, audio, visual, or physiological signals on unimodal affect recognition, effects of modality combinations and fusion strategies on multimodal affective analysis, the effects of ML-based and DL-based models on affective computing, effects of some potential factors on affective computing, and applications of affective computing in the real-life scenarios.

Although affective computing systems using either unimodal or multimodal data have made significant breakthroughs, there are only a few robust and effective algorithms to predict emotion and recognize sentiment under diverse and challenging scenes. Hence, we would like to conclude this review with many important recommendations for future research in affective computing:

1) It will be instrumental to develop new and more extended baseline databases, particularly multimodal affect databases, consisting of different modalities (textual, audio, visual and physiological). Conditions should include both spontaneous and non-spontaneous scenarios, with the provision of annotating data in both discrete and dimensional emotion models.
2) There are some challenging tasks of affective analysis to be solved including FER under partial occlusion or fake emotion expression, physiological emotion recognition based on various complex signals, and a baseline model specifically for both discrete emotion recognition and dimensional emotion prediction.
3) There is significant space for improving fusion strategies, particularly with rule-based or statistic-based knowledge, to implement a mutual fusion of different modalities that can consider the role and importance of each modality in affect recognition.

4) Zero/few-shot learning or unsupervised learning methods (e.g., self-supervised learning) need to be further explored, particularly thanks to their potential to enhance the robustness and stability of affective analysis under limited or biased databases.
5) A prominent application of affective analysis is robotics. Advances presented in this review make it possible to conceive robots equipped with emotional intelligence, which can appropriately imitate and promptly respond to mankind affect and the surrounding environment.

## References

[1] K.S. Fleckenstein, Defining Affect in Relation to Cognition: A Response to Susan McLeod, J. Adv. Compos. 11 (1991) 447–53.
[2] R.W. Picard, Affective Computing, Cambridge, MA, USA: MIT Press, 1997.
[3] R.W. Picard, E. Vyzas, J. Healey, Toward machine emotional intelligence: analysis of affective physiological state, IEEE Trans. Pattern Anal. Mach. Intell. 23 (2001) 1175–1191. https://doi.org/10.1109/34.954607.
[4] J. Park, J. Kim, Y. Oh, Feature vector classification based speech emotion recognition for service robots, IEEE Trans. Consum. Electron. 55 (2009) 1590–1596. https://doi.org/10.1109/TCE.2009.5278031.
[5] M. Scheutz, The Affect Dilemma for Artificial Agents: Should We Develop Affective Artificial Agents?, IEEE Trans. Affect. Comput. 3 (2012) 424–433. https://doi.org/10.1109/T-AFFC.2012.29.
[6] D. McColl, A. Hong, N. Hatakeyama, G. Nejat, B. Benhabib, A Survey of Autonomous Human Affect Detection Methods for Social Robots Engaged in Natural HRI, J. Intell. Robot. Syst. 82 (2016) 101–133. https://doi.org/10.1007/s10846-015-0259-2.
[7] J.A. Healey, R.W. Picard, Detecting stress during real-world driving tasks using physiological sensors, IEEE Trans. Intell. Transp. Syst. 6 (2005) 156–166. https://doi.org/10.1109/TITS.2005.848368.
[8] L.-P. Morency, J. Whitehill, J. Movellan, Generalized adaptive view-based appearance model: Integrated framework for monocular head pose estimation, in: 2008 8th IEEE Int. Conf. Autom. Face Gesture Recognit., IEEE, Amsterdam, Netherlands, 2008: pp. 1–8. https://doi.org/10.1109/AFGR.2008.4813429.
[9] J.A. Balazs, J.D. Velásquez, Opinion Mining and Information Fusion: A survey, Inf. Fusion. 27 (2016) 95–110. https://doi.org/10.1016/j.inffus.2015.06.002.
[10] E. Cambria, Affective Computing and Sentiment Analysis, IEEE Intell. Syst. 31 (2016) 102–107. https://doi.org/10.1109/MIS.2016.31.
[11] M. Munezero, C.S. Montero, E. Sutinen, J. Pajunen, Are They Different? Affect, Feeling, Emotion, Sentiment, and Opinion Detection in Text, IEEE Trans. Affect. Comput. 5 (2014) 101–111. https://doi.org/10.1109/TAFFC.2014.2317187.
[12] S. Poria, E. Cambria, R. Bajpai, A. Hussain, A review of affective computing: From unimodal analysis to multimodal fusion, Inf. Fusion. 37 (2017) 98–125. https://doi.org/10.1016/j.inffus.2017.02.003.
[13] P.V. Rouast, M. Adam, R. Chiong, Deep Learning for Human Affect Recognition: Insights and New Developments, IEEE Trans. Affect. Comput. (2019) 1–1. https://doi.org/10.1109/TAFFC.2018.2890471.
[14] N.J. Shoumy, L.-M. Ang, K.P. Seng, D.M.M. Rahaman, T. Zia, Multimodal big data affective analytics: A comprehensive survey using text, audio, visual and physiological signals, J. Netw. Comput. Appl. 149 (2020) 102447. https://doi.org/10.1016/j.jnca.2019.102447.
[15] E. Paul, Basic emotions, Wiley Online Library, New York: Sussex U.K, 1999.
[16] A. Mehrabian, Basic dimensions for a general psychological theory : implications for personality, social, environmental, and developmental studies, Cambridge : Oelgeschlager, Gunn & Hain, 1980. http://archive.org/details/basicdimensionsf0000mehr (accessed September 1, 2020).
[17] Y. Jiang, W. Li, M.S. Hossain, M. Chen, A. Alelaiwi, M. Al-Hammadi, A snapshot research and implementation of multimodal information fusion for data-driven emotion recognition, Inf. Fusion. 53 (2020) 209–221. https://doi.org/10.1016/j.inffus.2019.06.019.
[18] C.A. Corneanu, M.O. Simón, J.F. Cohn, S.E. Guerrero, Survey on RGB, 3D, Thermal, and Multimodal Approaches for Facial Expression Recognition: History, Trends, and Affect-Related Applications, IEEE Trans. Pattern Anal. Mach. Intell. 38 (2016) 1548–1568. https://doi.org/10.1109/TPAMI.2016.2515606.
[19] M. El Ayadi, M.S. Kamel, F. Karray, Survey on speech emotion recognition: Features, classification schemes, and databases, Pattern Recognit. 44 (2011) 572–587. https://doi.org/10.1016/j.patcog.2010.09.020.
[20] J. Zhang, Z. Yin, P. Chen, S. Nichele, Emotion recognition using multi-modal data and machine learning techniques: A tutorial and review, Inf. Fusion. 59 (2020) 103–126. https://doi.org/10.1016/j.inffus.2020.01.011.
[21] S. Poria, E. Cambria, A. Gelbukh, Aspect extraction for opinion mining with a deep convolutional neural network, Knowl.-Based Syst. 108 (2016) 42–49. https://doi.org/10.1016/j.knosys.2016.06.009.
[22] E. Cambria, R. Speer, C. Havasi, A. Hussain, SenticNet: A Publicly Available Semantic Resource for Opinion Mining, in: AAAI2010, 2010: pp. 14–18.
[23] M.S. Akhtar, D. Chauhan, D. Ghosal, S. Poria, A. Ekbal, P. Bhattacharyya, Multi-task Learning for Multi-modal Emotion Recognition and Sentiment Analysis, in: Proc. 2019 Conf. North Am. Chapter Assoc. Comput. Linguist. Hum. Lang. Technol. Vol. 1 Long Short Pap., Association for Computational Linguistics, Minneapolis, Minnesota, 2019: pp. 370–379. https://doi.org/10.18653/v1/N19-1034.
[24] A. Mehrabian, Communicating Without Words, Psychol. Today. (1968) 53–55.
[25] C.O. Alm, D. Roth, R. Sproat, Emotions from text: machine learning for text-based emotion prediction, in: Proc. Conf. Hum. Lang. Technol. Empir. Methods Nat. Lang. Process. - HLT 05, Association for Computational Linguistics, Vancouver, British Columbia, Canada, 2005: pp. 579–586. https://doi.org/10.3115/1220575.1220648.
[26] Z.-T. Liu, Q. Xie, M. Wu, W.-H. Cao, Y. Mei, J.-W. Mao, Speech emotion recognition based on an improved brain emotion learning model, Neurocomputing. 309 (2018) 145–156. https://doi.org/10.1016/j.neucom.2018.05.005.
[27] M. Sajjad, M. Nasir, F.U.M. Ullah, K. Muhammad, A.K. Sangaiah, S.W. Baik, Raspberry Pi assisted facial expression recognition framework for smart security in law-enforcement services, Inf. Sci. 479 (2019) 416–431. https://doi.org/10.1016/j.ins.2018.07.027.
[28] P. Sarkar, A. Etemad, Self-supervised ECG Representation Learning for Emotion Recognition, IEEE Trans. Affect. Comput. (2020) 1–1. https://doi.org/10.1109/TAFFC.2020.3014842.
[29] S.M. Alarcão, M.J. Fonseca, Emotions Recognition Using EEG Signals: A Survey, IEEE Trans. Affect. Comput. 10 (2019) 374–393. https://doi.org/10.1109/TAFFC.2017.2714671.
[30] J. Kim, E. Andre, Emotion recognition based on physiological changes in music listening, IEEE Trans. Pattern Anal. Mach. Intell. 30 (2008) 2067–2083. https://doi.org/10.1109/TPAMI.2008.26.
[31] P. Tzirakis, G. Trigeorgis, M.A. Nicolaou, B.W. Schuller, S. Zafeiriou, End-to-End Multimodal Emotion Recognition Using Deep Neural Networks, IEEE J. Sel. Top. Signal Process. 11 (2017) 1301–1309. https://doi.org/10.1109/JSTSP.2017.2764438.
[32] T. Baltrusaitis, P. Robinson, L. Morency, 3D Constrained Local Model for rigid and non-rigid facial tracking, in: 2012 IEEE Conf. Comput. Vis. Pattern Recognit., IEEE, Providence, RI, 2012: pp. 2610–2617. https://doi.org/10.1109/CVPR.2012.6247980.
[33] J.-C. Lin, C.-H. Wu, W.-L. Wei, Error Weighted Semi-Coupled Hidden Markov Model for Audio-Visual Emotion Recognition, IEEE Trans. Multimed. 14 (2012) 142–156. https://doi.org/10.1109/TMM.2011.2171334.

[34] X. Huang, J. Kortelainen, G. Zhao, X. Li, A. Moilanen, T. Seppänen, M. Pietikäinen, Multi-modal emotion analysis from facial expressions and electroencephalogram, Comput. Vis. Image Underst. 147 (2016) 114–124. https://doi.org/10.1016/j.cviu.2015.09.015.

[35] Z. Wang, X. Zhou, W. Wang, C. Liang, Emotion recognition using multimodal deep learning in multiple psychophysiological signals and video, Int. J. Mach. Learn. Cybern. 11 (2020) 923–934. https://doi.org/10.1007/s13042-019-01056-8.

[36] T. Meng, X. Jing, Z. Yan, W. Pedrycz, A survey on machine learning for data fusion, Inf. Fusion. 57 (2020) 115–129. https://doi.org/10.1016/j.inffus.2019.12.001.

[37] S. Zhang, S. Zhang, T. Huang, W. Gao, Q. Tian, Learning Affective Features With a Hybrid Deep Model for Audio–Visual Emotion Recognition, IEEE Trans. Circuits Syst. Video Technol. 28 (2018) 3030–3043. https://doi.org/10.1109/TCSVT.2017.2719043.

[38] B. Ko, A Brief Review of Facial Emotion Recognition Based on Visual Information, Sensors. 18 (2018) 401. https://doi.org/10.3390/s18020401.

[39] S. Li, W. Deng, Deep Facial Expression Recognition: A Survey, IEEE Trans. Affect. Comput. (2020) 1–1. https://doi.org/10.1109/TAFFC.2020.2981446.

[40] W. Merghani, A.K. Davison, M.H. Yap, A Review on Facial Micro-Expressions Analysis: Datasets, Features and Metrics, ArXiv180502397 Cs. (2018). http://arxiv.org/abs/1805.02397 (accessed November 21, 2019).

[41] G.R. Alexandre, J.M. Soares, G.A. Pereira Thé, Systematic review of 3D facial expression recognition methods, Pattern Recognit. 100 (2020) 107108. https://doi.org/10.1016/j.patcog.2019.107108.

[42] R. Liu, Y. Shi, C. Ji, M. Jia, A Survey of Sentiment Analysis Based on Transfer Learning, IEEE Access. 7 (2019) 85401–85412. https://doi.org/10.1109/ACCESS.2019.2925059.

[43] R.A. Khalil, E. Jones, M.I. Babar, T. Jan, M.H. Zafar, T. Alhussain, Speech Emotion Recognition Using Deep Learning Techniques: A Review, IEEE Access. 7 (2019) 117327–117345. https://doi.org/10.1109/ACCESS.2019.2936124.

[44] K. Patel, D. Mehta, C. Mistry, R. Gupta, S. Tanwar, N. Kumar, M. Alazab, Facial Sentiment Analysis Using AI Techniques: State-of-the-Art, Taxonomies, and Challenges, IEEE Access. 8 (2020) 90495–90519. https://doi.org/10.1109/ACCESS.2020.2993803.

[45] F. Noroozi, D. Kaminska, C. Corneanu, T. Sapinski, S. Escalera, G. Anbarjafari, Survey on Emotional Body Gesture Recognition, IEEE Trans. Affect. Comput. (2018) 1–1. https://doi.org/10.1109/TAFFC.2018.2874986.

[46] S. Poria, N. Majumder, R. Mihalcea, E. Hovy, Emotion Recognition in Conversation: Research Challenges, Datasets, and Recent Advances, IEEE Access. 7 (2019) 100943–100953. https://doi.org/10.1109/ACCESS.2019.2929050.

[47] L. Yue, W. Chen, X. Li, W. Zuo, M. Yin, A survey of sentiment analysis in social media, Knowl. Inf. Syst. 60 (2019) 617–663. https://doi.org/10.1007/s10115-018-1236-4.

[48] Z. Wang, S.-B. Ho, E. Cambria, A review of emotion sensing: categorization models and algorithms, Multimed. Tools Appl. 79 (2020) 35553–35582. https://doi.org/10.1007/s11042-019-08328-z.

[49] J. Han, Z. Zhang, N. Cummins, B. Schuller, Adversarial Training in Affective Computing and Sentiment Analysis: Recent Advances and Perspectives [Review Article], IEEE Comput. Intell. Mag. 14 (2019) 68–81. https://doi.org/10.1109/MCI.2019.2901088.

[50] P.J. Bota, C. Wang, A.L.N. Fred, H. Placido Da Silva, A Review, Current Challenges, and Future Possibilities on Emotion Recognition Using Machine Learning and Physiological Signals, IEEE Access. 7 (2019) 140990–141020. https://doi.org/10.1109/ACCESS.2019.2944001.

[51] B. Garcia-Martinez, A. Martinez-Rodrigo, R. Alcaraz, A. Fernandez-Caballero, A Review on Nonlinear Methods Using Electroencephalographic Recordings for Emotion Recognition, IEEE Trans. Affect. Comput. (2019) 1–1. https://doi.org/10.1109/TAFFC.2018.2890636.

[52] P. Ekman, Universals and cultural differences in facial expressions of emotion, Nebr. Symp. Motiv. 19 (1971) 207–283.

[53] J.L. Tracy, D. Randles, Four Models of Basic Emotions: A Review of Ekman and Cordaro, Izard, Levenson, and Panksepp and Watt, Emot. Rev. 3 (2011) 397–405. https://doi.org/10.1177/1754073911410747.

[54] J. Russell, A Circumplex Model of Affect, J. Pers. Soc. Psychol. 39 (1980) 1161–1178. https://doi.org/10.1037/h0077714.

[55] plutchik Robert, Emotion and Life: perspective from psychology biology and evolution, Am. Physiol. Assoc. (2003).

[56] E. Cambria, A. Livingstone, A. Hussain, The Hourglass of Emotions, in: A. Esposito, A.M. Esposito, A. Vinciarelli, R. Hoffmann, V.C. Müller (Eds.), Cogn. Behav. Syst., Springer Berlin Heidelberg, Berlin, Heidelberg, 2012: pp. 144–157. https://doi.org/10.1007/978-3-642-34584-5_11.

[57] Y. Susanto, A.G. Livingstone, B.C. Ng, E. Cambria, The Hourglass Model Revisited, IEEE Intell. Syst. 35 (2020) 96–102. https://doi.org/10.1109/MIS.2020.2992799.

[58] A.T. Lopes, E. de Aguiar, A.F. De Souza, T. Oliveira-Santos, Facial expression recognition with Convolutional Neural Networks: Coping with few data and the training sample order, Pattern Recognit. 61 (2017) 610–628. https://doi.org/10.1016/j.patcog.2016.07.026.

[59] Z. Ren, A. Baird, J. Han, Z. Zhang, B. Schuller, Generating and Protecting Against Adversarial Attacks for Deep Speech-Based Emotion Recognition Models, in: ICASSP 2020 - 2020 IEEE Int. Conf. Acoust. Speech Signal Process. ICASSP, 2020: pp. 7184–7188. https://doi.org/10.1109/ICASSP40776.2020.9054087.

[60] Z. Wang, S.-B. Ho, E. Cambria, Multi-Level Fine-Scaled Sentiment Sensing with Ambivalence Handling, Int. J. Uncertain. Fuzziness Knowl.-Based Syst. 28 (2020) 683–697. https://doi.org/10.1142/S0218488520500294.

[61] I. Bakker, T. van der Voordt, P. Vink, J. de Boon, Pleasure, Arousal, Dominance: Mehrabian and Russell revisited, Curr. Psychol. 33 (2014) 405–421. https://doi.org/10.1007/s12144-014-9219-4.

[62] J.A. Russell, A. Mehrabian, Evidence for a three-factor theory of emotions, J. Res. Personal. 11 (1977) 273–294. https://doi.org/10.1016/0092-6566(77)90037-X.

[63] H. Dabas, C. Sethi, C. Dua, M. Dalawat, D. Sethia, Emotion Classification Using EEG Signals, in: 2018: pp. 380–384. https://doi.org/10.1145/3297156.3297177.

[64] J. Blitzer, M. Dredze, F. Pereira, Biographies, Bollywood, Boom-boxes and Blenders: Domain Adaptation for Sentiment Classification, ACL 2007. (n.d.) 8.

[65] M. Dredze, K. Crammer, F. Pereira, Confidence-weighted linear classification, in: Proc. 25th Int. Conf. Mach. Learn. - ICML 08, ACM Press, Helsinki, Finland, 2008: pp. 264–271. https://doi.org/10.1145/1390156.1390190.

[66] A.L. Maas, R.E. Daly, P.T. Pham, D. Huang, A.Y. Ng, C. Potts, Learning Word Vectors for Sentiment Analysis, in: Proc. 49th Annu. Meet. Assoc. Comput. Linguist. Hum. Lang. Technol., Association for Computational Linguistics, Portland, Oregon, USA, 2011: pp. 142–150. https://www.aclweb.org/anthology/P11-1015 (accessed July 23, 2020).

[67] R. Socher, A. Perelygin, J. Wu, J. Chuang, C.D. Manning, A. Ng, C. Potts, Recursive Deep Models for Semantic Compositionality Over a Sentiment Treebank, (n.d.) 12.

[68] F. Burkhardt, A. Paeschke, M. Rolfes, W. Sendlmeier, B. Weiss, A Database of German Emotional Speech, (2005) 4.

[69] I. Sneddon, M. McRorie, G. McKeown, J. Hanratty, The Belfast Induced Natural Emotion Database, IEEE Trans. Affect. Comput. 3 (2012) 32–41. https://doi.org/10.1109/T-AFFC.2011.26.

[70] M. Lyons, S. Akamatsu, M. Kamachi, J. Gyoba, Coding facial expressions with Gabor wavelets, in: Proc. Third IEEE Int. Conf. Autom. Face Gesture Recognit., 1998: pp. 200–205. https://doi.org/10.1109/AFGR.1998.670949.

[71] P. Lucey, J.F. Cohn, T. Kanade, J. Saragih, Z. Ambadar, I. Matthews, The Extended Cohn-Kanade Dataset (CK+): A complete dataset for action unit and emotion-specified expression, in: 2010 IEEE Comput. Soc. Conf. Comput. Vis. Pattern Recognit. - Workshop, IEEE, San Francisco, CA, USA, 2010: pp. 94–101. https://doi.org/10.1109/CVPRW.2010.5543262.

[72] T. Kanade, J.F. Cohn, Yingli Tian, Comprehensive database for facial expression analysis, in: Proc. Fourth IEEE Int. Conf. Autom. Face Gesture Recognit. Cat No PR00580, 2000: pp. 46–53. https://doi.org/10.1109/AFGR.2000.840611.

[73] M.F. Valstar, M. Pantic, Induced Disgust, Happiness and Surprise: an Addition to the MMI Facial Expression Database, (2010) 6.

[74] G. Zhao, X. Huang, M. Taini, S.Z. Li, M. Pietikäinen, Facial expression recognition from near-infrared videos, Image Vis. Comput. 29 (2011) 607–619. https://doi.org/10.1016/j.imavis.2011.07.002.

[75] Lijun Yin, Xiaozhou Wei, Yi Sun, Jun Wang, M.J. Rosato, A 3D facial expression database for facial behavior research, in: 7th Int. Conf. Autom. Face Gesture Recognit. FGR06, 2006: pp. 211–216. https://doi.org/10.1109/FGR.2006.6.

[76] L. Yin, X. Chen, Y. Sun, T. Worm, M. Reale, A high-resolution 3D dynamic facial expression database, in: 2008 8th IEEE Int. Conf. Autom. Face Gesture Recognit., 2008: pp. 1–6. https://doi.org/10.1109/AFGR.2008.4813324.

[77] X. Zhang, L. Yin, J.F. Cohn, S. Canavan, M. Reale, A. Horowitz, P. Liu, J.M. Girard, BP4D-Spontaneous: a high-resolution spontaneous 3D dynamic facial expression database, Image Vis. Comput. 32 (2014) 692–706. https://doi.org/10.1016/j.imavis.2014.06.002.

[78] S. Cheng, I. Kotsia, M. Pantic, S. Zafeiriou, 4DFAB: A Large Scale 4D Database for Facial Expression Analysis and Biometric Applications, in: 2018 IEEECVF Conf. Comput. Vis. Pattern Recognit., IEEE, Salt Lake City, UT, USA, 2018: pp. 5117–5126. https://doi.org/10.1109/CVPR.2018.00537.

[79] X. Li, T. Pfister, X. Huang, G. Zhao, M. Pietikainen, A Spontaneous Micro-expression Database: Inducement, collection and baseline, in: 2013 10th IEEE Int. Conf. Workshop Autom. Face Gesture Recognit. FG, IEEE, Shanghai, China, 2013: pp. 1–6. https://doi.org/10.1109/FG.2013.6553717.

[80] W.-J. Yan, X. Li, S.-J. Wang, G. Zhao, Y.-J. Liu, Y.-H. Chen, X. Fu, CASME II: An Improved Spontaneous Micro-Expression Database and the Baseline Evaluation, PLoS ONE. 9 (2014). https://doi.org/10.1371/journal.pone.0086041.

[81] A.K. Davison, C. Lansley, N. Costen, K. Tan, M.H. Yap, SAMM: A Spontaneous Micro-Facial Movement Dataset, IEEE Trans. Affect. Comput. 9 (2018) 116–129. https://doi.org/10.1109/TAFFC.2016.2573832.

[82] I.J. Goodfellow, D. Erhan, P.L. Carrier, A. Courville, M. Mirza, B. Hamner, W. Cukierski, Y. Tang, D. Thaler, D.-H. Lee, Y. Zhou, C. Ramaiah, F. Feng, R. Li, X. Wang, D. Athanasakis, J. Shawe-Taylor, M. Milakov, J. Park, R. Ionescu, M. Popescu, C. Grozea, J. Bergstra, J. Xie, L. Romaszko, B. Xu, Z. Chuang, Y. Bengio, Challenges in Representation Learning: A Report on Three Machine Learning Contests, in: M. Lee, A. Hirose, Z.-G. Hou, R.M. Kil (Eds.), Neural Inf. Process., Springer, Berlin, Heidelberg, 2013: pp. 117–124. https://doi.org/10.1007/978-3-642-42051-1_16.

[83] A. Dhall, R. Goecke, S. Lucey, T. Gedeon, Static facial expression analysis in tough conditions: Data, evaluation protocol and benchmark, in: 2011 IEEE Int. Conf. Comput. Vis. Workshop ICCV Workshop, 2011: pp. 2106–2112. https://doi.org/10.1109/ICCVW.2011.6130508.

[84] C.F. Benitez-Quiroz, R. Srinivasan, A.M. Martinez, EmotioNet: An Accurate, Real-Time Algorithm for the Automatic Annotation of a Million Facial Expressions in the Wild, in: 2016 IEEE Conf. Comput. Vis. Pattern Recognit. CVPR, IEEE, Las Vegas, NV, USA, 2016: pp. 5562–5570. https://doi.org/10.1109/CVPR.2016.600.

[85] Z. Zhang, P. Luo, C.C. Loy, X. Tang, From Facial Expression Recognition to Interpersonal Relation Prediction, Int. J. Comput. Vis. 126 (2018) 550–569. https://doi.org/10.1007/s11263-017-1055-1.

[86] A. Mollahosseini, B. Hasani, M.H. Mahoor, AffectNet: A Database for Facial Expression, Valence, and Arousal Computing in the Wild, IEEE Trans. Affect. Comput. 10 (2019) 18–31. https://doi.org/10.1109/TAFFC.2017.2740923.

[87] S. Li, W. Deng, J. Du, Reliable Crowdsourcing and Deep Locality-Preserving Learning for Expression Recognition in the Wild, in: 2017 IEEE Conf. Comput. Vis. Pattern Recognit. CVPR, IEEE, Honolulu, HI, 2017: pp. 2584–2593. https://doi.org/10.1109/CVPR.2017.277.

[88] X. Jiang, Y. Zong, W. Zheng, C. Tang, W. Xia, C. Lu, J. Liu, DFEW: A Large-Scale Database for Recognizing Dynamic Facial Expressions in the Wild, in: Proc. 28th ACM Int. Conf. Multimed., ACM, Seattle WA USA, 2020: pp. 2881–2889. https://doi.org/10.1145/3394171.3413620.

[89] S. Abrilian, L. Devillers, S. Buisine, J.-C. Martin, EmoTV1: Annotation of real-life emotions for the specifications of multimodal a ective interfaces, in: 2005.

[90] H. Gunes, M. Piccardi, A Bimodal Face and Body Gesture Database for Automatic Analysis of Human Nonverbal Affective Behavior, in: 18th Int. Conf. Pattern Recognit. ICPR06, 2006: pp. 1148–1153. https://doi.org/10.1109/ICPR.2006.39.

[91] M. Kipp, J.-C. Martin, Gesture and Emotion: Can basic gestural form features discriminate emotions?, 2009. https://doi.org/10.1109/ACII.2009.5349544.

[92] A. Mehrabian, Pleasure-arousal-dominance: A general framework for describing and measuring individual differences in Temperament, Curr. Psychol. 14 (1996) 261–292. https://doi.org/10.1007/BF02686918.

[93] T. Bänziger, K. Scherer, Introducing the Geneva Multimodal Emotion Portrayal (GEMEP) corpus, Bluepr. Affect. Comput. Sourceb. (2010).

[94] M.F. Valstar, M. Mehu, B. Jiang, M. Pantic, K. Scherer, Meta-Analysis of the First Facial Expression Recognition Challenge, IEEE Trans. Syst. Man Cybern. Part B Cybern. 42 (2012) 966–979. https://doi.org/10.1109/TSMCB.2012.2200675.

[95] N. Fourati, C. Pelachaud, Emilya: Emotional body expression in daily actions database, in: Reykjavik, Iceland, 2014.

[96] S. Koelstra, C. Muhl, M. Soleymani, J.-S. Lee, A. Yazdani, T. Ebrahimi, T. Pun, A. Nijholt, I. Patras, DEAP: A Database for Emotion Analysis ;Using Physiological Signals, IEEE Trans. Affect. Comput. 3 (2012) 18–31. https://doi.org/10.1109/T-AFFC.2011.15.

[97] R.-N. Duan, J.-Y. Zhu, B.-L. Lu, Differential entropy feature for EEG-based emotion classification, in: 2013 6th Int. IEEEEMBS Conf. Neural Eng. NER, 2013: pp. 81–84. https://doi.org/10.1109/NER.2013.6695876.

[98] W.-L. Zheng, B.-L. Lu, Investigating Critical Frequency Bands and Channels for EEG-Based Emotion Recognition with Deep Neural Networks, IEEE Trans. Auton. Ment. Dev. 7 (2015) 162–175. https://doi.org/10.1109/TAMD.2015.2431497.

[99] J.A. Miranda Correa, M.K. Abadi, N. Sebe, I. Patras, AMIGOS: A Dataset for Affect, Personality and Mood Research on Individuals and Groups, IEEE Trans. Affect. Comput. (2018) 1–1. https://doi.org/10.1109/TAFFC.2018.2884461.

[100] P. Schmidt, A. Reiss, R. Duerichen, C. Marberger, K. Van Laerhoven, Introducing WESAD, a Multimodal Dataset for Wearable Stress and Affect Detection, in: Proc. 20th ACM Int. Conf. Multimodal Interact., Association for Computing Machinery, Boulder, CO, USA, 2018: pp. 400–408. https://doi.org/10.1145/3242969.3242985.

[101] C. Busso, M. Bulut, C.-C. Lee, A. Kazemzadeh, E. Mower, S. Kim, J.N. Chang, S. Lee, S.S. Narayanan, IEMOCAP: interactive emotional dyadic motion capture database, Lang. Resour. Eval. 42 (2008) 335. https://doi.org/10.1007/s10579-008-9076-6.

[102] A. Metallinou, C.-C. Lee, C. Busso, S. Carnicke, S. Narayanan, The USC CreativeIT Database: A Multimodal Database of Theatrical Improvisation, (n.d.) 4.

[103] A. Metallinou, Z. Yang, C. Lee, C. Busso, S. Carnicke, S. Narayanan, The USC CreativeIT database of multimodal dyadic interactions: from speech and full body motion capture to continuous emotional annotations, Lang. Resour. Eval. 50 (2016) 497–521. https://doi.org/10.1007/s10579-015-9300-0.

[104] L.-P. Morency, R. Mihalcea, P. Doshi, Towards multimodal sentiment analysis: harvesting opinions from the web, in: Proc. 13th Int. Conf. Multimodal Interfaces, Association for Computing Machinery, Alicante, Spain, 2011: pp. 169–176. https://doi.org/10.1145/2070481.2070509.

[105] M. Wollmer, F. Weninger, T. Knaup, B. Schuller, C. Sun, K. Sagae, L.-P. Morency, YouTube Movie Reviews: Sentiment Analysis in an Audio-Visual Context, IEEE Intell. Syst. 28 (2013) 46–53. https://doi.org/10.1109/MIS.2013.34.

[106] A. Bagher Zadeh, P.P. Liang, S. Poria, E. Cambria, L.-P. Morency, Multimodal Language Analysis in the Wild: CMU-MOSEI Dataset and Interpretable Dynamic Fusion Graph, in: Proc. 56th Annu. Meet. Assoc. Comput. Linguist. Vol. 1 Long Pap., Association for Computational Linguistics, Melbourne, Australia, 2018: pp. 2236–2246. https://doi.org/10.18653/v1/P18-1208.

[107] M. Soleymani, J. Lichtenauer, T. Pun, M. Pantic, A Multimodal Database for Affect Recognition and Implicit Tagging, IEEE Trans. Affect. Comput. 3 (2012) 42–55. https://doi.org/10.1109/T-AFFC.2011.25.

[108] F. Ringeval, A. Sonderegger, J. Sauer, D. Lalanne, Introducing the RECOLA multimodal corpus of remote collaborative and affective interactions, in: 2013 10th IEEE Int. Conf. Workshop Autom. Face Gesture Recognit. FG, 2013: pp. 1–8. https://doi.org/10.1109/FG.2013.6553805.

[109] M.K. Abadi, R. Subramanian, S.M. Kia, P. Avesani, I. Patras, N. Sebe, DECAF: MEG-Based Multimodal Database for Decoding Affective Physiological Responses, IEEE Trans. Affect. Comput. 6 (2015) 209–222. https://doi.org/10.1109/TAFFC.2015.2392932.

[110] F.A. Pozzi, E. Fersini, E. Messina, B. Liu, Chapter 1 - Challenges of Sentiment Analysis in Social Networks: An Overview, in: F.A. Pozzi, E. Fersini, E. Messina, B. Liu (Eds.), Sentim. Anal. Soc. Netw., Morgan Kaufmann, Boston, 2017: pp. 1–11. https://doi.org/10.1016/B978-0-12-804412-4.00001-2.

[111] P.J. Stone, E.B. Hunt, A computer approach to content analysis: studies using the General Inquirer system, in: Proc. May 21-23 1963 Spring Jt. Comput. Conf., Association for Computing Machinery, New York, NY, USA, 1963: pp. 241–256. https://doi.org/10.1145/1461551.1461583.

[112] B. Pang, L. Lee, S. Vaithyanathan, Thumbs up?: sentiment classification using machine learning techniques, in: Proc. ACL-02 Conf. Empir. Methods Nat. Lang. Process. - EMNLP 02, Association for Computational Linguistics, Not Known, 2002: pp. 79–86. https://doi.org/10.3115/1118693.1118704.

[113] L. Oneto, F. Bisio, E. Cambria, D. Anguita, Statistical Learning Theory and ELM for Big Social Data Analysis, IEEE Comput. Intell. Mag. 11 (2016) 45–55. https://doi.org/10.1109/MCI.2016.2572540.

[114] M. Taboada, J. Brooke, M. Tofiloski, K. Voll, M. Stede, Lexicon-Based Methods for Sentiment Analysis, Comput. Linguist. 37 (2011) 267–307. https://doi.org/10.1162/COLI_a_00049.

[115] X. Ding, B. Liu, P.S. Yu, A holistic lexicon-based approach to opinion mining, in: Proc. Int. Conf. Web Search Web Data Min. - WSDM 08, ACM Press, Palo Alto, California, USA, 2008: p. 231. https://doi.org/10.1145/1341531.1341561.

[116] P. Melville, W. Gryc, R.D. Lawrence, Sentiment analysis of blogs by combining lexical knowledge with text classification, in: Proc. 15th ACM SIGKDD Int. Conf. Knowl. Discov. Data Min. - KDD 09, ACM Press, Paris, France, 2009: p. 1275. https://doi.org/10.1145/1557019.1557156.

[117] S. Poria, E. Cambria, G. Winterstein, G.-B. Huang, Sentic patterns: Dependency-based rules for concept-level sentiment analysis, Knowl.-Based Syst. 69 (2014) 45–63. https://doi.org/10.1016/j.knosys.2014.05.005.

[118] E. Cambria, A. Hussain, C. Havasi, C. Eckl, Common Sense Computing: From the Society of Mind to Digital Intuition and beyond, in: J. Fierrez, J. Ortega-Garcia, A. Esposito, A. Drygajlo, M. Faundez-Zanuy (Eds.), Biom. ID Manag. Multimodal Commun., Springer Berlin Heidelberg, Berlin, Heidelberg, 2009: pp. 252–259. https://doi.org/10.1007/978-3-642-04391-8_33.

[119] L. Jia, C. Yu, W. Meng, The effect of negation on sentiment analysis and retrieval effectiveness, in: Proceeding 18th ACM Conf. Inf. Knowl. Manag. - CIKM 09, ACM Press, Hong Kong, China, 2009: p. 1827. https://doi.org/10.1145/1645953.1646241.

[120] I. Blekanov, M. Kukarkin, A. Maksimov, S. Bodrunova, Sentiment Analysis for Ad Hoc Discussions Using Multilingual Knowledge-Based Approach, in: Proc. 3rd Int. Conf. Appl. Inf. Technol. - ICAIT2018, ACM Press, Aizu-Wakamatsu, Japan, 2018: pp. 117–121. https://doi.org/10.1145/3274856.3274880.

[121] Ebru Aydogan, M. Ali Akcayol, A comprehensive survey for sentiment analysis tasks using machine learning techniques, Int. Symp. Innov. Intell. Syst. Appl. (2016). https://doi.org/10.1109/INISTA.2016.7571856.

[122] M. Ahmad, S. Aftab, S. Muhammad, S. Ahmad, Machine Learning Techniques for Sentiment Analysis: A Review, Int. J. Multidiscip. Sci. Eng. 8 (2017) 2045–7057.

[123] T. Mullen, N. Collier, Sentiment Analysis using Support Vector Machines with Diverse Information Sources, in: Proc. 2004 Conf. Empir. Methods Nat. Lang. Process., Association for Computational Linguistics, Barcelona, Spain, 2004: pp. 412–418. https://www.aclweb.org/anthology/W04-3253 (accessed October 29, 2020).

[124] A. Pak, P. Paroubek, Text Representation Using Dependency Tree Subgraphs for Sentiment Analysis, in: J. Xu, G. Yu, S. Zhou, R. Unland (Eds.), Database Syst. Adanced Appl., Springer, Berlin, Heidelberg, 2011: pp. 323–332. https://doi.org/10.1007/978-3-642-20244-5_31.

[125] J. Chen, H. Huang, S. Tian, Y. Qu, Feature selection for text classification with Naïve Bayes, Expert Syst. Appl. 36 (2009) 5432–5435. https://doi.org/10.1016/j.eswa.2008.06.054.

[126] R.S. Jagdale, V.S. Shirsat, S.N. Deshmukh, Sentiment Analysis on Product Reviews Using Machine Learning Techniques, in: P.K. Mallick, V.E. Balas, A.K. Bhoi, A.F. Zobaa (Eds.), Cogn. Inform. Soft Comput., Springer, Singapore, 2019: pp. 639–647. https://doi.org/10.1007/978-981-13-0617-4_61.

[127] Y. Xia, E. Cambria, A. Hussain, H. Zhao, Word Polarity Disambiguation Using Bayesian Model and Opinion-Level Features, Cogn. Comput. 7 (2015) 369–380. https://doi.org/10.1007/s12559-014-9298-4.

[128] A. Valdivia, M.V. Luzón, E. Cambria, F. Herrera, Consensus vote models for detecting and filtering neutrality in sentiment analysis, Inf. Fusion. 44 (2018) 126–135. https://doi.org/10.1016/j.inffus.2018.03.007.

[129] T. Le, A Hybrid Method for Text-Based Sentiment Analysis, in: 2019 Int. Conf. Comput. Sci. Comput. Intell. CSCI, 2019: pp. 1392–1397. https://doi.org/10.1109/CSCI49370.2019.00260.

[130] D. Li, R. Rzepka, M. Ptaszynski, K. Araki, A Novel Machine Learning-based Sentiment Analysis Method for Chinese Social Media Considering Chinese Slang Lexicon and Emoticons, in: Honolulu, Hawaii, USA, 2019.

[131] T. Mikolov, M. Karafiát, L. Burget, J. Cernocký, S. Khudanpur, Recurrent neural network based language model, in: 2010: pp. 1045–1048.

[132] A. Khatua, A. Khatua, E. Cambria, Predicting political sentiments of voters from Twitter in multi-party contexts, Appl. Soft Comput. 97 (2020) 106743. https://doi.org/10.1016/j.asoc.2020.106743.

[133] F. Liu, L. Zheng, J. Zheng, HieNN-DWE: A hierarchical neural network with dynamic word embeddings for document level sentiment classification, Neurocomputing. 403 (2020) 21–32. https://doi.org/10.1016/j.neucom.2020.04.084.

[134] Y. Kim, Convolutional Neural Networks for Sentence Classification, in: Proc. 2014 Conf. Empir. Methods Nat. Lang. Process. EMNLP, Association for Computational Linguistics, Doha, Qatar, 2014: pp. 1746–1751. https://doi.org/10.3115/v1/D14-1181.

[135] H.H. Do, P. Prasad, A. Maag, A. Alsadoon, Deep Learning for Aspect-Based Sentiment Analysis: A Comparative Review, Expert Syst. Appl. 118 (2019) 272–299. https://doi.org/10.1016/j.eswa.2018.10.003.

[136] R. Yin, P. Li, B. Wang, Sentiment Lexical-Augmented Convolutional Neural Networks for Sentiment Analysis, in: 2017 IEEE Second Int. Conf. Data Sci. Cyberspace DSC, 2017: pp. 630–635. https://doi.org/10.1109/DSC.2017.82.

[137] A. Conneau, H. Schwenk, L. Barrault, Y. Lecun, Very Deep Convolutional Networks for Text Classification, in: Proc. 15th Conf. Eur. Chapter Assoc. Comput. Linguist. Vol. 1 Long Pap., Association for Computational Linguistics, Valencia, Spain, 2017: pp. 1107–1116. https://www.aclweb.org/anthology/E17-1104 (accessed August 13, 2020).

[138] R. Johnson, T. Zhang, Deep Pyramid Convolutional Neural Networks for Text Categorization, in: Proc. 55th Annu. Meet. Assoc. Comput. Linguist. Vol. 1 Long Pap., Association for Computational Linguistics, Vancouver, Canada, 2017: pp. 562–570. https://doi.org/10.18653/v1/P17-1052.

[139] B. Huang, K. Carley, Parameterized Convolutional Neural Networks for Aspect Level Sentiment Classification, in: Proc. 2018 Conf. Empir. Methods Nat. Lang. Process., Association for Computational Linguistics, Brussels, Belgium, 2018: pp. 1091–1096. https://doi.org/10.18653/v1/D18-1136.

[140] A. Mousa, B. Schuller, Contextual Bidirectional Long Short-Term Memory Recurrent Neural Network Language Models: A Generative Approach to Sentiment Analysis, in: Proc. 15th Conf. Eur. Chapter Assoc. Comput. Linguist. Vol. 1 Long Pap., Association for Computational Linguistics, Valencia, Spain, 2017: pp. 1023–1032. https://www.aclweb.org/anthology/E17-1096 (accessed August 15, 2020).

[141] W. Wang, S.J. Pan, D. Dahlmeier, X. Xiao, Recursive Neural Conditional Random Fields for Aspect-based Sentiment Analysis, in: Proc. 2016 Conf. Empir. Methods Nat. Lang. Process., Association for Computational Linguistics, Austin, Texas, 2016: pp. 616–626. https://doi.org/10.18653/v1/D16-1059.

[142] P. Chen, Z. Sun, L. Bing, W. Yang, Recurrent Attention Network on Memory for Aspect Sentiment Analysis, in: Proc. 2017 Conf. Empir. Methods Nat. Lang. Process., Association for Computational Linguistics, Copenhagen, Denmark, 2017: pp. 452–461. https://doi.org/10.18653/v1/D17-1047.

[143] A. Mishra, S. Tamilselvam, R. Dasgupta, S. Nagar, K. Dey, Cognition-Cognizant Sentiment Analysis with Multitask Subjectivity Summarization based on Annotators' Gaze Behavior, in: 2018: p. 8.

[144] H. Chen, M. Sun, C. Tu, Y. Lin, Z. Liu, Neural Sentiment Classification with User and Product Attention, in: Proc. 2016 Conf. Empir. Methods Nat. Lang. Process., Association for Computational Linguistics, Austin, Texas, 2016: pp. 1650–1659. https://doi.org/10.18653/v1/D16-1171.

[145] Z.-Y. Dou, Capturing User and Product Information for Document Level Sentiment Analysis with Deep Memory Network, in: Proc. 2017 Conf. Empir. Methods Nat. Lang. Process., Association for Computational Linguistics, Copenhagen, Denmark, 2017: pp. 521–526. https://doi.org/10.18653/v1/D17-1054.

[146] Z. Wu, X.-Y. Dai, C. Yin, S. Huang, J. Chen, Improving Review Representations with User Attention and Product Attention for Sentiment Classification, in: Thirty-Second AAAI Conf. Artif. Intell. AAAI-18, 2018: pp. 5989–5996.

[147] A.K. J, T.E. Trueman, E. Cambria, A Convolutional Stacked Bidirectional LSTM with a Multiplicative Attention Mechanism for Aspect Category and Sentiment Detection, Cogn. Comput. (2021). https://doi.org/10.1007/s12559-021-09948-0.

[148] B. Liang, H. Su, L. Gui, E. Cambria, R. Xu, Aspect-based sentiment analysis via affective knowledge enhanced graph convolutional networks, Knowl.-Based Syst. 235 (2022) 107643. https://doi.org/10.1016/j.knosys.2021.107643.

[149] X. Li, L. Bing, W. Lam, B. Shi, Transformation Networks for Target-Oriented Sentiment Classification, in: Proc. 56th Annu. Meet. Assoc. Comput. Linguist. Vol. 1 Long Pap., Association for Computational Linguistics, Melbourne, Australia, 2018: pp. 946–956. https://doi.org/10.18653/v1/P18-1087.

[150] W. Li, L. Zhu, Y. Shi, K. Guo, E. Cambria, User reviews: Sentiment analysis using lexicon integrated two-channel CNN–LSTM family models, Appl. Soft Comput. 94 (2020) 106435. https://doi.org/10.1016/j.asoc.2020.106435.

[151] W. Xue, T. Li, Aspect Based Sentiment Analysis with Gated Convolutional Networks, in: Proc. 56th Annu. Meet. Assoc. Comput. Linguist. Vol. 1 Long Pap., Association for Computational Linguistics, Melbourne, Australia, 2018: pp. 2514–2523. https://doi.org/10.18653/v1/P18-1234.

[152] M.E. Basiri, S. Nemati, M. Abdar, E. Cambria, U.R. Acharya, ABCDM: An Attention-based Bidirectional CNN-RNN Deep Model for sentiment analysis, Future Gener. Comput. Syst. 115 (2021) 279–294. https://doi.org/10.1016/j.future.2020.08.005.

[153] M.S. Akhtar, A. Ekbal, E. Cambria, How Intense Are You? Predicting Intensities of Emotions and Sentiments using Stacked Ensemble, IEEE Comput. Intell. Mag. 15 (2020) 64–75. https://doi.org/10.1109/MCI.2019.2954667.

[154] T. Miyato, A.M. Dai, I. Goodfellow, Adversarial Training Methods for Semi-Supervised Text Classification, in: Int. Conf. Learn. Represent., 2017.

[155] Y. Ganin, E. Ustinova, H. Ajakan, P. Germain, H. Larochelle, F. Laviolette, M. March, V. Lempitsky, Domain-Adversarial Training of Neural Networks, J. Mach. Learn. Res. 17 (2016) 1–35.

[156] Q. Yang, Z. Li, Y. Zhang, Y. Wei, Y. Wu, End-to-End Adversarial Memory Network for Cross-domain Sentiment Classification, in: IJCAI 2017, 2017: pp. 2237–2243. https://www.ijcai.org/Proceedings/2017/311 (accessed August 14, 2020).

[157] Y. Li, Q. Pan, S. Wang, T. Yang, E. Cambria, A Generative Model for category text generation, Inf. Sci. 450 (2018) 301–315. https://doi.org/10.1016/j.ins.2018.03.050.

[158] I. Goodfellow, J. Pouget-Abadie, M. Mirza, B. Xu, D. Warde-Farley, S. Ozair, A. Courville, Y. Bengio, Generative Adversarial Nets, in: Z. Ghahramani, M. Welling, C. Cortes, N.D. Lawrence, K.Q. Weinberger (Eds.), Adv. Neural Inf. Process. Syst. 27, Curran Associates, Inc., 2014: pp. 2672–2680. http://papers.nips.cc/paper/5423-generative-adversarial-nets.pdf (accessed August 15, 2020).

[159] X. Chen, Y. Sun, B. Athiwaratkun, C. Cardie, K. Weinberger, Adversarial Deep Averaging Networks for Cross-Lingual Sentiment Classification, Trans. Assoc. Comput. Linguist. 6 (2018) 557–570. https://doi.org/10.1162/tacl_a_00039.

[160] A. Karimi, L. Rossi, A. Prati, Adversarial Training for Aspect-Based Sentiment Analysis with BERT, in: 2020 25th Int. Conf. Pattern Recognit. ICPR, IEEE, Milan, Italy, 2020. https://doi.org/10.1109/ICPR48806.2021.9412167.

[161] Chul Min Lee, S.S. Narayanan, Toward detecting emotions in spoken dialogs, IEEE Trans. Speech Audio Process. 13 (2005) 293–303. https://doi.org/10.1109/TSA.2004.838534.

[162] J. Pohjalainen, F. Fabien Ringeval, Z. Zhang, B. Schuller, Spectral and Cepstral Audio Noise Reduction Techniques in Speech Emotion Recognition, in: Proc. 2016 ACM Multimed. Conf. - MM 16, ACM Press, Amsterdam, The Netherlands, 2016: pp. 670–674. https://doi.org/10.1145/2964284.2967306.

[163] Z. Huang, M. Dong, Q. Mao, Y. Zhan, Speech Emotion Recognition Using CNN, in: Proc. ACM Int. Conf. Multimed. - MM 14, ACM Press, Orlando, Florida, USA, 2014: pp. 801–804. https://doi.org/10.1145/2647868.2654984.

[164] H.M. Fayek, M. Lech, L. Cavedon, Evaluating deep learning architectures for Speech Emotion Recognition, Neural Netw. 92 (2017) 60–68. https://doi.org/10.1016/j.neunet.2017.02.013.

[165] M.B. Akçay, K. Oğuz, Speech emotion recognition: Emotional models, databases, features, preprocessing methods, supporting modalities, and classifiers, Speech Commun. 116 (2020) 56–76. https://doi.org/10.1016/j.specom.2019.12.001.

[166] L.A. Low, N.C. Maddage, M. Lech, L.B. Sheeber, N.B. Allen, Detection of Clinical Depression in Adolescents' Speech During Family Interactions, IEEE Trans. Biomed. Eng. 58 (2011) 574–586. https://doi.org/10.1109/TBME.2010.2091640.

[167] F. Eyben, M. Wollmer, B. Schuller, OpenEAR — Introducing the munich open-source emotion and affect recognition toolkit, in: 2009 3rd Int. Conf. Affect. Comput. Intell. Interact. Workshop, IEEE, Amsterdam, 2009: pp. 1–6. https://doi.org/10.1109/ACII.2009.5349350.

[168] S. Ntalampiras, N. Fakotakis, Modeling the Temporal Evolution of Acoustic Parameters for Speech Emotion Recognition, IEEE Trans. Affect. Comput. 3 (2012) 116–125. https://doi.org/10.1109/T-AFFC.2011.31.

[169] L. Zhang, M. Song, N. Li, J. Bu, C. Chen, Feature selection for fast speech emotion recognition, in: Proc. 17th ACM Int. Conf. Multimed., Association for Computing Machinery, New York, NY, USA, 2009: pp. 753–756. https://doi.org/10.1145/1631272.1631405.

[170] D. Li, Y. Zhou, Z. Wang, D. Gao, Exploiting the Potentialities of Features for Speech Emotion Recognition, Inf. Sci. (2020). https://doi.org/10.1016/j.ins.2020.09.047.

[171] C. Busso, S. Lee, S. Narayanan, Analysis of Emotionally Salient Aspects of Fundamental Frequency for Emotion Detection, IEEE Trans. Audio Speech Lang. Process. 17 (2009) 582–596. https://doi.org/10.1109/TASL.2008.2009578.

[172] M. Lugger, B. Yang, The Relevance of Voice Quality Features in Speaker Independent Emotion Recognition, in: 2007 IEEE Int. Conf. Acoust. Speech Signal Process. - ICASSP 07, 2007: p. IV-17-IV–20. https://doi.org/10.1109/ICASSP.2007.367152.
[173] M.S. Likitha, S.R.R. Gupta, K. Hasitha, A.U. Raju, Speech based human emotion recognition using MFCC, in: 2017 Int. Conf. Wirel. Commun. Signal Process. Netw. WiSPNET, 2017: pp. 2257–2260. https://doi.org/10.1109/WiSPNET.2017.8300161.
[174] D. Bitouk, R. Verma, A. Nenkova, Class-level spectral features for emotion recognition, Speech Commun. 52 (2010) 613–625. https://doi.org/10.1016/j.specom.2010.02.010.
[175] P. Shen, Z. Changjun, X. Chen, Automatic Speech Emotion Recognition using Support Vector Machine, in: Proc. 2011 Int. Conf. Electron. Mech. Eng. Inf. Technol., 2011: pp. 621–625. https://doi.org/10.1109/EMEIT.2011.6023178.
[176] Y. Jin, P. Song, W. Zheng, L. Zhao, A feature selection and feature fusion combination method for speaker-independent speech emotion recognition, in: 2014 IEEE Int. Conf. Acoust. Speech Signal Process. ICASSP, 2014: pp. 4808–4812. https://doi.org/10.1109/ICASSP.2014.6854515.
[177] H. Atassi, A. Esposito, A Speaker Independent Approach to the Classification of Emotional Vocal Expressions, in: 2008 20th IEEE Int. Conf. Tools Artif. Intell., IEEE, Dayton, OH, USA, 2008: pp. 147–152. https://doi.org/10.1109/ICTAI.2008.158.
[178] K. Wang, N. An, B.N. Li, Y. Zhang, L. Li, Speech Emotion Recognition Using Fourier Parameters, IEEE Trans. Affect. Comput. 6 (2015) 69–75. https://doi.org/10.1109/TAFFC.2015.2392101.
[179] T.L. Nwe, S.W. Foo, L.C.D. Silva, Detection of stress and emotion in speech using traditional and FFT based log energy features, in: Fourth Int. Conf. Inf. Commun. Signal Process. 2003 Fourth Pac. Rim Conf. Multimed. Proc. 2003 Jt., 2003: pp. 1619–1623 vol.3. https://doi.org/10.1109/ICICS.2003.1292741.
[180] E. Navas, I. Hernaez, Iker Luengo, An objective and subjective study of the role of semantics and prosodic features in building corpora for emotional TTS, IEEE Trans. Audio Speech Lang. Process. 14 (2006) 1117–1127. https://doi.org/10.1109/TASL.2006.876121.
[181] A. Milton, S. Sharmy Roy, S. Tamil Selvi, SVM Scheme for Speech Emotion Recognition using MFCC Feature, Int. J. Comput. Appl. 69 (2013) 34–39. https://doi.org/10.5120/11872-7667.
[182] Y. Pan, P. Shen, L. Shen, Speech Emotion Recognition Using Support Vector Machine, Int. J. Smart Home. 6 (2012) 101–108.
[183] T. Seehapoch, S. Wongthanavasu, Speech emotion recognition using Support Vector Machines, in: 2013 5th Int. Conf. Knowl. Smart Technol. KST, 2013: pp. 86–91. https://doi.org/10.1109/KST.2013.6512793.
[184] E. Yüncü, H. Hacihabiboglu, C. Bozsahin, Automatic Speech Emotion Recognition Using Auditory Models with Binary Decision Tree and SVM, in: 2014 22nd Int. Conf. Pattern Recognit., 2014: pp. 773–778. https://doi.org/10.1109/ICPR.2014.143.
[185] A. Bhavan, P. Chauhan, Hitkul, R.R. Shah, Bagged support vector machines for emotion recognition from speech, Knowl.-Based Syst. 184 (2019) 104886. https://doi.org/10.1016/j.knosys.2019.104886.
[186] L. Chen, W. Su, Y. Feng, M. Wu, J. She, K. Hirota, Two-layer fuzzy multiple random forest for speech emotion recognition in human-robot interaction, Inf. Sci. 509 (2020) 150–163. https://doi.org/10.1016/j.ins.2019.09.005.
[187] Q. Mao, M. Dong, Z. Huang, Y. Zhan, Learning Salient Features for Speech Emotion Recognition Using Convolutional Neural Networks, IEEE Trans. Multimed. 16 (2014) 2203–2213. https://doi.org/10.1109/TMM.2014.2360798.
[188] J. Lee, I. Tashev, High-Level Feature Representation Using Recurrent Neural Network for Speech Emotion Recognition, in: Dresden, Germany, 2015: pp. 1537–1540.
[189] F. Eyben, M. Wöllmer, A. Graves, B. Schuller, E. Douglas-Cowie, R. Cowie, On-line emotion recognition in a 3-D activation-valence-time continuum using acoustic and linguistic cues, J. Multimodal User Interfaces. 3 (2010) 7–19. https://doi.org/10.1007/s12193-009-0032-6.
[190] B.T. Atmaja, M. Akagi, Speech Emotion Recognition Based on Speech Segment Using LSTM with Attention Model, in: 2019 IEEE Int. Conf. Signals Syst. ICSigSys, 2019: pp. 40–44. https://doi.org/10.1109/ICSIGSYS.2019.8811080.
[191] M. Neumann, N.T. Vu, Improving Speech Emotion Recognition with Unsupervised Representation Learning on Unlabeled Speech, in: ICASSP 2019 - 2019 IEEE Int. Conf. Acoust. Speech Signal Process. ICASSP, 2019: pp. 7390–7394. https://doi.org/10.1109/ICASSP.2019.8682541.
[192] M. Abdelwahab, C. Busso, Domain Adversarial for Acoustic Emotion Recognition, IEEEACM Trans. Audio Speech Lang. Process. 26 (2018) 2423–2435. https://doi.org/10.1109/TASLP.2018.2867099.
[193] A.M. Badshah, J. Ahmad, N. Rahim, S.W. Baik, Speech Emotion Recognition from Spectrograms with Deep Convolutional Neural Network, in: 2017 Int. Conf. Platf. Technol. Serv. PlatCon, 2017: pp. 1–5. https://doi.org/10.1109/PlatCon.2017.7883728.
[194] S. Zhang, S. Zhang, T. Huang, W. Gao, Speech Emotion Recognition Using Deep Convolutional Neural Network and Discriminant Temporal Pyramid Matching, IEEE Trans. Multimed. 20 (2018) 1576–1590. https://doi.org/10.1109/TMM.2017.2766843.
[195] D. Bertero, F.B. Siddique, C.-S. Wu, Y. Wan, R.H.Y. Chan, P. Fung, Real-Time Speech Emotion and Sentiment Recognition for Interactive Dialogue Systems, in: Proc. 2016 Conf. Empir. Methods Nat. Lang. Process., Association for Computational Linguistics, Austin, Texas, 2016: pp. 1042–1047. https://doi.org/10.18653/v1/D16-1110.
[196] G.-B. Huang, Q.-Y. Zhu, C.-K. Siew, Extreme learning machine: Theory and applications, Neurocomputing. 70 (2006) 489–501. https://doi.org/10.1016/j.neucom.2005.12.126.
[197] S. Ghosh, E. Laksana, L.-P. Morency, S. Scherer, Representation Learning for Speech Emotion Recognition, in: San Francisco, USA, 2016: pp. 3603–3607. https://doi.org/10.21437/Interspeech.2016-692.
[198] S. Mirsamadi, E. Barsoum, C. Zhang, Automatic speech emotion recognition using recurrent neural networks with local attention, in: 2017 IEEE Int. Conf. Acoust. Speech Signal Process. ICASSP, 2017: pp. 2227–2231. https://doi.org/10.1109/ICASSP.2017.7952552.
[199] M. Chen, X. He, J. Yang, H. Zhang, 3-D Convolutional Recurrent Neural Networks With Attention Model for Speech Emotion Recognition, IEEE Signal Process. Lett. 25 (2018) 1440–1444. https://doi.org/10.1109/LSP.2018.2860246.
[200] G. Trigeorgis, F. Ringeval, R. Brueckner, E. Marchi, M.A. Nicolaou, B. Schuller, S. Zafeiriou, Adieu features? End-to-end speech emotion recognition using a deep convolutional recurrent network, in: 2016 IEEE Int. Conf. Acoust. Speech Signal Process. ICASSP, 2016: pp. 5200–5204. https://doi.org/10.1109/ICASSP.2016.7472669.
[201] P. Tzirakis, J. Zhang, B.W. Schuller, End-to-End Speech Emotion Recognition Using Deep Neural Networks, in: 2018 IEEE Int. Conf. Acoust. Speech Signal Process. ICASSP, 2018: pp. 5089–5093. https://doi.org/10.1109/ICASSP.2018.8462677.
[202] X. Wu, S. Liu, Y. Cao, X. Li, J. Yu, D. Dai, X. Ma, S. Hu, Z. Wu, X. Liu, H. Meng, Speech Emotion Recognition Using Capsule Networks, in: ICASSP 2019 - 2019 IEEE Int. Conf. Acoust. Speech Signal Process. ICASSP, 2019: pp. 6695–6699. https://doi.org/10.1109/ICASSP.2019.8683163.
[203] Z. Zhao, Y. Zhao, Z. Bao, H. Wang, Z. Zhang, C. Li, Deep Spectrum Feature Representations for Speech Emotion Recognition, in: ASMMC-MMAC18, Association for Computing Machinery, New York, NY, USA, 2018: pp. 27–33. https://doi.org/10.1145/3267935.3267948.
[204] S. Sahu, R. Gupta, G. Sivaraman, W. AbdAlmageed, C. Espy-Wilson, Adversarial Auto-Encoders for Speech Based Emotion Recognition, in: Interspeech 2017, ISCA, 2017: pp. 1243–1247. https://doi.org/10.21437/Interspeech.2017-1421.
[205] J. Han, Z. Zhang, Z. Ren, F. Ringeval, B. Schuller, Towards Conditional Adversarial Training for Predicting Emotions from Speech, in: 2018 IEEE Int. Conf. Acoust. Speech Signal Process. ICASSP, 2018: pp. 6822–6826. https://doi.org/10.1109/ICASSP.2018.8462579.
[206] S. Sahu, R. Gupta, C. Espy-Wilson, Modeling Feature Representations for Affective Speech using Generative Adversarial Networks, IEEE Trans. Affect. Comput. (2020). https://doi.org/10.1109/TAFFC.2020.2998118.

[207] F. Bao, M. Neumann, N.T. Vu, CycleGAN-Based Emotion Style Transfer as Data Augmentation for Speech Emotion Recognition, in: Interspeech 2019, ISCA, 2019: pp. 2828–2832. https://doi.org/10.21437/Interspeech.2019-2293.

[208] J. Zhu, T. Park, P. Isola, A.A. Efros, Unpaired Image-to-Image Translation Using Cycle-Consistent Adversarial Networks, in: 2017 IEEE Int. Conf. Comput. Vis. ICCV, 2017: pp. 2242–2251. https://doi.org/10.1109/ICCV.2017.244.

[209] Z. Zeng, M. Pantic, G.I. Roisman, T.S. Huang, A Survey of Affect Recognition Methods: Audio, Visual, and Spontaneous Expressions, IEEE Trans. Pattern Anal. Mach. Intell. 31 (2009) 39–58. https://doi.org/10.1109/TPAMI.2008.52.

[210] Y.-I. Tian, T. Kanade, J.F. Cohn, Recognizing action units for facial expression analysis, IEEE Trans. Pattern Anal. Mach. Intell. 23 (2001) 97–115. https://doi.org/10.1109/34.908962.

[211] E. Sariyanidi, H. Gunes, A. Cavallaro, Automatic Analysis of Facial Affect: A Survey of Registration, Representation, and Recognition, IEEE Trans. Pattern Anal. Mach. Intell. 37 (2015) 1113–1133. https://doi.org/10.1109/TPAMI.2014.2366127.

[212] P. Liu, S. Han, Z. Meng, Y. Tong, Facial Expression Recognition via a Boosted Deep Belief Network, in: 2014 IEEE Conf. Comput. Vis. Pattern Recognit., 2014: pp. 1805–1812. https://doi.org/10.1109/CVPR.2014.233.

[213] H. Jung, S. Lee, J. Yim, S. Park, J. Kim, Joint Fine-Tuning in Deep Neural Networks for Facial Expression Recognition, in: 2015 IEEE Int. Conf. Comput. Vis. ICCV, IEEE, Santiago, Chile, 2015: pp. 2983–2991. https://doi.org/10.1109/ICCV.2015.341.

[214] B. Xia, W. Wang, S. Wang, E. Chen, Learning from Macro-expression: a Micro-expression Recognition Framework, in: Proc. 28th ACM Int. Conf. Multimed., ACM, Seattle WA USA, 2020: pp. 2936–2944. https://doi.org/10.1145/3394171.3413774.

[215] X. Ben, Y. Ren, J. Zhang, S.-J. Wang, K. Kpalma, W. Meng, Y.-J. Liu, Video-based Facial Micro-Expression Analysis: A Survey of Datasets, Features and Algorithms, IEEE Trans. Pattern Anal. Mach. Intell. (2021) 1–1. https://doi.org/10.1109/TPAMI.2021.3067464.

[216] Y.-J. Liu, J.-K. Zhang, W.-J. Yan, S.-J. Wang, G. Zhao, X. Fu, A Main Directional Mean Optical Flow Feature for Spontaneous Micro-Expression Recognition, IEEE Trans. Affect. Comput. 7 (2016) 299–310. https://doi.org/10.1109/TAFFC.2015.2485205.

[217] H. Zheng, J. Zhu, Z. Yang, Z. Jin, Effective micro-expression recognition using relaxed K-SVD algorithm, Int. J. Mach. Learn. Cybern. 8 (2017) 2043–2049. https://doi.org/10.1007/s13042-017-0684-6.

[218] B. Sun, S. Cao, D. Li, J. He, L. Yu, Dynamic Micro-Expression Recognition Using Knowledge Distillation, IEEE Trans. Affect. Comput. (2020) 1–1. https://doi.org/10.1109/TAFFC.2020.2986962.

[219] A. Majumder, L. Behera, V.K. Subramanian, Automatic Facial Expression Recognition System Using Deep Network-Based Data Fusion, IEEE Trans. Cybern. 48 (2018) 103–114. https://doi.org/10.1109/TCYB.2016.2625419.

[220] A. Barman, P. Dutta, Facial expression recognition using distance and texture signature relevant features, Appl. Soft Comput. 77 (2019) 88–105. https://doi.org/10.1016/j.asoc.2019.01.011.

[221] G. Wen, T. Chang, H. Li, L. Jiang, Dynamic Objectives Learning for Facial Expression Recognition, IEEE Trans. Multimed. (2020) 1–1. https://doi.org/10.1109/TMM.2020.2966858.

[222] K. Yurtkan, H. Demirel, Feature selection for improved 3D facial expression recognition, Pattern Recognit. Lett. 38 (2014) 26–33. https://doi.org/10.1016/j.patrec.2013.10.026.

[223] Q. Zhen, D. Huang, Y. Wang, L. Chen, Muscular Movement Model-Based Automatic 3D/4D Facial Expression Recognition, IEEE Trans. Multimed. 18 (2016) 1438–1450. https://doi.org/10.1109/TMM.2016.2557063.

[224] M. Behzad, N. Vo, X. Li, G. Zhao, Automatic 4D Facial Expression Recognition via Collaborative Cross-domain Dynamic Image Network, in: Proc. Br. Mach. Vis. Conf. BMVC, BMVA Press, 2019: p. 149.1--149.12.

[225] Z. Yu, C. Zhang, Image based Static Facial Expression Recognition with Multiple Deep Network Learning, in: Proc. 2015 ACM Int. Conf. Multimodal Interact., Association for Computing Machinery, New York, NY, USA, 2015: pp. 435–442. https://doi.org/10.1145/2818346.2830595.

[226] D.K. Jain, P. Shamsolmoali, P. Sehdev, Extended deep neural network for facial emotion recognition, Pattern Recognit. Lett. 120 (2019) 69–74. https://doi.org/10.1016/j.patrec.2019.01.008.

[227] A. Yao, D. Cai, P. Hu, S. Wang, L. Sha, Y. Chen, HoloNet: towards robust emotion recognition in the wild, in: Proc. 18th ACM Int. Conf. Multimodal Interact., Association for Computing Machinery, New York, NY, USA, 2016: pp. 472–478. https://doi.org/10.1145/2993148.2997639.

[228] Z. Zhang, Feature-based facial expression recognition: sensitivity analysis and experiments with a multilayer perceptron, Int. J. Pattern Recognit. Artif. Intell. 13 (1999) 893–911. https://doi.org/10.1142/S0218001499000495.

[229] N. Sun, Q. Li, R. Huan, J. Liu, G. Han, Deep spatial-temporal feature fusion for facial expression recognition in static images, Pattern Recognit. Lett. 119 (2019) 49–61. https://doi.org/10.1016/j.patrec.2017.10.022.

[230] D. Ghimire, J. Lee, Geometric Feature-Based Facial Expression Recognition in Image Sequences Using Multi-Class AdaBoost and Support Vector Machines, Sensors. 13 (2013) 7714–7734. https://doi.org/10.3390/s130607714.

[231] Sujono, A.A.S. Gunawan, Face Expression Detection on Kinect Using Active Appearance Model and Fuzzy Logic, Procedia Comput. Sci. 59 (2015) 268–274. https://doi.org/10.1016/j.procs.2015.07.558.

[232] T.F. Cootes, G.J. Edwards, C.J. Taylor, Active appearance models, IEEE Trans. Pattern Anal. Mach. Intell. 23 (2001) 681–685. https://doi.org/10.1109/34.927467.

[233] P. Ekman, J.C. Hager, W.V. Friesen, Facial Action Coding System: The Manual on CD ROM, Salt Lake City, 2002.

[234] F. Makhmudkhujaev, M. Abdullah-Al-Wadud, M.T.B. Iqbal, B. Ryu, O. Chae, Facial expression recognition with local prominent directional pattern, Signal Process. Image Commun. 74 (2019) 1–12. https://doi.org/10.1016/j.image.2019.01.002.

[235] Y. Yan, Z. Zhang, S. Chen, H. Wang, Low-resolution facial expression recognition: A filter learning perspective, Signal Process. 169 (2020) 107370. https://doi.org/10.1016/j.sigpro.2019.107370.

[236] Y. Yao, D. Huang, X. Yang, Y. Wang, L. Chen, Texture and Geometry Scattering Representation-Based Facial Expression Recognition in 2D+3D Videos, ACM Trans. Multimed. Comput. Commun. Appl. 14 (2018) 18:1-18:23. https://doi.org/10.1145/3131345.

[237] Y. Zong, X. Huang, W. Zheng, Z. Cui, G. Zhao, Learning from Hierarchical Spatiotemporal Descriptors for Micro-Expression Recognition, IEEE Trans. Multimed. 20 (2018) 3160–3172. https://doi.org/10.1109/TMM.2018.2820321.

[238] K. Zhang, Y. Huang, Y. Du, L. Wang, Facial Expression Recognition Based on Deep Evolutional Spatial-Temporal Networks, IEEE Trans. Image Process. 26 (2017) 4193–4203. https://doi.org/10.1109/TIP.2017.2689999.

[239] F. Zhang, T. Zhang, Q. Mao, C. Xu, Joint Pose and Expression Modeling for Facial Expression Recognition, in: 2018 IEEECVF Conf. Comput. Vis. Pattern Recognit., IEEE, Salt Lake City, UT, USA, 2018: pp. 3359–3368. https://doi.org/10.1109/CVPR.2018.00354.

[240] B. Fasel, J. Luettin, Automatic facial expression analysis: a survey, Pattern Recognit. 36 (2003) 259–275. https://doi.org/10.1016/S0031-3203(02)00052-3.

[241] J. Hamm, C.G. Kohler, R.C. Gur, R. Verma, Automated Facial Action Coding System for dynamic analysis of facial expressions in neuropsychiatric disorders, J. Neurosci. Methods. 200 (2011) 237–256. https://doi.org/10.1016/j.jneumeth.2011.06.023.

[242] C. Shan, S. Gong, P.W. McOwan, Facial expression recognition based on Local Binary Patterns: A comprehensive study, Image Vis. Comput. 27 (2009) 803–816. https://doi.org/10.1016/j.imavis.2008.08.005.

[243] W. Gu, C. Xiang, Y.V. Venkatesh, D. Huang, H. Lin, Facial expression recognition using radial encoding of local Gabor features and classifier synthesis, Pattern Recognit. 45 (2012) 80–91. https://doi.org/10.1016/j.patcog.2011.05.006.

[244] G. Zhao, M. Pietikainen, Dynamic Texture Recognition Using Local Binary Patterns with an Application to Facial Expressions, IEEE Trans. Pattern Anal. Mach. Intell. 29 (2007) 915–928. https://doi.org/10.1109/TPAMI.2007.1110.

[245] Y. Wang, J. See, R.C.-W. Phan, Y.-H. Oh, Efficient Spatio-Temporal Local Binary Patterns for Spontaneous Facial Micro-Expression Recognition, PLOS ONE. (2015). https://doi.org/10.1371/journal.pone.0124674.

[246] A.K. Davison, M.H. Yap, N. Costen, K. Tan, C. Lansley, D. Leightley, Micro-Facial Movements: An Investigation on Spatio-Temporal Descriptors, in: L. Agapito, M.M. Bronstein, C. Rother (Eds.), Comput. Vis. - ECCV 2014 Workshop, Springer International Publishing, Cham, 2015: pp. 111–123. https://doi.org/10.1007/978-3-319-16181-5_8.

[247] S.-T. Liong, J. See, R.C.-W. Phan, K. Wong, S.-W. Tan, Hybrid Facial Regions Extraction for Micro-expression Recognition System, J. Signal Process. Syst. 90 (2018) 601–617. https://doi.org/10.1007/s11265-017-1276-0.

[248] S. Zhang, B. Feng, Z. Chen, X. Huang, Micro-Expression Recognition by Aggregating Local Spatio-Temporal Patterns, Springer International Publishing, 2017. https://doi.org/10.1007/978-3-319-51811-4_52.

[249] Q. Zhen, D. Huang, H. Drira, B.B. Amor, Y. Wang, M. Daoudi, Magnifying Subtle Facial Motions for Effective 4D Expression Recognition, IEEE Trans. Affect. Comput. 10 (2019) 524–536. https://doi.org/10.1109/TAFFC.2017.2747553.

[250] A. Moeini, K. Faez, H. Sadeghi, H. Moeini, 2D facial expression recognition via 3D reconstruction and feature fusion, J. Vis. Commun. Image Represent. 35 (2016) 1–14. https://doi.org/10.1016/j.jvcir.2015.11.006.

[251] A.C. Le Ngo, S.-T. Liong, J. See, R.C.-W. Phan, Are subtle expressions too sparse to recognize?, in: 2015 IEEE Int. Conf. Digit. Signal Process. DSP, 2015: pp. 1246–1250. https://doi.org/10.1109/ICDSP.2015.7252080.

[252] W. Zheng, Multi-View Facial Expression Recognition Based on Group Sparse Reduced-Rank Regression, IEEE Trans. Affect. Comput. 5 (2014) 71–85. https://doi.org/10.1109/TAFFC.2014.2304712.

[253] A. Azazi, S. Lebai Lutfi, I. Venkat, F. Fernández-Martínez, Towards a robust affect recognition: Automatic facial expression recognition in 3D faces, Expert Syst. Appl. 42 (2015) 3056–3066. https://doi.org/10.1016/j.eswa.2014.10.042.

[254] A. Savran, B. Sankur, Non-rigid registration based model-free 3D facial expression recognition, Comput. Vis. Image Underst. 162 (2017) 146–165. https://doi.org/10.1016/j.cviu.2017.07.005.

[255] M. Chen, H.T. Ma, J. Li, H. Wang, Emotion recognition using fixed length micro-expressions sequence and weighting method, in: 2016 IEEE Int. Conf. Real-Time Comput. Robot. RCAR, 2016: pp. 427–430. https://doi.org/10.1109/RCAR.2016.7784067.

[256] K. Simonyan, A. Zisserman, Very Deep Convolutional Networks for Large-Scale Image Recognition, in: Int. Conf. Learn. Represent., 2015.

[257] O.M. Parkhi, A. Vedaldi, A. Zisserman, Deep Face Recognition, in: Procedings Br. Mach. Vis. Conf. 2015, British Machine Vision Association, Swansea, 2015: p. 41.1-41.12. https://doi.org/10.5244/C.29.41.

[258] K. He, X. Zhang, S. Ren, J. Sun, Deep Residual Learning for Image Recognition, in: 2016 IEEE Conf. Comput. Vis. Pattern Recognit. CVPR, 2016: pp. 770–778. https://doi.org/10.1109/CVPR.2016.90.

[259] C. Szegedy, Wei Liu, Yangqing Jia, P. Sermanet, S. Reed, D. Anguelov, D. Erhan, V. Vanhoucke, A. Rabinovich, Going deeper with convolutions, in: 2015 IEEE Conf. Comput. Vis. Pattern Recognit. CVPR, 2015: pp. 1–9. https://doi.org/10.1109/CVPR.2015.7298594.

[260] H. Yang, U. Ciftci, L. Yin, Facial Expression Recognition by De-expression Residue Learning, in: 2018 IEEECVF Conf. Comput. Vis. Pattern Recognit., IEEE, Salt Lake City, UT, 2018: pp. 2168–2177. https://doi.org/10.1109/CVPR.2018.00231.

[261] C. Wang, M. Peng, T. Bi, T. Chen, Micro-attention for micro-expression recognition, Neurocomputing. 410 (2020) 354–362. https://doi.org/10.1016/j.neucom.2020.06.005.

[262] X. Liu, B.V.K. Vijaya Kumar, P. Jia, J. You, Hard negative generation for identity-disentangled facial expression recognition, Pattern Recognit. 88 (2019) 1–12. https://doi.org/10.1016/j.patcog.2018.11.001.

[263] Z. Meng, P. Liu, J. Cai, S. Han, Y. Tong, Identity-Aware Convolutional Neural Network for Facial Expression Recognition, in: 2017 12th IEEE Int. Conf. Autom. Face Gesture Recognit. FG 2017, 2017: pp. 558–565. https://doi.org/10.1109/FG.2017.140.

[264] Z. Wang, F. Zeng, S. Liu, B. Zeng, OAENet: Oriented attention ensemble for accurate facial expression recognition, Pattern Recognit. 112 (2021) 107694. https://doi.org/10.1016/j.patcog.2020.107694.

[265] H. Li, N. Wang, Y. Yu, X. Yang, X. Gao, LBAN-IL: A novel method of high discriminative representation for facial expression recognition, Neurocomputing. 432 (2021) 159–169. https://doi.org/10.1016/j.neucom.2020.12.076.

[266] P.D.M. Fernandez, F.A.G. Pena, T.I. Ren, A. Cunha, FERAtt: Facial Expression Recognition With Attention Net, in: 2019 IEEECVF Conf. Comput. Vis. Pattern Recognit. Workshop CVPRW, IEEE, Long Beach, CA, USA, 2019: pp. 837–846. https://doi.org/10.1109/CVPRW.2019.00112.

[267] S. Xie, H. Hu, Y. Wu, Deep multi-path convolutional neural network joint with salient region attention for facial expression recognition, Pattern Recognit. 92 (2019) 177–191. https://doi.org/10.1016/j.patcog.2019.03.019.

[268] K. Wang, X. Peng, J. Yang, S. Lu, Y. Qiao, Suppressing Uncertainties for Large-Scale Facial Expression Recognition, in: 2020 IEEECVF Conf. Comput. Vis. Pattern Recognit. CVPR, IEEE, Seattle, WA, USA, 2020: pp. 6896–6905. https://doi.org/10.1109/CVPR42600.2020.00693.

[269] K. Zhu, Z. Du, W. Li, D. Huang, Y. Wang, L. Chen, Discriminative Attention-based Convolutional Neural Network for 3D Facial Expression Recognition, in: 2019 14th IEEE Int. Conf. Autom. Face Gesture Recognit. FG 2019, 2019: pp. 1–8. https://doi.org/10.1109/FG.2019.8756524.

[270] D. Gera, S. Balasubramanian, Landmark guidance independent spatio-channel attention and complementary context information based facial expression recognition, Pattern Recognit. Lett. 145 (2021) 58–66. https://doi.org/10.1016/j.patrec.2021.01.029.

[271] L. Chen, M. Zhou, W. Su, M. Wu, J. She, K. Hirota, Softmax regression based deep sparse autoencoder network for facial emotion recognition in human-robot interaction, Inf. Sci. 428 (2018) 49–61. https://doi.org/10.1016/j.ins.2017.10.044.

[272] Z. Chen, D. Huang, Y. Wang, L. Chen, Fast and Light Manifold CNN based 3D Facial Expression Recognition across Pose Variations, in: Proc. 26th ACM Int. Conf. Multimed., Association for Computing Machinery, New York, NY, USA, 2018: pp. 229–238. https://doi.org/10.1145/3240508.3240568.

[273] Huibin Li, Jian Sun, Zongben Xu, Multimodal 2D+3D Facial Expression Recognition With Deep Fusion Convolutional Neural Network, IEEE Trans. Multimed. (2017). https://doi.org/10.1109/TMM.2017.2713408.

[274] S. Li, W. Deng, A Deeper Look at Facial Expression Dataset Bias, IEEE Trans. Affect. Comput. (2020) 1–1. https://doi.org/10.1109/TAFFC.2020.2973158.

[275] H. Li, N. Wang, X. Ding, X. Yang, X. Gao, Adaptively Learning Facial Expression Representation via C-F Labels and Distillation, IEEE Trans. Image Process. 30 (2021) 2016–2028. https://doi.org/10.1109/TIP.2021.3049955.

[276] D. Tran, L. Bourdev, R. Fergus, L. Torresani, M. Paluri, Learning Spatiotemporal Features with 3D Convolutional Networks, in: 2015 IEEE Int. Conf. Comput. Vis. ICCV, 2015: pp. 4489–4497. https://doi.org/10.1109/ICCV.2015.510.

[277] D.A.A. CHANTI, A. Caplier, Deep Learning for Spatio-Temporal Modeling of Dynamic Spontaneous Emotions, IEEE Trans. Affect. Comput. (2018) 1–1. https://doi.org/10.1109/TAFFC.2018.2873600.

[278] D. Nguyen, K. Nguyen, S. Sridharan, A. Ghasemi, D. Dean, C. Fookes, Deep Spatio-Temporal Features for Multimodal Emotion Recognition, in: 2017 IEEE Winter Conf. Appl. Comput. Vis. WACV, 2017: pp. 1215–1223. https://doi.org/10.1109/WACV.2017.140.

[279] Y. Wang, H. Ma, X. Xing, Z. Pan, Eulerian Motion Based 3DCNN Architecture for Facial Micro-Expression Recognition, in: Y.M. Ro, W.-H. Cheng, J. Kim, W.-T. Chu, P. Cui, J.-W. Choi, M.-C. Hu, W. De Neve (Eds.), Multimed. Model., Springer International Publishing, Cham, 2020: pp. 266–277. https://doi.org/10.1007/978-3-030-37731-1_22.

[280] L. Lo, H.-X. Xie, H.-H. Shuai, W.-H. Cheng, MER-GCN: Micro-Expression Recognition Based on Relation Modeling with Graph Convolutional Networks, in: 2020 IEEE Conf. Multimed. Inf. Process. Retr. MIPR, 2020: pp. 79–84. https://doi.org/10.1109/MIPR49039.2020.00023.

[281] D.H. Kim, W.J. Baddar, J. Jang, Y.M. Ro, Multi-Objective Based Spatio-Temporal Feature Representation Learning Robust to Expression Intensity Variations for Facial Expression Recognition, IEEE Trans. Affect. Comput. 10 (2019) 223–236. https://doi.org/10.1109/TAFFC.2017.2695999.

[282] M. Behzad, N. Vo, X. Li, G. Zhao, Towards Reading Beyond Faces for Sparsity-aware 3D/4D Affect Recognition, Neurocomputing. 458 (2021) 297–307. https://doi.org/10.1016/j.neucom.2021.06.023.

[283] D.H. Kim, W.J. Baddar, Y.M. Ro, Micro-Expression Recognition with Expression-State Constrained Spatio-Temporal Feature Representations, in: Proc. 24th ACM Int. Conf. Multimed., Association for Computing Machinery, New York, NY, USA, 2016: pp. 382–386. https://doi.org/10.1145/2964284.2967247.

[284] Z. Xia, X. Hong, X. Gao, X. Feng, G. Zhao, Spatiotemporal Recurrent Convolutional Networks for Recognizing Spontaneous Micro-Expressions, IEEE Trans. Multimed. 22 (2020) 626–640. https://doi.org/10.1109/TMM.2019.2931351.

[285] D. Kollias, S.P. Zafeiriou, Exploiting multi-CNN features in CNN-RNN based Dimensional Emotion Recognition on the OMG in-the-wild Dataset, IEEE Trans. Affect. Comput. (2020) 1–1. https://doi.org/10.1109/TAFFC.2020.3014171.

[286] D. Liu, X. Ouyang, S. Xu, P. Zhou, K. He, S. Wen, SAANet: Siamese action-units attention network for improving dynamic facial expression recognition, Neurocomputing. 413 (2020) 145–157. https://doi.org/10.1016/j.neucom.2020.06.062.

[287] F. Zhang, T. Zhang, Q. Mao, C. Xu, Geometry Guided Pose-Invariant Facial Expression Recognition, IEEE Trans. Image Process. 29 (2020) 4445–4460. https://doi.org/10.1109/TIP.2020.2972114.

[288] H. Yang, Z. Zhang, L. Yin, Identity-Adaptive Facial Expression Recognition through Expression Regeneration Using Conditional Generative Adversarial Networks, in: 2018 13th IEEE Int. Conf. Autom. Face Gesture Recognit. FG 2018, 2018: pp. 294–301. https://doi.org/10.1109/FG.2018.00050.

[289] Y. Fu, X. Wu, X. Li, Z. Pan, D. Luo, Semantic Neighborhood-Aware Deep Facial Expression Recognition, IEEE Trans. Image Process. 29 (2020) 6535–6548. https://doi.org/10.1109/TIP.2020.2991510.

[290] K. Ali, C.E. Hughes, Facial Expression Recognition Using Disentangled Adversarial Learning, ArXiv190913135 Cs. (2019). http://arxiv.org/abs/1909.13135 (accessed September 23, 2020).

[291] J. Yu, C. Zhang, Y. Song, W. Cai, ICE-GAN: Identity-Aware and Capsule-Enhanced GAN with Graph-Based Reasoning for Micro-Expression Recognition and Synthesis, in: 2021 Int. Jt. Conf. Neural Netw. IJCNN, 2021: pp. 1–8. https://doi.org/10.1109/IJCNN52387.2021.9533988.

[292] S. Piana, A. Staglianò, F. Odone, A. Camurri, Adaptive Body Gesture Representation for Automatic Emotion Recognition, ACM Trans. Interact. Intell. Syst. 6 (2016) 1–31. https://doi.org/10.1145/2818740.

[293] D. Stoeva, M. Gelautz, Body Language in Affective Human-Robot Interaction, in: Companion 2020 ACMIEEE Int. Conf. Hum.-Robot Interact., ACM, Cambridge United Kingdom, 2020: pp. 606–608. https://doi.org/10.1145/3371382.3377432.

[294] Y. Yang, D. Ramanan, Articulated Human Detection with Flexible Mixtures of Parts, IEEE Trans. Pattern Anal. Mach. Intell. 35 (2013) 2878–2890. https://doi.org/10.1109/TPAMI.2012.261.

[295] S. Ren, K. He, R. Girshick, J. Sun, Faster R-CNN: Towards Real-Time Object Detection with Region Proposal Networks, IEEE Trans. Pattern Anal. Mach. Intell. 39 (2017) 1137–1149. https://doi.org/10.1109/TPAMI.2016.2577031.

[296] W. Weiyi, E. Valentin, S. Hichem, Adaptive Real-Time Emotion Recognition from Body Movements, ACM Trans. Interact. Intell. Syst. TiiS. (2015). https://dl.acm.org/doi/abs/10.1145/2738221 (accessed November 2, 2020).

[297] A. Kleinsmith, N. Bianchi-Berthouze, Recognizing Affective Dimensions from Body Posture, in: A.C.R. Paiva, R. Prada, R.W. Picard (Eds.), Affect. Comput. Intell. Interact., Springer, Berlin, Heidelberg, 2007: pp. 48–58. https://doi.org/10.1007/978-3-540-74889-2_5.

[298] G. Castellano, S.D. Villalba, Recognising Human Emotions from Body Movement and Gesture Dynamics, in: Affect. Comput. Intell. Interact., Springer, Berlin, Heidelberg, 2007: pp. 71–82. https://doi.org/10.1007/978-3-540-74889-2_7.

[299] S. Saha, S. Datta, A. Konar, R. Janarthanan, A study on emotion recognition from body gestures using Kinect sensor, in: 2014 Int. Conf. Commun. Signal Process., 2014: pp. 056–060. https://doi.org/10.1109/ICCSP.2014.6949798.

[300] Y. Maret, D. Oberson, M. Gavrilova, 11111, in: L. Rutkowski, R. Scherer, M. Korytkowski, W. Pedrycz, R. Tadeusiewicz, J.M. Zurada (Eds.), Artif. Intell. Soft Comput., Springer International Publishing, Cham, 2018: pp. 474–485. https://doi.org/10.1007/978-3-319-91253-0_44.

[301] S. Senecal, L. Cuel, A. Aristidou, N. Magnenat-Thalmann, Continuous body emotion recognition system during theater performances, Comput. Animat. Virtual Worlds. 27 (2016) 311–320. https://doi.org/10.1002/cav.1714.

[302] D. Glowinski, N. Dael, A. Camurri, G. Volpe, M. Mortillaro, K. Scherer, Toward a Minimal Representation of Affective Gestures, IEEE Trans. Affect. Comput. 2 (2011) 106–118. https://doi.org/10.1109/T-AFFC.2011.7.

[303] M.A. Razzaq, J. Bang, S.S. Kang, S. Lee, UnSkEm: Unobtrusive Skeletal-based Emotion Recognition for User Experience, in: 2020 Int. Conf. Inf. Netw. ICOIN, 2020: pp. 92–96. https://doi.org/10.1109/ICOIN48656.2020.9016601.

[304] R. Santhoshkumar, M. Kalaiselvi Geetha, Vision-Based Human Emotion Recognition Using HOG-KLT Feature, in: P.K. Singh, W. Pawłowski, S. Tanwar, N. Kumar, J.J.P.C. Rodrigues, M.S. Obaidat (Eds.), Proc. First Int. Conf. Comput. Commun. Cyber-Secur. IC4S 2019, Springer, Singapore, 2020: pp. 261–272. https://doi.org/10.1007/978-981-15-3369-3_20.

[305] R. Santhoshkumar, M. Kalaiselvi Geetha, Human Emotion Recognition Using Body Expressive Feature, in: A. Chaudhary, C. Choudhary, M.K. Gupta, C. Lal, T. Badal (Eds.), Microservices Big Data Anal., Springer, Singapore, 2020: pp. 141–149. https://doi.org/10.1007/978-981-15-0128-9_13.

[306] A. Kapur, A. Kapur, N. Virji-Babul, G. Tzanetakis, P.F. Driessen, Gesture-Based Affective Computing on Motion Capture Data, in: J. Tao, T. Tan, R.W. Picard (Eds.), Affect. Comput. Intell. Interact., Springer, Berlin, Heidelberg, 2005: pp. 1–7. https://doi.org/10.1007/11573548_1.

[307] A. Kleinsmith, N. Bianchi-Berthouze, A. Steed, Automatic Recognition of Non-Acted Affective Postures, IEEE Trans. Syst. Man Cybern. Part B Cybern. 41 (2011) 1027–1038. https://doi.org/10.1109/TSMCB.2010.2103557.

[308] E.P. Volkova, B.J. Mohler, T.J. Dodds, J. Tesch, H.H. Bülthoff, Emotion categorization of body expressions in narrative scenarios, Front. Psychol. 5 (2014). https://doi.org/10.3389/fpsyg.2014.00623.

[309] N. Fourati, C. Pelachaud, Multi-level classification of emotional body expression, in: 2015 11th IEEE Int. Conf. Workshop Autom. Face Gesture Recognit. FG, 2015: pp. 1–8. https://doi.org/10.1109/FG.2015.7163145.

[310] Z. Shen, J. Cheng, X. Hu, Q. Dong, Emotion Recognition Based on Multi-View Body Gestures, in: 2019 IEEE Int. Conf. Image Process. ICIP, 2019: pp. 3317–3321. https://doi.org/10.1109/ICIP.2019.8803460.

[311] R. Santhoshkumar, M.K. Geetha, Deep Learning Approach for Emotion Recognition from Human Body Movements with Feedforward Deep Convolution Neural Networks, Procedia Comput. Sci. 152 (2019) 158–165. https://doi.org/10.1016/j.procs.2019.05.038.

[312] S.T. Ly, G.-S. Lee, S.-H. Kim, H.-J. Yang, Emotion Recognition via Body Gesture: Deep Learning Model Coupled with Keyframe Selection, in: Proc. 2018 Int. Conf. Mach. Learn. Mach. Intell., Association for Computing Machinery, New York, NY, USA, 2018: pp. 27–31. https://doi.org/10.1145/3278312.3278313.

[313] J. Wu, Y. Zhang, S. Sun, Q. Li, X. Zhao, Generalized zero-shot emotion recognition from body gestures, Appl. Intell. (2021). https://doi.org/10.1007/s10489-021-02927-w.

[314] D. Avola, L. Cinque, A. Fagioli, G.L. Foresti, C. Massaroni, Deep Temporal Analysis for Non-Acted Body Affect Recognition, IEEE Trans. Affect. Comput. (2020) 1–1. https://doi.org/10.1109/TAFFC.2020.3003816.

[315] L. Wang, Y. Xiong, Z. Wang, Y. Qiao, D. Lin, X. Tang, L.V. Gool, Temporal Segment Networks for Action Recognition in Videos, IEEE Trans. Pattern Anal. Mach. Intell. 41 (2019) 2740–2755. https://doi.org/10.1109/TPAMI.2018.2868668.

[316] S. Yan, Y. Xiong, D. Lin, Spatial Temporal Graph Convolutional Networks for Skeleton-Based Action Recognition, in: Proc. Thirty-Second AAAI Conf. Artif. Intell., AAAI Press, 2018.

[317] C.H. Lampert, H. Nickisch, S. Harmeling, Learning to detect unseen object classes by between-class attribute transfer, in: 2009 IEEE Conf. Comput. Vis. Pattern Recognit., 2009: pp. 951–958. https://doi.org/10.1109/CVPR.2009.5206594.

[318] A. Banerjee, U. Bhattacharya, A. Bera, Learning Unseen Emotions from Gestures via Semantically-Conditioned Zero-Shot Perception with Adversarial Autoencoders, ArXiv200908906 Cs. (2020). http://arxiv.org/abs/2009.08906 (accessed October 28, 2020).

[319] M. Egger, M. Ley, S. Hanke, Emotion Recognition from Physiological Signal Analysis: A Review, Electron. Notes Theor. Comput. Sci. 343 (2019) 35–55. https://doi.org/10.1016/j.entcs.2019.04.009.

[320] K. Shirahama, M. Grzegorzek, Emotion Recognition Based on Physiological Sensor Data Using Codebook Approach, in: E. Piętka, P. Badura, J. Kawa, W. Wieclawek (Eds.), Inf. Technol. Med., Springer International Publishing, Cham, 2016: pp. 27–39. https://doi.org/10.1007/978-3-319-39904-1_3.

[321] S. Katsigiannis, N. Ramzan, DREAMER: A Database for Emotion Recognition Through EEG and ECG Signals From Wireless Low-cost Off-the-Shelf Devices, IEEE J. Biomed. Health Inform. 22 (2018) 98–107. https://doi.org/10.1109/JBHI.2017.2688239.

[322] C. Li, Z. Zhang, R. Song, J. Cheng, Y. Liu, X. Chen, EEG-based Emotion Recognition via Neural Architecture Search, IEEE Trans. Affect. Comput. (2021) 1–1. https://doi.org/10.1109/TAFFC.2021.3130387.

[323] H.J. Yoon, S.Y. Chung, EEG-based emotion estimation using Bayesian weighted-log-posterior function and perceptron convergence algorithm, Comput. Biol. Med. 43 (2013) 2230–2237. https://doi.org/10.1016/j.compbiomed.2013.10.017.

[324] Z. Yin, Y. Wang, L. Liu, W. Zhang, J. Zhang, Cross-Subject EEG Feature Selection for Emotion Recognition Using Transfer Recursive Feature Elimination, Front. Neurorobotics. 11 (2017). https://doi.org/10.3389/fnbot.2017.00019.

[325] Z. Yin, L. Liu, L. Liu, J. Zhang, Y. Wang, Dynamical recursive feature elimination technique for neurophysiological signal-based emotion recognition, Cogn. Technol. Work. 19 (2017) 667–685. https://doi.org/10.1007/s10111-017-0450-2.

[326] K.M. Puk, S. Wang, J. Rosenberger, K.C. Gandy, H.N. Harris, Y.B. Peng, A. Nordberg, P. Lehmann, J. Tommerdahl, J.-C. Chiao, Emotion Recognition and Analysis Using ADMM-Based Sparse Group Lasso, IEEE Trans. Affect. Comput. (2019) 1–1. https://doi.org/10.1109/TAFFC.2019.2943551.

[327] H. He, Y. Tan, J. Ying, W. Zhang, Strengthen EEG-based emotion recognition using firefly integrated optimization algorithm, Appl. Soft Comput. 94 (2020) 106426. https://doi.org/10.1016/j.asoc.2020.106426.

[328] J. Atkinson, D. Campos, Improving BCI-based emotion recognition by combining EEG feature selection and kernel classifiers, Expert Syst. Appl. 47 (2016) 35–41. https://doi.org/10.1016/j.eswa.2015.10.049.

[329] Y. Gao, H.J. Lee, R.M. Mehmood, Deep learninig of EEG signals for emotion recognition, in: 2015 IEEE Int. Conf. Multimed. Expo Workshop ICMEW, 2015: pp. 1–5. https://doi.org/10.1109/ICMEW.2015.7169796.

[330] Y. Li, W. Zheng, Z. Cui, T. Zhang, Y. Zong, A Novel Neural Network Model based on Cerebral Hemispheric Asymmetry for EEG Emotion Recognition, in: Proc. Twenty-Seventh Int. Jt. Conf. Artif. Intell., International Joint Conferences on Artificial Intelligence Organization, Stockholm, Sweden, 2018: pp. 1561–1567. https://doi.org/10.24963/ijcai.2018/216.

[331] T. Song, S. Liu, W. Zheng, Y. Zong, Z. Cui, Instance-Adaptive Graph for EEG Emotion Recognition, Proc. AAAI Conf. Artif. Intell. 34 (2020) 2701–2708. https://doi.org/10.1609/aaai.v34i03.5656.

[332] T. Zhang, Z. Cui, C. Xu, W. Zheng, J. Yang, Variational Pathway Reasoning for EEG Emotion Recognition, Proc. AAAI Conf. Artif. Intell. 34 (2020) 2709–2716. https://doi.org/10.1609/aaai.v34i03.5657.

[333] P. Zhong, D. Wang, C. Miao, EEG-Based Emotion Recognition Using Regularized Graph Neural Networks, IEEE Trans. Affect. Comput. (2020) 1–1. https://doi.org/10.1109/TAFFC.2020.2994159.

[334] Z. Gao, X. Wang, Y. Yang, Y. Li, K. Ma, G. Chen, A Channel-fused Dense Convolutional Network for EEG-based Emotion Recognition, IEEE Trans. Cogn. Dev. Syst. (2020) 1–1. https://doi.org/10.1109/TCDS.2020.2976112.

[335] H. Cui, A. Liu, X. Zhang, X. Chen, K. Wang, X. Chen, EEG-based emotion recognition using an end-to-end regional-asymmetric convolutional neural network, Knowl.-Based Syst. 205 (2020) 106243. https://doi.org/10.1016/j.knosys.2020.106243.

[336] Y.-L. Hsu, J.-S. Wang, W.-C. Chiang, C.-H. Hung, Automatic ECG-Based Emotion Recognition in Music Listening, IEEE Trans. Affect. Comput. 11 (2020) 85–99. https://doi.org/10.1109/TAFFC.2017.2781732.

[337] S.Z. Bong, M. Murugappan, S. Yaacob, Analysis of Electrocardiogram (ECG) Signals for Human Emotional Stress Classification, in: S.G. Ponnambalam, J. Parkkinen, K.C. Ramanathan (Eds.), Trends Intell. Robot. Autom. Manuf., Springer, Berlin, Heidelberg, 2012: pp. 198–205. https://doi.org/10.1007/978-3-642-35197-6_22.

[338] S. Jerritta, M. Murugappan, K. Wan, S. Yaacob, Emotion recognition from electrocardiogram signals using Hilbert Huang Transform, in: 2012 IEEE Conf. Sustain. Util. Dev. Eng. Technol. Stud., 2012: pp. 82–86. https://doi.org/10.1109/STUDENT.2012.6408370.

[339] Z. Cheng, L. Shu, J. Xie, C.L.P. Chen, A novel ECG-based real-time detection method of negative emotions in wearable applications, in: 2017 Int. Conf. Secur. Pattern Anal. Cybern. SPAC, 2017: pp. 296–301. https://doi.org/10.1109/SPAC.2017.8304293.

[340] L. Zhang, S. Walter, X. Ma, P. Werner, A. Al-Hamadi, H.C. Traue, S. Gruss, "BioVid Emo DB": A multimodal database for emotion analyses validated by subjective ratings, in: 2016 IEEE Symp. Ser. Comput. Intell. SSCI, 2016: pp. 1–6. https://doi.org/10.1109/SSCI.2016.7849931.

[341] J. Selvaraj, M. Murugappan, K. Wan, S. Yaacob, Classification of emotional states from electrocardiogram signals: a non-linear approach based on hurst, Biomed. Eng. OnLine. 12 (2013) 44. https://doi.org/10.1186/1475-925X-12-44.

[342] H. Ferdinando, T. Seppänen, E. Alasaarela, Enhancing Emotion Recognition from ECG Signals using Supervised Dimensionality Reduction:, in: Proc. 6th Int. Conf. Pattern Recognit. Appl. Methods, SCITEPRESS - Science and Technology Publications, Porto, Portugal, 2017: pp. 112–118. https://doi.org/10.5220/0006147801120118.

[343] G. Chen, Y. Zhu, Z. Hong, Z. Yang, EmotionalGAN: Generating ECG to Enhance Emotion State Classification, in: Proc. 2019 Int. Conf. Artif. Intell. Comput. Sci., Association for Computing Machinery, New York, NY, USA, 2019: pp. 309–313. https://doi.org/10.1145/3349341.3349422.

[344] P. Sarkar, A. Etemad, Self-Supervised Learning for ECG-Based Emotion Recognition, in: ICASSP 2020 - 2020 IEEE Int. Conf. Acoust. Speech Signal Process. ICASSP, 2020: pp. 3217–3221. https://doi.org/10.1109/ICASSP40776.2020.9053985.

[345] G. Caridakis, G. Castellano, L. Kessous, A. Raouzaiou, L. Malatesta, S. Asteriadis, K. Karpouzis, Multimodal emotion recognition from expressive faces, body gestures and speech, in: C. Boukis, A. Pnevmatikakis, L. Polymenakos (Eds.), Artif. Intell. Innov. 2007 Theory Appl., Springer US, Boston, MA, 2007: pp. 375–388. https://doi.org/10.1007/978-0-387-74161-1_41.

[346] C. Sarkar, S. Bhatia, A. Agarwal, J. Li, Feature Analysis for Computational Personality Recognition Using YouTube Personality Data set, in: Proc. 2014 ACM Multi Media Workshop Comput. Personal. Recognit. - WCPR 14, ACM Press, Orlando, Florida, USA, 2014: pp. 11–14. https://doi.org/10.1145/2659522.2659528.

[347] G.K. Verma, U.S. Tiwary, Multimodal fusion framework: A multiresolution approach for emotion classification and recognition from physiological signals, NeuroImage. 102 (2014) 162–172. https://doi.org/10.1016/j.neuroimage.2013.11.007.

[348] X. Zhang, J. Liu, J. Shen, S. Li, K. Hou, B. Hu, J. Gao, T. Zhang, B. Hu, Emotion Recognition From Multimodal Physiological Signals Using a Regularized Deep Fusion of Kernel Machine, IEEE Trans. Cybern. (2020) 1–14. https://doi.org/10.1109/TCYB.2020.2987575.

[349] M.S. Hossain, G. Muhammad, Emotion recognition using deep learning approach from audio–visual emotional big data, Inf. Fusion. 49 (2019) 69–78. https://doi.org/10.1016/j.inffus.2018.09.008.

[350] N. Sebe, I. Cohen, T. Gevers, T.S. Huang, Emotion Recognition Based on Joint Visual and Audio Cues, in: 18th Int. Conf. Pattern Recognit. ICPR06, IEEE, Hong Kong, China, 2006: pp. 1136–1139. https://doi.org/10.1109/ICPR.2006.489.

[351] B. Schuller, G. Rigoll, M. Lang, Speech emotion recognition combining acoustic features and linguistic information in a hybrid support vector machine-belief network architecture, in: 2004 IEEE Int. Conf. Acoust. Speech Signal Process., 2004: p. I–577. https://doi.org/10.1109/ICASSP.2004.1326051.

[352] J. Sebastian, P. Pierucci, Fusion Techniques for Utterance-Level Emotion Recognition Combining Speech and Transcripts, in: Interspeech 2019, ISCA, 2019: pp. 51–55. https://doi.org/10.21437/Interspeech.2019-3201.

[353] S. Poria, E. Cambria, A. Hussain, G.-B. Huang, Towards an intelligent framework for multimodal affective data analysis, Neural Netw. 63 (2015) 104–116. https://doi.org/10.1016/j.neunet.2014.10.005.

[354] S. Poria, I. Chaturvedi, E. Cambria, A. Hussain, Convolutional MKL Based Multimodal Emotion Recognition and Sentiment Analysis, in: 2016 IEEE 16th Int. Conf. Data Min. ICDM, 2016: pp. 439–448. https://doi.org/10.1109/ICDM.2016.0055.

[355] S.E. Eskimez, R.K. Maddox, C. Xu, Z. Duan, Noise-Resilient Training Method for Face Landmark Generation From Speech, IEEEACM Trans. Audio Speech Lang. Process. 28 (2020) 27–38. https://doi.org/10.1109/TASLP.2019.2947741.

[356] Mingli Song, Jiajun Bu, Chun Chen, Nan Li, Audio-visual based emotion recognition-a new approach, in: Proc. 2004 IEEE Comput. Soc. Conf. Comput. Vis. Pattern Recognit. 2004 CVPR 2004, IEEE, Washington, DC, USA, 2004: pp. 1020–1025. https://doi.org/10.1109/CVPR.2004.1315276.

[357] K. Nickel, T. Gehrig, R. Stiefelhagen, J. McDonough, A joint particle filter for audio-visual speaker tracking, in: Proc. 7th Int. Conf. Multimodal Interfaces - ICMI 05, ACM Press, Toronto, Italy, 2005: p. 61. https://doi.org/10.1145/1088463.1088477.

[358] Z. Zeng, Y. Hu, M. Liu, Y. Fu, T.S. Huang, Training combination strategy of multi-stream fused hidden Markov model for audio-visual affect recognition, in: Proc. 14th Annu. ACM Int. Conf. Multimed. - Multimed. 06, ACM Press, Santa Barbara, CA, USA, 2006: p. 65. https://doi.org/10.1145/1180639.1180661.

[359] G. Caridakis, L. Malatesta, L. Kessous, N. Amir, A. Raouzaiou, K. Karpouzis, Modeling naturalistic affective states via facial and vocal expressions recognition, in: Proc. 8th Int. Conf. Multimodal Interfaces - ICMI 06, ACM Press, Banff, Alberta, Canada, 2006: p. 146. https://doi.org/10.1145/1180995.1181029.

[360] J. Chen, Z. Chen, Z. Chi, H. Fu, Facial Expression Recognition in Video with Multiple Feature Fusion, IEEE Trans. Affect. Comput. 9 (2018) 38–50. https://doi.org/10.1109/TAFFC.2016.2593719.

[361] D. Priyasad, T. Fernando, S. Denman, S. Sridharan, C. Fookes, Attention Driven Fusion for Multi-Modal Emotion Recognition, in: ICASSP 2020 - 2020 IEEE Int. Conf. Acoust. Speech Signal Process. ICASSP, 2020: pp. 3227–3231. https://doi.org/10.1109/ICASSP40776.2020.9054441.

[362] T. Mittal, U. Bhattacharya, R. Chandra, A. Bera, D. Manocha, M3ER: Multiplicative Multimodal Emotion Recognition using Facial, Textual, and Speech Cues, Proc. AAAI Conf. Artif. Intell. 34 (2020) 1359–1367. https://doi.org/10.1609/aaai.v34i02.5492.

[363] H.-C. Yang, C.-C. Lee, An Attribute-invariant Variational Learning for Emotion Recognition Using Physiology, in: ICASSP 2019 - 2019 IEEE Int. Conf. Acoust. Speech Signal Process. ICASSP, 2019: pp. 1184–1188. https://doi.org/10.1109/ICASSP.2019.8683290.

[364] S. Zhao, Y. Ma, Y. Gu, J. Yang, T. Xing, P. Xu, R. Hu, H. Chai, K. Keutzer, An End-to-End Visual-Audio Attention Network for Emotion Recognition in User-Generated Videos, Proc. AAAI Conf. Artif. Intell. 34 (2020) 303–311. https://doi.org/10.1609/aaai.v34i01.5364.

[365] Y. Zhang, Z.-R. Wang, J. Du, Deep Fusion: An Attention Guided Factorized Bilinear Pooling for Audio-video Emotion Recognition, in: Int. Jt. Conf. Neural Netw., 2019. https://doi.org/10.1109/IJCNN.2019.8851942.

[366] K. Hara, H. Kataoka, Y. Satoh, Can Spatiotemporal 3D CNNs Retrace the History of 2D CNNs and ImageNet?, in: 2018 IEEECVF Conf. Comput. Vis. Pattern Recognit., IEEE, Salt Lake City, UT, USA, 2018: pp. 6546–6555. https://doi.org/10.1109/CVPR.2018.00685.

[367] M. Hao, W.-H. Cao, Z.-T. Liu, M. Wu, P. Xiao, Visual-audio emotion recognition based on multi-task and ensemble learning with multiple features, Neurocomputing. 391 (2020) 42–51. https://doi.org/10.1016/j.neucom.2020.01.048.

[368] O. Martin, I. Kotsia, B. Macq, I. Pitas, The eNTERFACE'05 Audio-Visual Emotion Database, in: Proc. 22nd Int. Conf. Data Eng. Workshop, IEEE Computer Society, USA, 2006: p. 8. https://doi.org/10.1109/ICDEW.2006.145.

[369] M. Glodek, S. Reuter, M. Schels, K. Dietmayer, F. Schwenker, Kalman Filter Based Classifier Fusion for Affective State Recognition, in: Z.-H. Zhou, F. Roli, J. Kittler (Eds.), Mult. Classif. Syst., Springer Berlin Heidelberg, Berlin, Heidelberg, 2013: pp. 85–94. https://doi.org/10.1007/978-3-642-38067-9_8.

[370] W. Xue, W. Zhou, T. Li, Q. Wang, MTNA: A Neural Multi-task Model for Aspect Category Classification and Aspect Term Extraction On Restaurant Reviews, in: Proc. Eighth Int. Jt. Conf. Nat. Lang. Process. Vol. 2 Short Pap., Asian Federation of Natural Language Processing, Taipei, Taiwan, 2017: pp. 151–156. https://www.aclweb.org/anthology/I17-2026 (accessed August 17, 2020).

[371] B. Zhang, S. Khorram, E.M. Provost, Exploiting Acoustic and Lexical Properties of Phonemes to Recognize Valence from Speech, in: ICASSP 2019 - 2019 IEEE Int. Conf. Acoust. Speech Signal Process. ICASSP, IEEE, Brighton, United Kingdom, 2019: pp. 5871–5875. https://doi.org/10.1109/ICASSP.2019.8683190.

[372] S. Yoon, S. Byun, K. Jung, Multimodal Speech Emotion Recognition Using Audio and Text, in: 2018 IEEE Spok. Lang. Technol. Workshop SLT, 2018: pp. 112–118. https://doi.org/10.1109/SLT.2018.8639583.

[373] L. Cai, Y. Hu, J. Dong, S. Zhou, Audio-Textual Emotion Recognition Based on Improved Neural Networks, Math. Probl. Eng. 2019 (2019) 1–9. https://doi.org/10.1155/2019/2593036.

[374] C.-H. Wu, W.-B. Liang, Emotion Recognition of Affective Speech Based on Multiple Classifiers Using Acoustic-Prosodic Information and Semantic Labels, IEEE Trans. Affect. Comput. 2 (2011) 10–21. https://doi.org/10.1109/T-AFFC.2010.16.

[375] Q. Jin, C. Li, S. Chen, H. Wu, Speech emotion recognition with acoustic and lexical features, in: 2015 IEEE Int. Conf. Acoust. Speech Signal Process. ICASSP, 2015: pp. 4749–4753. https://doi.org/10.1109/ICASSP.2015.7178872.

[376] L. Pepino, P. Riera, L. Ferrer, A. Gravano, Fusion Approaches for Emotion Recognition from Speech Using Acoustic and Text-Based Features, in: ICASSP 2020 - 2020 IEEE Int. Conf. Acoust. Speech Signal Process. ICASSP, IEEE, Barcelona, Spain, 2020: pp. 6484–6488. https://doi.org/10.1109/ICASSP40776.2020.9054709.

[377] A. Metallinou, M. Wollmer, A. Katsamanis, F. Eyben, B. Schuller, S. Narayanan, Context-Sensitive Learning for Enhanced Audiovisual Emotion Classification, IEEE Trans. Affect. Comput. 3 (2012) 184–198. https://doi.org/10.1109/T-AFFC.2011.40.

[378] E. Cambria, D. Hazarika, S. Poria, A. Hussain, R.B.V. Subramanyam, Benchmarking Multimodal Sentiment Analysis, in: A. Gelbukh (Ed.), Comput. Linguist. Intell. Text Process., Springer International Publishing, 2018: pp. 166–179. https://doi.org/10.1007/978-3-319-77116-8_13.

[379] V. Perez Rosas, R. Mihalcea, L.-P. Morency, Multimodal Sentiment Analysis of Spanish Online Videos, IEEE Intell. Syst. 28 (2013) 38–45. https://doi.org/10.1109/MIS.2013.9.

[380] J. Arguello, C. Rosé, Topic-Segmentation of Dialogue, in: Proc. Anal. Conversat. Text Speech, Association for Computational Linguistics, New York City, New York, 2006: pp. 42–49. https://www.aclweb.org/anthology/W06-3407 (accessed August 23, 2020).

[381] H. Peng, Y. Ma, S. Poria, Y. Li, E. Cambria, Phonetic-enriched text representation for Chinese sentiment analysis with reinforcement learning, Inf. Fusion. 70 (2021) 88–99. https://doi.org/10.1016/j.inffus.2021.01.005.

[382] S. Poria, E. Cambria, A. Gelbukh, Deep Convolutional Neural Network Textual Features and Multiple Kernel Learning for Utterance-level Multimodal Sentiment Analysis, in: Proc. 2015 Conf. Empir. Methods Nat. Lang. Process., Association for Computational Linguistics, Lisbon, Portugal, 2015: pp. 2539–2544. https://doi.org/10.18653/v1/D15-1303.

[383] S. Poria, E. Cambria, D. Hazarika, N. Majumder, A. Zadeh, L.-P. Morency, Context-Dependent Sentiment Analysis in User-Generated Videos, in: Proc. 55th Annu. Meet. Assoc. Comput. Linguist. Vol. 1 Long Pap., Association for Computational Linguistics, Vancouver, Canada, 2017: pp. 873–883. https://doi.org/10.18653/v1/P17-1081.

[384] P. Schmidt, A. Reiss, R. Dürichen, K. Van Laerhoven, Wearable-Based Affect Recognition—A Review, Sensors. 19 (2019) 4079. https://doi.org/10.3390/s19194079.

[385] C. Li, C. Xu, Z. Feng, Analysis of physiological for emotion recognition with the IRS model, Neurocomputing. 178 (2016) 103–111. https://doi.org/10.1016/j.neucom.2015.07.112.

[386] B. Nakisa, M.N. Rastgoo, A. Rakotonirainy, F. Maire, V. Chandran, Long Short Term Memory Hyperparameter Optimization for a Neural Network Based Emotion Recognition Framework, IEEE Access. 6 (2018) 49325–49338. https://doi.org/10.1109/ACCESS.2018.2868361.

[387] M.M. Hassan, Md.G.R. Alam, Md.Z. Uddin, S. Huda, A. Almogren, G. Fortino, Human emotion recognition using deep belief network architecture, Inf. Fusion. 51 (2019) 10–18. https://doi.org/10.1016/j.inffus.2018.10.009.

[388] J. Ma, H. Tang, W.-L. Zheng, B.-L. Lu, Emotion Recognition using Multimodal Residual LSTM Network, in: Proc. 27th ACM Int. Conf. Multimed., ACM, Nice France, 2019: pp. 176–183. https://doi.org/10.1145/3343031.3350871.

[389] W. Wei, Q. Jia, Y. Feng, G. Chen, Emotion Recognition Based on Weighted Fusion Strategy of Multichannel Physiological Signals, Comput. Intell. Neurosci. 2018 (2018) e5296523. https://doi.org/10.1155/2018/5296523.

[390] C. Li, Z. Bao, L. Li, Z. Zhao, Exploring temporal representations by leveraging attention-based bidirectional LSTM-RNNs for multi-modal emotion recognition, Inf. Process. Manag. 57 (2020) 102185. https://doi.org/10.1016/j.ipm.2019.102185.

[391] M.N. Dar, M.U. Akram, S.G. Khawaja, A.N. Pujari, CNN and LSTM-Based Emotion Charting Using Physiological Signals, Sensors. 20 (2020) 4551. https://doi.org/10.3390/s20164551.

[392] Z. Yin, M. Zhao, Y. Wang, J. Yang, J. Zhang, Recognition of emotions using multimodal physiological signals and an ensemble deep learning model, Comput. Methods Programs Biomed. 140 (2017) 93–110. https://doi.org/10.1016/j.cmpb.2016.12.005.

[393] L. He, D. Jiang, L. Yang, E. Pei, P. Wu, H. Sahli, Multimodal Affective Dimension Prediction Using Deep Bidirectional Long Short-Term Memory Recurrent Neural Networks, in: Proc. 5th Int. Workshop Audiov. Emot. Chall. - AVEC 15, ACM Press, Brisbane, Australia, 2015: pp. 73–80. https://doi.org/10.1145/2808196.2811641.

[394] H. Ranganathan, S. Chakraborty, S. Panchanathan, Multimodal emotion recognition using deep learning architectures, in: 2016 IEEE Winter Conf. Appl. Comput. Vis. WACV, 2016: pp. 1–9. https://doi.org/10.1109/WACV.2016.7477679.

[395] Y. Lu, W.-L. Zheng, B. Li, B.-L. Lu, Combining Eye Movements and EEG to Enhance Emotion Recognition, Proc. Twenty-Fourth Int. Jt. Conf. Artif. Intell. IJCAI 2015. (n.d.) 7.

[396] B. Xing, H. Zhang, K. Zhang, L. Zhang, X. Wu, X. Shi, S. Yu, S. Zhang, Exploiting EEG Signals and Audiovisual Feature Fusion for Video Emotion Recognition, IEEE Access. 7 (2019) 59844–59861. https://doi.org/10.1109/ACCESS.2019.2914872.

[397] G. Yin, S. Sun, D. Yu, D. Li, K. Zhang, A Efficient Multimodal Framework for Large Scale Emotion Recognition by Fusing Music and Electrodermal Activity Signals, ArXiv200809743 Cs. (2021). http://arxiv.org/abs/2008.09743 (accessed December 13, 2021).

[398] W. Liu, W.-L. Zheng, B.-L. Lu, Emotion Recognition Using Multimodal Deep Learning, in: Proc. 23rd Int. Conf. Neural Inf. Process. - Vol. 9948, Springer-Verlag, Berlin, Heidelberg, 2016: pp. 521–529. https://doi.org/10.1007/978-3-319-46672-9_58.

[399] M. Soleymani, S. Asghari-Esfeden, Y. Fu, M. Pantic, Analysis of EEG Signals and Facial Expressions for Continuous Emotion Detection, IEEE Trans. Affect. Comput. 7 (2016) 17–28. https://doi.org/10.1109/TAFFC.2015.2436926.

[400] D. Wu, J. Zhang, Q. Zhao, Multimodal Fused Emotion Recognition About Expression-EEG Interaction and Collaboration Using Deep Learning, IEEE Access. 8 (2020) 133180–133189. https://doi.org/10.1109/ACCESS.2020.3010311.

[401] K. Zhang, H. Zhang, S. Li, C. Yang, L. Sun, The PMEmo Dataset for Music Emotion Recognition, in: Proc. 2018 ACM Int. Conf. Multimed. Retr., Association for Computing Machinery, New York, NY, USA, 2018: pp. 135–142. https://doi.org/10.1145/3206025.3206037.

[402] K. Shuang, Q. Yang, J. Loo, R. Li, M. Gu, Feature distillation network for aspect-based sentiment analysis, Inf. Fusion. 61 (2020) 13–23. https://doi.org/10.1016/j.inffus.2020.03.003.

[403] H. Fang, N. Mac Parthaláin, A.J. Aubrey, G.K.L. Tam, R. Borgo, P.L. Rosin, P.W. Grant, D. Marshall, M. Chen, Facial expression recognition in dynamic sequences: An integrated approach, Pattern Recognit. 47 (2014) 1271–1281. https://doi.org/10.1016/j.patcog.2013.09.023.

[404] J. Chen, B. Hu, P. Moore, X. Zhang, X. Ma, Electroencephalogram-based emotion assessment system using ontology and data mining techniques, Appl. Soft Comput. 30 (2015) 663–674. https://doi.org/10.1016/j.asoc.2015.01.007.

[405] S. Poria, D. Hazarika, N. Majumder, G. Naik, E. Cambria, R. Mihalcea, MELD: A Multimodal Multi-Party Dataset for Emotion Recognition in Conversations, in: Proc. 57th Annu. Meet. Assoc. Comput. Linguist., Association for Computational Linguistics, Florence, Italy, 2019: pp. 527–536. https://doi.org/10.18653/v1/P19-1050.

[406] S.A. Abdu, A.H. Yousef, A. Salem, Multimodal Video Sentiment Analysis Using Deep Learning Approaches, a Survey, Inf. Fusion. 76 (2021) 204–226. https://doi.org/10.1016/j.inffus.2021.06.003.

[407] S. Petridis, M. Pantic, Audiovisual Discrimination Between Speech and Laughter: Why and When Visual Information Might Help, IEEE Trans. Multimed. 13 (2011) 216–234. https://doi.org/10.1109/TMM.2010.2101586.

[408] Devangini Patel, X. Hong, G. Zhao, Selective deep features for micro-expression recognition, in: 2016 23rd Int. Conf. Pattern Recognit. ICPR, 2016: pp. 2258–2263. https://doi.org/10.1109/ICPR.2016.7899972.

[409] A.C.L. Ngo, C.W. Phan, J. See, Spontaneous Subtle Expression Recognition: Imbalanced Databases and Solutions, in: Asian Conf. Comput. Vis., 2014. https://doi.org/10.1007/978-3-319-16817-3_3.

[410] X. Jia, X. Ben, H. Yuan, K. Kpalma, W. Meng, Macro-to-micro transformation model for micro-expression recognition, J. Comput. Sci. 25 (2018) 289–297. https://doi.org/10.1016/j.jocs.2017.03.016.

[411] H. Zhang, W. Su, J. Yu, Z. Wang, Weakly Supervised Local-Global Relation Network for Facial Expression Recognition, in: Proc. Twenty-Ninth Int. Jt. Conf. Artif. Intell., International Joint Conferences on Artificial Intelligence Organization, Yokohama, Japan, 2020: pp. 1040–1046. https://doi.org/10.24963/ijcai.2020/145.

[412] B. Schuller, M. Valstar, F. Eyben, G. McKeown, R. Cowie, M. Pantic, AVEC 2011–The First International Audio/Visual Emotion Challenge, in: S. D'Mello, A. Graesser, B. Schuller, J.-C. Martin (Eds.), Affect. Comput. Intell. Interact., Springer, Berlin, Heidelberg, 2011: pp. 415–424. https://doi.org/10.1007/978-3-642-24571-8_53.

[413] F. Ringeval, A. Michaud, E. Ciftçi, H. Güleç, A.A. Salah, M. Pantic, B. Schuller, M. Valstar, R. Cowie, H. Kaya, M. Schmitt, S. Amiriparian, N. Cummins, D. Lalanne, AVEC 2018 Workshop and Challenge: Bipolar Disorder and Cross-Cultural Affect Recognition, in: Proc. 2018 Audiov. Emot. Chall. Workshop - AVEC18, ACM Press, Seoul, Republic of Korea, 2018: pp. 3–13. https://doi.org/10.1145/3266302.3266316.

[414] Ercheng Pei, Xiaohan Xia, Le Yang, Dongmei Jiang, H. Sahli, Deep neural network and switching Kalman filter based continuous affect recognition, in: 2016 IEEE Int. Conf. Multimed. Expo Workshop ICMEW, 2016: pp. 1–6. https://doi.org/10.1109/ICMEW.2016.7574729.

[415] Y. Song, L.-P. Morency, R. Davis, Learning a sparse codebook of facial and body microexpressions for emotion recognition, in: Proc. 15th ACM Int. Conf. Multimodal Interact., Association for Computing Machinery, New York, NY, USA, 2013: pp. 237–244. https://doi.org/10.1145/2522848.2522851.

[416] M. Valstar, J. Gratch, B. Schuller, F. Ringeval, D. Lalanne, M. Torres Torres, S. Scherer, G. Stratou, R. Cowie, M. Pantic, AVEC 2016: Depression, Mood, and Emotion Recognition Workshop and Challenge, in: Proc. 6th Int. Workshop Audiov. Emot. Chall., Association for Computing Machinery, New York, NY, USA, 2016: pp. 3–10. https://doi.org/10.1145/2988257.2988258.

[417] Y. Zhang, D. Song, X. Li, P. Zhang, P. Wang, L. Rong, G. Yu, B. Wang, A Quantum-Like multimodal network framework for modeling interaction dynamics in multiparty conversational sentiment analysis, Inf. Fusion. 62 (2020) 14–31. https://doi.org/10.1016/j.inffus.2020.04.003.

[418] P. Tzirakis, J. Chen, S. Zafeiriou, B. Schuller, End-to-end multimodal affect recognition in real-world environments, Inf. Fusion. 68 (2021) 46–53. https://doi.org/10.1016/j.inffus.2020.10.011.

[419] D. Camacho, M.V. Luzón, E. Cambria, New trends and applications in social media analytics, Future Gener. Comput. Syst. 114 (2021) 318–321. https://doi.org/10.1016/j.future.2020.08.007.

[420] E. Cambria, C. Havasi, A. Hussain, SenticNet 2: A Semantic and Affective Resource for Opinion Mining and Sentiment Analysis, in: Fla. Artif. Intell. Res. Soc. Conf., 2012: pp. 202–207.

[421] E. Cambria, D. Olsher, D. Rajagopal, SenticNet 3: a common and common-sense knowledge base for cognition-driven sentiment analysis, in: Proc. Twenty-Eighth AAAI Conf. Artif. Intell., AAAI Press, Québec City, Québec, Canada, 2014: pp. 1515–1521.

[422] E. Cambria, S. Poria, R. Bajpai, B. Schuller, SenticNet 4: A Semantic Resource for Sentiment Analysis Based on Conceptual Primitives, in: COLING2016, Osaka, Japan, 2016: pp. 2666–2677.

[423] E. Cambria, S. Poria, D. Hazarika, K. Kwok, SenticNet 5: Discovering Conceptual Primitives for Sentiment Analysis by Means of Context Embeddings, in: Proc. Twenty-Eighth AAAI Conf. Artif. Intell., 2018: pp. 1795–1802.

[424] E. Cambria, Y. Li, F.Z. Xing, S. Poria, K. Kwok, SenticNet 6: Ensemble Application of Symbolic and Subsymbolic AI for Sentiment Analysis, in: Proc. 29th ACM Int. Conf. Inf. Knowl. Manag., ACM, Virtual Event Ireland, 2020: pp. 105–114. https://doi.org/10.1145/3340531.3412003.

[425] Y. Ma, H. Peng, E. Cambria, Targeted Aspect-Based Sentiment Analysis via Embedding Commonsense Knowledge into an Attentive LSTM, in: Proc. AAAI Conf. Artif. Intell., 2018: pp. 5876–5883.

[426] F.Z. Xing, E. Cambria, R.E. Welsch, Intelligent Asset Allocation via Market Sentiment Views, IEEE Comput. Intell. Mag. 13 (2018) 25–34. https://doi.org/10.1109/MCI.2018.2866727.

[427] A. Picasso, S. Merello, Y. Ma, L. Oneto, E. Cambria, Technical analysis and sentiment embeddings for market trend prediction, Expert Syst. Appl. 135 (2019) 60–70. https://doi.org/10.1016/j.eswa.2019.06.014.

[428] Y. Qian, Y. Zhang, X. Ma, H. Yu, L. Peng, EARS: Emotion-aware recommender system based on hybrid information fusion, Inf. Fusion. 46 (2019) 141–146. https://doi.org/10.1016/j.inffus.2018.06.004.

[429] D. Xu, Z. Tian, R. Lai, X. Kong, Z. Tan, W. Shi, Deep learning based emotion analysis of microblog texts, Inf. Fusion. 64 (2020) 1–11. https://doi.org/10.1016/j.inffus.2020.06.002.

[430] D. Yang, A. Alsadoon, P.W.C. Prasad, A.K. Singh, A. Elchouemi, An Emotion Recognition Model Based on Facial Recognition in Virtual Learning Environment, Procedia Comput. Sci. 125 (2018) 2–10. https://doi.org/10.1016/j.procs.2017.12.003.

[431] C. Zuheros, E. Martínez-Cámara, E. Herrera-Viedma, F. Herrera, Sentiment Analysis based Multi-Person Multi-criteria Decision Making methodology using natural language processing and deep learning for smarter decision aid. Case study of restaurant choice using TripAdvisor reviews, Inf. Fusion. 68 (2021) 22–36. https://doi.org/10.1016/j.inffus.2020.10.019.

[432] L. Zhang, B. Verma, D. Tjondronegoro, V. Chandran, Facial Expression Analysis under Partial Occlusion: A Survey, ACM Comput. Surv. 51 (2018) 1–49. https://doi.org/10.1145/3158369.

[433] N. Savva, A. Scarinzi, N. Bianchi-Berthouze, Continuous Recognition of Player's Affective Body Expression as Dynamic Quality of Aesthetic Experience, IEEE Trans. Comput. Intell. AI Games. 4 (2012) 199–212. https://doi.org/10.1109/TCIAIG.2012.2202663.

[434] K. Kaza, A. Psaltis, K. Stefanidis, K.C. Apostolakis, S. Thermos, K. Dimitropoulos, P. Daras, Body Motion Analysis for Emotion Recognition in Serious Games, in: M. Antona, C. Stephanidis (Eds.), Univers. Access Hum.-Comput. Interact. Interact. Tech. Environ., Springer International Publishing, Cham, 2016: pp. 33–42. https://doi.org/10.1007/978-3-319-40244-4_4.

[435] W. Dong, L. Yang, R. Gravina, G. Fortino, ANFIS fusion algorithm for eye movement recognition via soft multi-functional electronic skin, Inf. Fusion. 71 (2021) 99–108. https://doi.org/10.1016/j.inffus.2021.02.003.

[436] H. Cai, Z. Qu, Z. Li, Y. Zhang, X. Hu, B. Hu, Feature-level fusion approaches based on multimodal EEG data for depression recognition, Inf. Fusion. 59 (2020) 127–138. https://doi.org/10.1016/j.inffus.2020.01.008.

[437] S. Piana, A. Staglianò, A. Camurri, A set of Full-Body Movement Features for Emotion Recognition to Help Children affected by Autism Spectrum Condition, in: 2013: p. 7.

[438] L.O. Sawada, L.Y. Mano, J.R. Torres Neto, J. Ueyama, A module-based framework to emotion recognition by speech: a case study in clinical simulation, J. Ambient Intell. Humaniz. Comput. (2019). https://doi.org/10.1007/s12652-019-01280-8.

[439] E. Kanjo, E.M.G. Younis, C.S. Ang, Deep learning analysis of mobile physiological, environmental and location sensor data for emotion detection, Inf. Fusion. 49 (2019) 46–56. https://doi.org/10.1016/j.inffus.2018.09.001.

[440] D.H. Kim, M.K. Lee, D.Y. Choi, B.C. Song, Multi-modal emotion recognition using semi-supervised learning and multiple neural networks in the wild, in: Proc. 19th ACM Int. Conf. Multimodal Interact., Association for Computing Machinery, New York, NY, USA, 2017: pp. 529–535. https://doi.org/10.1145/3136755.3143005.

[441] A. Tsiami, P. Koutras, N. Efthymiou, P.P. Filntisis, G. Potamianos, P. Maragos, Multi3: Multi-Sensory Perception System for Multi-Modal Child Interaction with Multiple Robots, in: 2018 IEEE Int. Conf. Robot. Autom. ICRA, 2018: pp. 4585–4592. https://doi.org/10.1109/ICRA.2018.8461210.